\documentclass[aps,prx,reprint,superscriptaddress,nofootinbib,longbibliography,floatfix]{revtex4-2}

\usepackage{amsmath,amssymb,mathtools}
\usepackage{amsthm}

\usepackage{booktabs,multirow} %
\usepackage{microtype}
\usepackage{braket}
\usepackage{xcolor}
\usepackage{graphicx}
\usepackage{capt-of}
\usepackage{quantikz}
\usepackage[colorlinks=true,linkcolor=blue,citecolor=blue,urlcolor=blue,filecolor=blue]{hyperref}
\usepackage[nameinlink]{cleveref}

\crefname{section}{Sec.}{Secs.}%
\Crefname{section}{Sec.}{Secs.}%
\crefname{subsection}{Sec.}{Secs.}%
\Crefname{subsection}{Sec.}{Secs.}%
\crefname{subsubsection}{Sec.}{Secs.}%
\Crefname{subsubsection}{Sec.}{Secs.}%
\crefname{appendix}{App.}{Apps.}%
\Crefname{appendix}{App.}{Apps.}%
\crefname{theorem}{Thm.}{Thms.}%
\Crefname{theorem}{Thm.}{Thms.}%
\crefname{equation}{Eq.}{Eqs.}%
\Crefname{equation}{Eq.}{Eqs.}%
\crefname{figure}{Fig.}{Figs.}%
\Crefname{figure}{Fig.}{Figs.}%
\crefname{table}{Tab.}{Tabs.}%
\Crefname{table}{Tab.}{Tabs.}%
\newskip\classtextfloatsep
\newcommand{\relaxfloatfractions}{\let\classtopfraction\topfraction\let\classtextfraction\textfraction\def\topfraction{.99}\def\textfraction{.01}\classtextfloatsep=\textfloatsep\textfloatsep=16pt\relax}
\newcommand{\restorefloatfractions}{\let\topfraction\classtopfraction\let\textfraction\classtextfraction\textfloatsep=\classtextfloatsep}

\newcommand{\blk}[1]{\ensuremath{\mathsf{#1}}}           %
\newcommand{\wb}{\ensuremath{\omega}}                    %
\newcommand{\carry}[1]{\ensuremath{\gamma_{#1}}}         %
\newcommand{\rep}{\ensuremath{\beta}}                    %

\newcommand{\Ttot}{\ensuremath{\mathrm{T}}}                          %
\newcommand{\Tsel}{\ensuremath{\mathrm{T}_{\mathrm{sel}}}}            %
\newcommand{\Tact}{\ensuremath{\mathrm{T}_{\mathrm{act}}}}

\newcommand{\Tref}{\ensuremath{\mathrm{T}_{\mathrm{ref}}}}
\newcommand{\Tperm}{\ensuremath{\mathrm{T}_{\mathrm{perm}}}}
\newcommand{\Tsynth}{\ensuremath{\mathrm{T}_{\mathrm{synth}}}}

\newcommand{\Trot}{\ensuremath{\mathrm{T}_{\mathrm{rot}}}}              %
\newcommand{\Apeak}{\ensuremath{\mathrm{A}_{\mathrm{peak}}}}
\newcommand{\Api}{\ensuremath{\mathrm{A}_{\mathsf{\Pi}}}}             %
\newcommand{\Nopen}{\ensuremath{\mathrm{N}_{\mathrm{open}}}}
\newcommand{\Nrep}{\ensuremath{\mathrm{N}_{\mathrm{rep}}}}
\newcommand{\Nrot}{\ensuremath{\mathrm{N}_{\mathrm{rot}}}}
\newcommand{\Nbr}{\ensuremath{\mathrm{N}_{\mathrm{br}}}}              %
\newcommand{\Npred}{\ensuremath{\mathrm{N}_{\mathrm{pred}}}}          %
\newcommand{\Npre}{\ensuremath{\mathrm{N}_{\mathrm{pre}}}}            %
\newcommand{\Sw}{\ensuremath{\mathrm{S}}}                             %
\newcommand{\epssynth}{\ensuremath{\varepsilon_{\mathrm{synth}}}}
\newcommand{\Lnorm}{\ensuremath{\Lambda}}                             %
\newcommand{\Rrows}{\ensuremath{R}}                                   %

\newcommand{\onb}{\ensuremath{\mathsf{B}}}      %
\newcommand{\cld}{\ensuremath{\mathrm{cl}}}     %
\newcommand{\jev}{\ensuremath{j^{(\mathrm{e})}}}   %
\newcommand{\jod}{\ensuremath{j^{(\mathrm{o})}}}   %
\newcommand{\sgn}{\ensuremath{\varsigma}}        %

\newcommand{\Perm}{\ensuremath{\mathsf{\Pi}}}                         %
\newcommand{\PermC}{\ensuremath{\widetilde{\mathsf{\Pi}}}}            %
\newcommand{\Pact}{\blk{P}}                                           %
\newcommand{\PH}{\ensuremath{\mathsf{P}_{\mathrm{H}}}}                %
\newcommand{\Pb}{\ensuremath{\mathsf{P}_{\onb}}}                      %
\newcommand{\Pup}{\ensuremath{\mathsf{P}_{\uparrow}}}
\newcommand{\Pdn}{\ensuremath{\mathsf{P}_{\downarrow}}}
\newcommand{\Tog}{\blk{Tog}}                                          %

\newcommand{\Unif}[1]{\ensuremath{\mathsf{Unif}_{#1}}}

\newcommand{\PREPARE}{\textsc{Prepare}}
\newcommand{\SELECT}{\textsc{Select}}
\newcommand{\AND}{\ensuremath{\mathrm{AND}}}
\newcommand{\CNOT}{\ensuremath{\mathrm{CNOT}}}

\newcommand{\CZ}{\ensuremath{\mathrm{CZ}}}
\newcommand{\SWAP}{\ensuremath{\mathrm{SWAP}}}
\newcommand{\CSWAP}{\ensuremath{\mathrm{CSWAP}}}
\newcommand{\iv}[1]{\ensuremath{[\,#1\,]}}                            %
\newcommand{\ketbra}[2]{\ensuremath{\ket{#1}\!\bra{#2}}}              %

\definecolor{CrayonYellow}{HTML}{fcb42c} %
\definecolor{CrayonBlue}{HTML}{253598}   %
\definecolor{CrayonRed}{HTML}{C51717}    %
\definecolor{CrayonGreen}{HTML}{5B9940}  %
\definecolor{CrayonViolet}{HTML}{6D2183} %
\definecolor{CrayonRose}{HTML}{F1558E}   %
\definecolor{CrayonYellowFill}{HTML}{faf3eb} %
\definecolor{CrayonBlueFill}{HTML}{F0EEF9}
\definecolor{CrayonRedFill}{HTML}{FBECE9}
\definecolor{CrayonGreenFill}{HTML}{EBF1E7}
\definecolor{CrayonVioletFill}{HTML}{F5EDF6}
\definecolor{CrayonRoseFill}{HTML}{FAECEF}
\definecolor{CrayonYellowWash}{HTML}{F5D9B4} %
\definecolor{CrayonBlueWash}{HTML}{DEDAF4}
\definecolor{CrayonRedWash}{HTML}{F9D5CD}
\definecolor{CrayonGreenWash}{HTML}{D2E1C9}
\definecolor{CrayonVioletWash}{HTML}{EAD7EE}
\definecolor{CrayonRoseWash}{HTML}{F8D4DD}
\newcommand{\crayontitle}[2]{\tikz[baseline=(crayon.base)]{%
  \node[draw=#1,fill=#1Fill,rounded corners=2.5pt,inner xsep=4.5pt,%
        inner ysep=2.5pt,line width=0.5pt](crayon){#2};}}
\newcommand{\crayonswatch}[1]{\tikz[baseline=-0.3em]{%
  \node[draw=#1,fill=#1Fill,circle,minimum size=8pt,inner sep=0pt,%
        line width=0.5pt]{};}}

\newcommand{\plotkey}[2]{\tikz[baseline=-0.3em,x=3.6pt,y=3.6pt,line cap=round,line join=round]{%
  \draw[#1,line width=1.1pt](-1.16,0)--(1.16,0);%
  \filldraw[#1,line width=0.7pt]#2;}}
\newcommand{\plotmarkercircle}{(0,0) circle[radius=0.5]}                         %
\newcommand{\plotmarkersquare}{(-0.5,-0.5) rectangle (0.5,0.5)}                   %
\newcommand{\plotmarkerdiamond}{(0,-0.7071)--(0.7071,0)--(0,0.7071)--(-0.7071,0)--cycle} %
\newcommand{\plotmarkertriangle}{(0,0.5)--(-0.5,-0.5)--(0.5,-0.5)--cycle}         %
\newcommand{\plotmarkerplus}{(-1/6,-0.5)--(1/6,-0.5)--(1/6,-1/6)--(0.5,-1/6)--(0.5,1/6)--(1/6,1/6)--(1/6,0.5)--(-1/6,0.5)--(-1/6,1/6)--(-0.5,1/6)--(-0.5,-1/6)--(-1/6,-1/6)--cycle} %

\newcommand{\tocheck}[1]{#1}
\newcommand{\newedit}[1]{#1}

\begin{document}

\title{Efficient Block Encoding of Structured Hamiltonians by Separating Where and What}

\author{Alessandro Summer}
\email{alessandro.summer@alice-bob.com}
\affiliation{Alice \& Bob, 53 Bd du Général Martial Valin, 75015 Paris, France}

\author{François Jamet}
\email{francois.jamet@alice-bob.com}
\affiliation{Alice \& Bob, 53 Bd du Général Martial Valin, 75015 Paris, France}

\date{\today}

\begin{abstract}
\newedit{In fault-tolerant block encodings of structured Hamiltonians, the support geometry can set how the non-Clifford cost of \SELECT{} scales with the system size, however many operators act on each support.
We construct permute--act--unpermute circuits: a \CSWAP{} network fixed by this geometry brings a selected support to a fixed target register, a shared local circuit acts there, and the permutation is undone.
The operators add only an $XZ$ mask on this register, at a cost independent of the system size, and one query can mix several support geometries.}
For two-site supports, we establish and attain the minimum \CSWAP{} count within address-controlled networks of site transpositions.
Compiled for Pauli terms, these circuits use \newedit{$8\Sw N+o(N)$} $T$ gates with $O(\log N)$ permutation work qubits, where $N$ is the number of system qubits and $\Sw$ is determined by the support geometry, with $\Sw=1$ for nearest-neighbour interactions and $\Sw=3/2$ for all-to-all pairs\newedit{; in both cases the correction is $O(\log N)$}.
We also introduce a bridged \CSWAP{}, which remembers the logical value of its temporary \AND{} across the target action and repairs it from the prepared labels, so that the inverse \CSWAP{} undoes the exchange without recomputing the value.
When the repair is Clifford, it halves the $T$ count of a matched forward--inverse \CSWAP{} pair at the cost of one retained work qubit.
For a Heisenberg ring, the complete block-encoding query reduces the $T$ count by a factor approaching $3$ as the system size grows\newedit{, relative to flat and product-wise Pauli walks}.
For a five-orbital Anderson impurity model, which combines nearest-neighbour and all-to-all support geometries, the reduction is about $1.7$ at large bath sizes\newedit{, relative to the cheapest of three comparators, each built from the methods of the references it follows}.
\end{abstract}

\maketitle

\section{Introduction}
\label{sec:intro}

\relaxfloatfractions %
\suppressfloats[t]
\begin{figure}[t]
\centering
\includegraphics[width=\columnwidth]{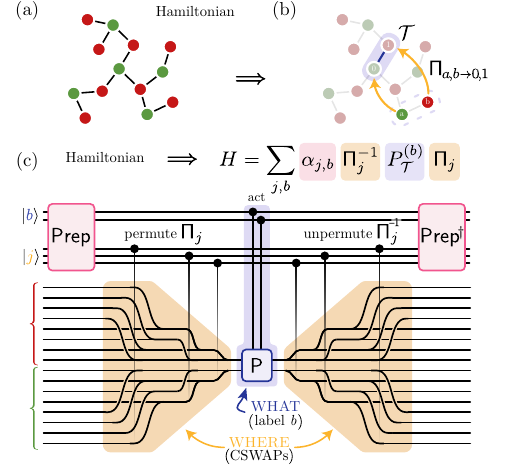}
\vspace{-.6cm}
\caption{
\textbf{Permute--act--unpermute block encoding of a structured Hamiltonian.}
(a)~\newedit{The sites of a Hamiltonian are coloured} into groups (red and green); no support holds two sites of one group. (b)~$\Perm_j$ brings a selected support to $\mathcal T$. (c)~Circuit, with one swap network per group.
\PREPARE{} encodes the support address $j$ and operator label $b$.
The address-controlled permutation $\Perm_j$ brings the selected support
to $\mathcal T$, where the shared target action
$\Pact$ applies the operator labelled by $b$. Finally $\Perm_j^{-1}$
restores the system register.
Thus, the permutation selects \emph{where} the term acts, while the
target action selects \emph{what} is applied.
}
\label{fig:bridge-overview}
\end{figure}

Simulating quantum many-body systems has motivated quantum computation since Refs.~\cite{Feynman1982,Lloyd1996}, and it remains among the most promising applications of fault-tolerant machines. In the quantum singular value transformation of Ref.~\cite{Gilyen2019}, which generalises the quantum signal processing of Ref.~\cite{LowChuang2017QSP}, state preparation, time evolution and spectral functions, Refs.~\cite{AbramsLloyd1999,AspuruGuzik2005,Haah2018,JametAPLQuantum2025,Summer2024}, are all obtained as polynomials of a Hamiltonian block-encoded as in Ref.~\cite{LowChuang2019}, so the non-Clifford cost of the block encoding is paid at every query. In the linear-combination-of-unitaries (LCU) approach of Refs.~\cite{ChildsWiebe2012,Berry2014Sparse,Berry2015LCU,ChakrabortyGilyenJeffery2019,MartynRossiTanChuang2021}, \PREPARE{} loads the normalised term weights into a label register and \SELECT{} applies the unitary labelled there. Generic constructions, such as the unary iteration of Ref.~\cite{Babbush2018}, treat the terms as a flat list and pay a $T$ count proportional to their number. \newedit{The lattice-gauge block encodings of Refs.~\cite{RajputRoggeroWiebe2022,RhodesKreshchukPathak2024,SakamotoMorisakiHarunaItouFujiiMitarai2024,KanLemieuxOkrutSahinoglu2025} instead iterate over the sites once per term type, so \SELECT{} is linear in the number of sites for a fixed number of types.}

Many Hamiltonians of interest have more structure than a list of their terms. A small repertoire of local operators, such as fields, hopping or exchange terms, plaquette operators and density--density interactions, recurs over many supports arranged by a geometry. Examples are the nearest-neighbour or finite-range bonds of the lattice models of Refs.~\cite{KanSymons2025,ChouDasSarma2026}, the plaquettes of the lattice gauge theory of Ref.~\cite{Draper2026}, and the all-to-all power-law couplings of the Rydberg-atom arrays reviewed in Ref.~\cite{Browaeys2020}. Many models combine several such families, as an impurity model attaches bath chains to an all-to-all core. A flat list selects the support again for every operator that acts on it and writes the operator again for every support that carries it, so its cost grows with the product of the two.

We separate these two roles by writing each term as a local operator conjugated by a support-dependent permutation, $H=\sum_{j,b}\alpha_{j,b}\Perm_j^{-1}P^{(b)}_{\mathcal T}\Perm_j$. The address $j$ fixes \emph{where} the term acts, and the label $b$ fixes \emph{what} is applied on a fixed target register $\mathcal T$ of $n_{\mathcal T}$ qubits. The coefficients may depend on both. \SELECT{} then becomes permute--act--unpermute, as in \Cref{fig:bridge-overview}. The address-controlled permutation $\Perm_j$ brings the selected support to $\mathcal T$, and the operator labelled $b$ acts there. The inverse permutation then restores the system register. \\
Moving selected degrees of freedom to common work positions has precedents in real-space simulation, Ref.~\cite{Kivlichan2017RealSpace}, fermionic \SELECT{}, Ref.~\cite{Wan2021}, and first-quantised simulation, Refs.~\cite{SuBerryWiebeRubinBabbush2021,PocrnicLoaizaArrazolaWiebeMotlagh2026}. Related chemistry block encodings exchange the two spin sectors of the system register under the control of a single prepared bit, Refs.~\cite{Lee2021THC,Low2026DenserSurfaceCode}, while Ref.~\cite{Draper2026} expresses a lattice-gauge block encoding in this form.
What differs is the cost that the permutation is built to minimise. Refs.~\cite{Wan2021,SuBerryWiebeRubinBabbush2021,PocrnicLoaizaArrazolaWiebeMotlagh2026} select the sites of a term independently, with one swap network per site or register. Ref.~\cite{Wan2021} builds this network ancilla-free, at logarithmic depth, and Ref.~\cite{SuBerryWiebeRubinBabbush2021} uses it to run the arithmetic once, in ancilla registers. Here, instead, the permutation is built from as few \CSWAP{}s as the structure of the supports allows.

This split makes the costs of where and what add instead of multiplying. \newedit{\SELECT{} circuits whose $T$ count is linear in $N$ already exist, for spin terms, Ref.~\cite{YOSKM2024}, and for fermionic ones, Ref.~\cite{Wan2021}. In the product-wise walk of Ref.~\cite{YOSKM2024}, however, each operator class makes its own pass over the sites, so the coefficient of $N$ grows with the number of classes. Here the two costs have separate sources. The geometry of the supports fixes the permutation, the only stage that touches all $N$ system qubits, and thereby how the $T$ count of \SELECT{} scales with $N$. The permutation depends on the address alone, so every operator class reuses it. For the Pauli LCUs of this paper, the operator is an $XZ$ mask on $\mathcal T$, and further operator classes change only this mask, at a cost independent of $N$: an explicit mask adds $8n_{\mathcal T}\,T$ to \PREPARE{} however many classes there are. The register-wide mask of Ref.~\cite{DellaChiara2025} instead spans all $N$ qubits and needs $2N$ mask qubits. The number of terms enters the query only through the coefficients that \PREPARE{} loads. For the Heisenberg ring, six Pauli terms per site make flat and product-wise walks cost $24N$ $T$ gates at leading order, against $8N$ for the permutation. Families with different geometries, such as the bath chains and the all-to-all core of an impurity model, combine in one query, \Cref{sec:interface}. Each keeps its own permutation, and the prepared label makes every inactive family act as the identity, so their costs add.}

\newedit{We build the permutation} from the \CSWAP{} network of the \textsc{SelectSwap} lookup of Ref.~\cite{Low2024}. That lookup swaps classical data words into an output register. \newedit{Such lookups also enter recent low-$T$ block encodings, Ref.~\cite{LiuZhuLowLinYang2025}, and analyses of the optimal $T$ count of block encodings in the worst case over families of coefficients, Refs.~\cite{LiOuWangYaoYuanZhang2026,MaJovenLiu2026}.} Our networks move system qubits instead, so the support geometry\newedit{, not the coefficients,} sets their structure. \SELECT{} then costs \newedit{$8\Sw N+o(N)$} $T$ gates, where the \CSWAP{} density $\Sw$ depends on the geometry alone. It is $1$ for finite-range bonds and plaquettes in any dimension, and $3/2$ for all-to-all pairs\newedit{, and in these cases the correction is $O(\log N)$}. For pair supports, we prove that this \CSWAP{} count is the minimum over address-controlled networks of site transpositions. \newedit{For supports of any fixed size, any network that delivers every support can be controlled with $O(\log N)$ control qubits when \PREPARE{} writes one address per support, but the $T$ count of its controls depends on the form of that address. Delivering the selected qubits and computing the controls that deliver them are thus separate resource problems.} \newedit{For all-to-all pairs, selecting the two sites jointly, instead of with one swap network per site, also lowers the \CSWAP{} count. Replacing each \CSWAP{} by a controlled fermionic transposition permutes fermionic modes at unchanged $T$ count and ancilla count, because the Jordan--Wigner signs it adds need only \CZ{} gates.}

The matched forward and inverse permutations allow a second saving, in time rather than across terms. Each \CSWAP{} forms a temporary \AND{} of its control and the parity of its two qubits, at $4\,T$, as reconstructible from Refs.~\cite{Jones2013,Babbush2018,Gidney2018}. The inverse \CSWAP{} needs the same bit, and the usual schedule treats it as scratch: it is erased before the target action and recomputed from the system afterwards. \newedit{The bit can instead be kept across the action. When the prepared labels fix how the action changes the parity of the two qubits, they also fix the update that the kept bit needs, and this \emph{repair} replaces the recomputation.} We call the result a \emph{bridged} \CSWAP{}. It halves the cost of a matched pair when the repair is Clifford, at the price of one work bit kept live across the action.

Permute--act--unpermute, thus, is based on the compilation of the permutation from the structure of the Hamiltonian, and a \PREPARE{} that writes the address in the form the \CSWAP{} network consumes.

We apply this approach to two Hamiltonians to test the two ends of the structure geometry: the Heisenberg ring, with nearest-neighbour bonds, and the Anderson impurity model (AIM), which attaches bath chains to an all-to-all impurity core. As the ring grows, and as the bath grows at a fixed number of impurity orbitals, their complete queries asymptotically need \newedit{one third of the $T$ gates of flat and product-wise Pauli walks, and three fifths of those of the cheapest of three comparators, each built from the methods of the references it follows}, respectively, at comparable peak workspace.

\Cref{sec:permutation} builds \SELECT{}, the bridged \CSWAP{} and the resource law, and \Cref{sec:cases} applies them to the two models. \Cref{app:permutations} constructs the permutation networks and proves their \CSWAP{} bound, and \Cref{app:target-actions} gives the target actions and the coefficient sampler. \Cref{app:heisenberg,app:aim-complete-query} assemble and count the complete queries.

\restorefloatfractions
\section{Permute--act--unpermute \texorpdfstring{\SELECT{}}{SELECT}}
\label{sec:permutation}

This section constructs a \SELECT{} for structured Hamiltonians \newedit{in which the geometry of the supports sets how its $T$ count scales with $N$. Further operator classes on the same supports change only the target action, and families with different geometries use separate circuits whose costs add}. Three ideas achieve this. First, each term is split into \emph{where} it acts and \emph{what} it applies. Second, the geometry of the supports fixes the cost of \emph{where}. Third, a temporary value of the forward permutation is kept and repaired for the inverse permutation, instead of being recomputed.

The subsections follow the questions that this construction raises. \Cref{sec:where-what} explains why a flat list of terms is costly. It then splits each term into a permutation and a target action. Counting the resulting cost needs a unit. \Cref{sec:cswap-unit} fixes it as the temporary-\AND{} opening and prices a \CSWAP{} in these units. \Cref{sec:pau-resources} counts the \CSWAP{}s of the permutation. It shows that the geometry of the supports fixes their number. The \CSWAP{}s and the target action also need controls. \Cref{sec:interface} specifies what \PREPARE{} writes for them and what the remaining control logic costs. \Cref{sec:closed-cost} combines these pieces into the cost with every \CSWAP{} pair closed. This baseline pays two openings per pair. \Cref{sec:cswaps} removes one of them by repairing the work bit of each \CSWAP{} instead of recomputing it. The price is one work bit kept live per pair treated this way.

\subsection{Separating where from what}
\label{sec:where-what}

A Pauli LCU writes the Hamiltonian as
\begin{equation}
H=\sum_{j,b}\alpha_{j,b}P^{(b)}_j .
\label{eq:pauli-lcu}
\end{equation}
The index $j$ names a \emph{support}, a set of qubits on which terms act. The label $b$ names the local operator applied there, which may act on only some of these qubits. In the Heisenberg ring of \Cref{case:heisenberg}, for instance, $j$ names a bond, and a field term acts on one of its two sites. In a block encoding, \PREPARE{} writes the state $\sum_{j,b}\sqrt{|\alpha_{j,b}|/\Lnorm}\ket{j,b}$ on an ancillary register, with $\Lnorm=\sum_{j,b}|\alpha_{j,b}|$. \SELECT{} then applies the term named by that register,
\begin{equation}
\SELECT{} = \sum_{j,b} \ketbra{j,b}{j,b}\otimes P^{(b)}_j ,
\end{equation}
where the signs of the coefficients can be later supplied through just Clifford gates on the label register.

The cost of \SELECT{} is the problem this section addresses. A generic \SELECT{} treats the $L$ terms as a flat list, which leads to the unary iteration of Ref.~\cite{Babbush2018} at a cost of \newedit{$4(L-1)\,T$ gates for a controlled \SELECT{}, and, following Ref.~\cite{KhattarGidney2025}, $4(L-2)\,T$ gates for an uncontrolled one}. Here, the skew tree of Ref.~\cite{KhattarGidney2025} can lower the \newedit{uncontrolled} cost to $4(L-1-\log_2L)\,T$ gates for $L$ a power of two. Both costs grow with the number of terms. Yet in most Hamiltonians a few local operators recur on many supports, while the flat-list implementation pays a price that grows as the product of the number of operators and the number of supports.

We remove this repetition by writing every term as one fixed operator moved onto its support. A family of terms is \emph{structured} when it can be written as
\begin{equation}
H=\sum_{j,b}\alpha_{j,b}\,\Perm_j^{-1}P^{(b)}_{\mathcal T}\Perm_j .
\label{eq:structured-family}
\end{equation}
The \emph{target register} $\mathcal T$ is a fixed set of $n_{\mathcal T}$ system qubits. The permutation $\Perm_j$ brings support $j$ onto $\mathcal T$. The operator $P^{(b)}_{\mathcal T}$ then acts on $\mathcal T$ alone, whatever the support. The address $j$ thus carries \emph{where} a term acts, and the label $b$ carries \emph{what} it applies. The coefficients $\alpha_{j,b}$ may still depend on both.

For the Pauli LCUs of this paper, the target operator is a Pauli string on $\mathcal T$ up to a phase,
\begin{equation}
P^{(b)}_{\mathcal T}\propto\prod_{q\in\mathcal T}X_q^{\xi_q(b)}Z_q^{\zeta_q(b)} .
\label{eq:target-xz}
\end{equation}
Every Pauli string can be written in this form. The bits $\xi_q(b)$ and $\zeta_q(b)$ are the $X$ and $Z$ exponents on target qubit $q$, and they depend on the label alone. They play the role of the register-wide $XZ$ mask of Ref.~\cite{DellaChiara2025}. Here, however, their number is set by the size of the target register rather than of the system. \Cref{eq:structured-family} itself does not require a Pauli target. $P^{(b)}_{\mathcal T}$ may act on any number of qubits of $\mathcal T$ and need not be a Pauli string. The same target circuit then serves every support that carries the label $b$.

\Cref{eq:structured-family} turns \SELECT{} into three steps, shown in \Cref{fig:bridge-overview}. The permutation $\Perm_j$ brings the selected support to $\mathcal T$. The \emph{target action} $\Pact$ then applies $P^{(b)}_{\mathcal T}$ on $\mathcal T$. Finally, $\Perm_j^{-1}$ restores the system register. We call this construction permute--act--unpermute.

Its benefit is that the cost of \emph{where} and the cost of \emph{what} add instead of multiplying. Now, only the permutation touches all $N$ system qubits, and it depends on the $j$ alone, while the target action touches only $\mathcal T$. It depends on the label and, at most, on the low address bits that record where within $\mathcal T$ the support arrived. New operator classes on the same supports therefore change the target action but leave the permutation unchanged. \newedit{Families with different support geometries are combined as described in \Cref{sec:interface}.}

\subsection{Counting cost in temporary ANDs}
\label{sec:cswap-unit}

All non-Clifford gates of \SELECT{} can be written as temporary logical \AND{}s, so we count its cost in \emph{openings}. An opening writes the product of two bits into a clean ancilla at a cost of $4\,T$ gates, following Refs.~\cite{Jones2013,Gidney2018}. Closing that ancilla uses a measurement and a Clifford correction, at no $T$ cost. Following Ref.~\cite{Gidney2018}, every circuit draws a temporary \AND{} as a blank corner with two controls, both where it is opened and where it is closed.

Permutations are built from \CSWAP{}s, each needing one opening and having a control $c$ and data qubits $q_0,q_1$. A \CSWAP{} can be implemented starting with a $\CNOT_{q_1\to q_0}$ to store the parity $q_0\oplus q_1$ on $q_0$. Then, the opening writes the \emph{work bit}
\begin{equation}
\wb=c\,(q_0\oplus q_1).
\label{eq:work-bit}
\end{equation}
A $\CNOT_{\wb\to q_1}$ then exchanges the data when $c=1$. It leaves $q_0$ untouched, so $\wb$ still equals the product of its two controls and can be closed here, at no $T$ cost. A final $\CNOT_{q_1\to q_0}$ completes the exchange, which is phase exact, see \Cref{fig:bridge-pair}(a).

Every \CSWAP{} runs twice, once in $\Perm_j$ and once in $\Perm_j^{-1}$. The standard schedule closes the work bit inside the forward \CSWAP{}, as above, and reopens it for the inverse one. We call such a matched pair \emph{closed}. It costs two openings, or $8\,T$. \Cref{sec:cswaps} shows when one of the two openings can be saved.

\subsection{The support geometry fixes the permutation}
\label{sec:pau-resources}

The permutation is the only stage that touches every system qubit. Its number of \CSWAP{}s therefore sets the leading cost of \SELECT{}.

\begin{figure}[t]
\centering
\includegraphics{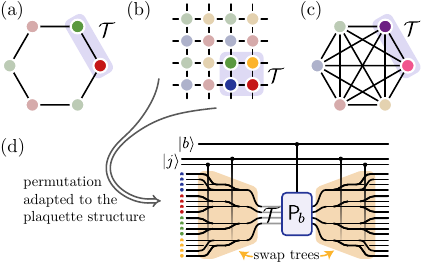}
\caption{\textbf{Site groups that fix the permutation.} Sites of one group share a colour, and one selected support is highlighted as the target register $\mathcal T$. (a)~Nearest-neighbour ring, $\chi=2$ and $n_{\mathcal T}=2$. (b)~Square-lattice plaquettes, $\chi=4$ and $n_{\mathcal T}=4$. (c)~All-to-all pairs, $\chi=N$ and $n_{\mathcal T}=2$, drawn for $N=6$. (d)~One swap tree per group gathers the selected site of that group at its root. For the four groups of (b), the roots form $\mathcal T$ itself, where $\Pact$ applies $P^{(b)}_{\mathcal T}$. When $\chi$ exceeds the support size, a second network moves the roots that hold the selected support onto $\mathcal T$.}
\label{fig:colour-trees}
\end{figure}

The basic building block is a \emph{swap tree}, the \CSWAP{} network of Ref.~\cite{Low2024}. A swap tree acts on $M$ sites, held at its \emph{leaves}, which are numbered by the address that selects them. With $M-1$ \CSWAP{}s, it brings the site at the selected leaf to leaf zero, the \emph{root} of the tree, so the root is one of the $M$ sites and not an extra qubit. The \CSWAP{}s are arranged in $\lceil\log_2M\rceil$ layers, one per address bit. Each layer exchanges the roots of pairs of \emph{subtrees}, which halves the number of subtree roots at which the selected site can sit. The \CSWAP{}s that act on the roots of the subtrees holding a leaf, one per layer, form the \emph{path} of that leaf, and the selected site travels along the path of its leaf. For $M$ a power of two, the layers thus hold $M/2+M/4+\dots+1=M-1$ \CSWAP{}s. For other $M$, the \CSWAP{}s whose upper leaf would lie beyond the $M$ leaves are omitted, and the count stays $M-1$. Fewer \CSWAP{}s cannot suffice. Run backwards from the root, each \CSWAP{} adds at most one site to those that can reach it. Every \CSWAP{} is controlled by a single address bit, as sketched in \Cref{fig:colour-trees}(d) and drawn in full in \Cref{fig:swap-tree}. If every support is a single site, $\mathcal T$ is one qubit and $\Perm_j$ is a single swap tree over all $N$ sites, with that qubit as its root. A two-site support needs one site on each qubit of $\mathcal T$, so the permutation must combine several trees such that all sites of the selected support arrive together.

The support geometry sets how many \CSWAP{}s the permutation needs, and the choice of trees can lower that number. For nearest-neighbour bonds, selecting the two ends of a bond independently uses one tree per end. A first tree over all $N$ sites brings site $j$ to $q_0$. A second tree over all sites brings site $j+1$ to $q_1$. Every site then sits in both trees, and $\Perm_j$ holds about $2N$ \CSWAP{}s. The bond structure makes half of this work unnecessary. Every bond joins an even site to an odd site. A tree over the even sites alone, rooted at $q_0$, can therefore bring the even end of the bond there. A tree over the odd sites alone, rooted at $q_1$, brings the odd end there. Each site now belongs to a single tree, and the two trees hold $N-2$ \CSWAP{}s, about half as many as before.

The only price is that the two ends of the bond can arrive swapped. Which end lands on $q_0$ now depends on the parity of $j$. More generally, a network brings the sites of the selected support to $\mathcal T$, but which site lands on which qubit of $\mathcal T$ may depend on the address. The \CSWAP{} counts of this subsection leave that assignment free. An exchange on $\mathcal T$ controlled by address bits can fix it. Alternatively, the target action can read the low address bits that record where each site landed and act on the right qubits directly, as the Anderson impurity model does in \Cref{case:aim}. Either way the extra cost acts on $\mathcal T$ alone and does not grow with $N$.

The same idea applies whenever the sites split into groups such that no support holds two sites of the same group. Let $\chi$ be the least number of such groups. \newedit{The groups depend only on which sites share a support, not on the coefficients or on a lattice, so the \CSWAP{} counts below hold for any family of supports. Once the groups are fixed, the rest of the structure of the family enters through the controls of these \CSWAP{}s, as discussed below.} One tree per group again costs about one \CSWAP{} per site. A second network then moves the tree roots that hold the selected support onto $\mathcal T$ as in \Cref{fig:colour-trees}. It is small when $\chi$ is bounded, and it becomes the whole permutation at $\chi=N$, as shown below. For pair supports the total is
\begin{equation}
\Bigl\lceil N+\frac{\chi}{2}-3\Bigr\rceil\ \text{\CSWAP{}s}.
\label{eq:pau-chi-switches}
\end{equation}
\Cref{app:cswap-count} proves that no network of address-controlled \CSWAP{}s does better. 

We write this count as \newedit{$\Sw N+o(N)$}, with $\Sw$ the \emph{\CSWAP{} density}. For nearest-neighbour bonds, $\chi=2$ and \Cref{eq:pau-chi-switches} gives the $N-2$ \CSWAP{}s of the two trees, so $\Sw_{\mathrm{nn}}=1$. Every $\chi$ that does not grow with $N$ gives the same density, because the network on the roots then holds a number of \CSWAP{}s independent of $N$. Finite-range bonds in any fixed dimension are of this kind, but locality is not required: bonds of any length between the two halves of a bipartite graph still give $\chi=2$. All-to-all pairs are the opposite extreme. Any two sites may form a support, so every group holds a single site and $\chi=N$. The network on the roots is then the whole permutation. The sites are paired off, and each pair gives one leaf to a swap tree rooted at $q_0$ and one to a swap tree rooted at $q_1$. One orientation \CSWAP{} per pair, applied first, decides which of its sites goes to which tree, so that the lower selected site reaches $q_0$. This network gives $\Sw_{\mathrm{aa}}=3/2$. 
\Cref{app:aa-network} analyses this structure more thoroughly.  Selecting the two sites with two swap trees over all $N$ sites would need about $2N$ \CSWAP{}s, against about $3N/2$ for the pair network. The two case studies of \Cref{sec:cases} sit at these two extremes. These densities are for pairs. For supports of $n_{\mathcal T}$ sites, \Cref{app:cswap-count} brackets the minimum between its lower bound and an explicit network. This bracket gives $\Sw=1$ whenever $\chi$ does not grow with $N$. At the all-to-all extreme, where every set of $n_{\mathcal T}$ sites is a support and $\chi=N$, the density lies between $3/2$ and $n_{\mathcal T}$ itself. The \CSWAP{} count thus stays linear in $N$, although the number of maximum supports grows as $N^{n_{\mathcal T}}$. The upper end is reached by $n_{\mathcal T}$ swap trees in sequence, each of which brings one selected site to $\mathcal T$. Each tree moves the other selected sites, whose leaves must then be recomputed. \Cref{app:sequential-tree-controls} computes them from $O(\log N)$ nonlinear products so the \CSWAP{}s still set the cost. \newedit{For three-site supports it lies at the lower end. A network that first sorts the selected sites into distinct bins reaches $\Sw=3/2$, and the minimum exceeds $3N/2$ by $\sqrt N/2+O(N^{1/4})$ \CSWAP{}s, \Cref{app:all-triples}. For four or more sites at $\chi=N$, the density remains open.} \newedit{Any network that delivers every support, a minimum one included, can be controlled coherently with $O(\log N)$ control qubits for supports of fixed size with one address each, but its number of control openings can grow with the number of supports, \Cref{app:generic-controls}. The logarithmic opening counts of our networks follow from the address structure that \PREPARE{} writes, \Cref{sec:interface}.}

\subsection{What \texorpdfstring{\PREPARE{}}{PREPARE} supplies to \texorpdfstring{\SELECT{}}{SELECT}}
\label{sec:interface}

The permutation and the target action read their controls from registers that \PREPARE{} writes. The form of these registers decides how much control logic \SELECT{} must still compute. We fix that form by three choices.

First, \PREPARE{} writes the address in the form that the network reads. A swap tree reads the leaf of the selected site, numbered within the tree of its group, not its index in the system. Converting one into the other may require a few products of address bits. A network that ranges over many tree roots may need further products, such as the retained products of all-to-all pairs, whose number \Cref{app:aa-network} establishes at every $N$. We call all these products \emph{nonlinear controls} and write $\Npred$ for their number. Each costs one opening. A \emph{control generator} $\mathsf G$ computes them first, and $\mathsf G^\dagger$ erases them at the end. 

Second, the target action reads the exponents of \Cref{eq:target-xz} from label bits. \PREPARE{} can write the label in two forms. A \emph{compressed} label is as short as the binary logarithm of the number of operator classes. An \emph{explicit mask} holds all $2n_{\mathcal T}$ exponents, which adds $8n_{\mathcal T}\,T$ to coefficient sampling. \Cref{app:target-repairs} traces this charge to coherent alias sampling, which spends one opening on each output bit. When each exponent is a label bit or a parity of label bits, it controls one \CNOT{} or one \CZ{} on a target qubit, and $\Pact$ is Clifford. An explicit mask always gives this case\newedit{; the phase of its $Y$ factors is loaded with it, \Cref{app:sign-diagonal}}. A compressed label may also need a few products of label bits, at a cost independent of $N$. Either way, the cost of \emph{what} does not grow with $N$.

Third, one block encoding can combine several structured families, each with its own permutation, target register and target action. The prepared label flags exactly one family in each branch and sets all exponents read by $\Pact$ to zero for the others. Each inactive family then acts as the identity: its target action is trivial, and its permutation and inverse cancel for any address. The family circuits can therefore run sequentially without outer controls, although their costs add because every branch executes all permutations.
This latter construction resembles the one of Ref.~\cite{LandahlMorrison2021}.
The Anderson impurity model of \Cref{case:aim} combines its bath chains and all-to-all impurity core in this way.

\subsection{Cost with closed \texorpdfstring{\CSWAP{}}{CSWAP} pairs}
\label{sec:closed-cost}

With every \CSWAP{} pair closed, the previous subsections give the cost of \SELECT{}. The \newedit{$\Sw N+o(N)$} \CSWAP{}s cost one opening in each direction. The $\Npred$ nonlinear controls cost one opening each. Hence
\begin{equation}
\begin{split}
\Tsel^{\cld} &= 8\Sw\,N + 4\,\Npred + o(N), \\
\Api^{\cld} &= \Npred+1 = O(\log N).
\end{split}
\label{eq:pau-closed-scaling}
\end{equation}
\newedit{When the correction to $\Sw N$ is $O(1)$, as for every row of \Cref{tab:pau-geometries}, $\Tsel^{\cld}=8\Sw N+O(\log N)$. The ancilla count assumes that every nonlinear product stays live from $\mathsf G$ until $\mathsf G^\dagger$ erases it. The bound $\Npred=O(\log N)$ holds for the networks of this paper, whose controls read the address in the form that \PREPARE{} writes, \Cref{sec:interface}; for an arbitrary network and address encoding, the number of control openings can grow with the number of supports, \Cref{app:generic-controls}.} Here $\Api^{\cld}$ counts the ancillae of the permutation. It excludes the address and label registers written by \PREPARE{}.

\begin{table*}[t]
\caption{\textbf{Resources with closed \CSWAP{} pairs for common geometries.}
Counts are for one uncontrolled \SELECT{} with an explicit $XZ$ mask, following \Cref{eq:pau-closed-scaling}. The mask is prepared by \PREPARE{}, so the target action is Clifford and contributes no $T$ gates. The results hold for $N=2^b$, and for side length $2^b$ on periodic lattices. The ring and all-to-all rows correspond to the geometries used in \Cref{case:heisenberg,case:aim}, while the remaining lattice constructions are given in \Cref{app:lattice-addresses}. At other sizes, borrow chains introduce additional openings and ancillae. \newedit{For all-to-all connectivity at any $N$, the count is $\lceil3(N-2)/2\rceil$ \CSWAP{}s and the $\Npre(N)$ retained products of \Cref{eq:aim-pair-network}.} \newedit{The Heisenberg query of \Cref{app:heisenberg-ring-complete-query} samples a three-bit compressed label instead. Its target action spends four openings, two on the bond exponents and two on the orientation exchanges, so its \SELECT{} costs $16\,T$ more than the ring row, which labels field terms by tree root, \Cref{app:lattice-addresses}, and leaves the mask to \PREPARE{}.} For fermionic modes, the same resource counts hold with controlled fermionic swaps because the Jordan--Wigner signs require only Clifford corrections, \Cref{eq:fermionic-transposition}.
}
\label{tab:pau-geometries}
\begin{ruledtabular}
\begin{tabular}{llcccccc}
Structure & Examples & $\chi$ & $\Sw$ & $n_{\mathcal T}$ & \CSWAP{}s & $\Tsel^{\cld}$ & $\Api^{\cld}$\\
\hline
Nearest-neighbour ring & Heisenberg, XXZ, transverse-field Ising & $2$ & $1$ & $2$ & $N-2$ & $8N+4\log_2N-24$ & $\log_2N-1$\\
Square lattice, fermionic & Fermi--Hubbard & $2$ & $1$ & $2$ & $N-2$ & $8N+4\log_2N-28$ & $\log_2N-2$\\
Square plaquettes & Wen plaquette model, ring exchange & $4$ & $1$ & $4$ & $N-4$ & $8N+4\log_2N-48$ & $\log_2N-3$\\
All-to-all pairs & General pair couplings, impurity interactions & $N$ & $\tfrac32$ & $2$ & $\tfrac32N-3$ & \newedit{$12N+4\log_2N-28$} & \newedit{$\log_2N$}\\
\end{tabular}
\end{ruledtabular}
\end{table*}

\Cref{tab:pau-geometries} evaluates \Cref{eq:pau-closed-scaling} for common structures. On a lattice, all swap trees share one address register per axis. \Cref{app:lattice-addresses} obtains this by shifting the leaves of the trees, so that each tree reads the shared field decremented by a prepared bit.

\newedit{The same networks permute fermionic modes under Jordan--Wigner. Each \CSWAP{} becomes a controlled fermionic transposition, whose sign has two parts: the exchange sign of two occupied modes, and the string sign of carrying one fermion past the occupied modes between them. Both are products of \CZ{} gates from the control and from the work bit $\wb$ that the \CSWAP{} already holds open, \Cref{eq:fermionic-transposition}. A transposition thus costs the $4\,T$ of its \CSWAP{} and no further qubit, and every row of \Cref{tab:pau-geometries} holds for fermionic modes with the same $T$ count and ancillae. On a lattice, a Jordan--Wigner order that places the tree roots next to each other leaves the target operators $XX$, $YY$, $Z$ and $ZZ$ without strings. In the all-to-all network, the strings of each tree layer span about $N$ modes, so one \CZ{} per mode would cost $O(N\log N)$ Clifford gates. A parity rule avoids this, in a Jordan--Wigner order that follows the leaves of the swap trees. Each tree \CSWAP{} acts after both of its subtrees have finished, so the modes between its two endpoints are the finished lower subtree and idle modes of the other tree. Its string parity is the parity of the lower subtree without its root XOR that of the idle modes between the endpoints, and \CNOT{}s keep both parities in place. Each string then takes at most three \CZ{} gates, and the signs of the whole network cost a number of Clifford gates linear in $N$, \Cref{app:aa-network}.}

For the model of Ref.~\cite{Hubbard1963}, the Hubbard row thus has the leading term $8N$ as achieved also in Ref.~\cite{LandahlMorrison2021} by reindexing the Hubbard terms by Majorana operators and iterating over every second site. All three use $O(\log N)$ ancillae.  For all-to-all pairs, the chemistry \SELECT{} of \newedit{Ref.~\cite{Babbush2018}}, fitted to the hopping and density terms of the plane-wave dual basis of Ref.~\cite{Babbush2018LowDepth}, costs $12N+O(\log N)$. There $\Sw_{\mathrm{aa}}=3/2$ gives the same leading term. That circuit, however, is built for the hopping and density terms, whereas the pair network reaches $12N$ for any Pauli operator on the pair. Its leading-order gain is over two independent swap trees on all $N$ sites, which would cost $16N$. 
With \newedit{controlled fermionic transpositions}, the all-to-all row also covers the quadratic fermionic model of Ref.~\cite{Wan2021}, where the published circuit costs $48(N-1)$ $T$ gates, so the pair network needs four times fewer. That circuit, however, uses no ancillae and has logarithmic $T$ depth, whereas the pair network reuses one work bit, so its \CSWAP{}s run in sequence.

These counts cover \SELECT{} and its workspace only. Coefficient loading in \PREPARE{} follows established constructions. \Cref{sec:cases} reports complete block encodings for the two case studies.

\begin{figure}[t]
\centering
\includegraphics[width=\columnwidth]{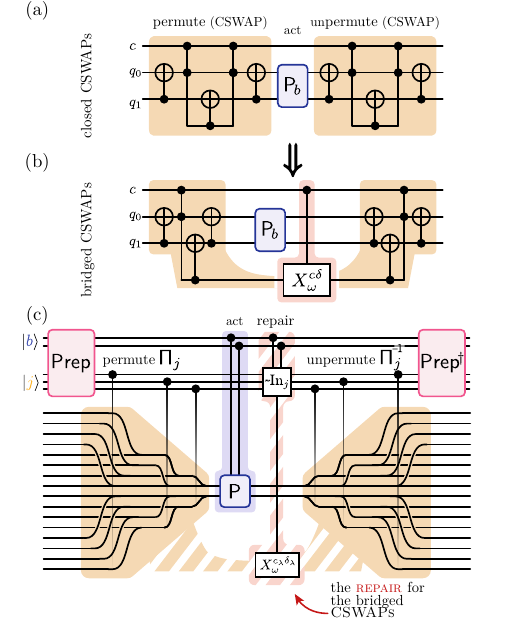}
\vspace{-.5cm}
\caption{
\textbf{Closed and bridged \CSWAP{} pairs.} (a)~In a closed pair, the temporary work bit is erased after the forward \CSWAP{} and recomputed for the inverse one, requiring two $4\,T$ openings.
(b)~In a bridged pair, the work bit is retained across the target action and updated by the repair $X_\wb^{c\delta}$ before being reused by the inverse \CSWAP{}. The pair therefore requires only one opening when the repair is Clifford.
(c)~Permute--act--unpermute \SELECT{} with bridged \CSWAP{}s. The forward permutation leaves its work bits live across the target action. The repair predicates, derived from the prepared address and label, update them before the inverse permutation consumes them.
}
\label{fig:bridge-pair}
\end{figure}

\subsection{The bridged \texorpdfstring{\CSWAP{}}{CSWAP} remembers and repairs its work bit}
\label{sec:cswaps}
A closed pair spends half of its cost on recomputation. The forward \CSWAP{} computes the work bit $\wb=c\,(q_0\oplus q_1)$ and closes it, and the inverse \CSWAP{} computes the same product again. \newedit{The inverse \CSWAP{} needs only the value of this product, not a fresh computation of it.} An exchange leaves the pair parity $q_0\oplus q_1$ unchanged, so after the forward \CSWAP{} the \newedit{work bit} is still valid. Between the two occurrences, only the target action can change that parity. Suppose that it flips the parity by a bit $\delta$. The inverse \CSWAP{} then requires
\begin{equation}
\wb'=c\,(q_0\oplus q_1\oplus\delta)=\wb\oplus c\,\delta .
\label{eq:repair-bit}
\end{equation}
The forward work bit can therefore stay live across the target action, and the \emph{repair} $X_\wb^{c\delta}$ updates its logical value before the inverse \CSWAP{}, in place of the reopening. When the repair is a Clifford gate, the pair costs one opening instead of two. We call this implementation a \emph{bridged} \CSWAP{}, \Cref{fig:bridge-pair}(b), drawn compactly as in \Cref{fig:aim-repair-application}(a).

The bridge requires only that $\delta$ be fixed by the prepared registers, not by the system state. Every Pauli target satisfies this condition, because only the $X$ exponents of \Cref{eq:target-xz} flip data bits, and they depend on the label alone. For a \CSWAP{} next to the target action, $\delta$ is the parity of the $X$ exponents that land on its two qubits. A $Z$ factor flips no bit, so a diagonal target needs no repair at all. An operator that superposes the two parity sectors has no fixed $\delta$, and its \CSWAP{}s keep the closed schedule.

Inside a network, other \CSWAP{}s run between the two occurrences of a given \CSWAP{} $\lambda$. They carry the $X$ exponents of the target back onto the system, so the parity change $\delta_\lambda(j,b)$ seen by $\lambda$ depends on the address as well as on the label. So does the control $c_\lambda$ of $\lambda$. The work bit of $\lambda$ is then repaired by $X_\wb^{c_\lambda\delta_\lambda}$. We call $c_\lambda\delta_\lambda$ the \emph{repair predicate} of $\lambda$, and the list of these predicates over a network its \emph{repair word}. \Cref{app:repair-count} evaluates them for Pauli targets. There $\delta_\lambda$ is the parity of the $X$-mask bits that land on the two positions of $\lambda$, which fixes how many repairs need an opening.

A repair predicate that is affine in the prepared bits is a parity, applied by \CNOT{}s at no $T$ cost. Otherwise the control generator $\mathsf G$ computes it. A predicate of higher degree costs a single opening only when it can reuse products computed earlier, as the repair word of the Anderson impurity model does. \Cref{app:bridge-instances} builds that word from nine such products, each at one opening, and obtains their negated address factors as free parities. We write $\Nrep$ for these openings. Several work bits that share a repair predicate receive it by \CNOT{} fan-out. This gives the rule applied in \Cref{sec:cases}. A \CSWAP{} is bridged whenever repairing its work bit costs less than reopening it. A Clifford repair always passes this test. A repair that needs an opening of its own passes it only when several work bits share it.

The bridged \CSWAP{} extends earlier reuse of temporary values. Refs.~\cite{AmyRoss2021,KhattarGidney2025} retain temporary values and exploit logically constrained ancillae. 
Automatic uncomputation compilers~\cite{Ding2020SQUARE,Paradis2021Unqomp,Paradis2024Reqomp,SharmaAchour2025SPARE} can decide which temporary values to erase, which to compute again later, or which can be merged, but not which to repair.
\newedit{The bridge instead keeps the work bit through the target action, which changes its inputs, derives its update from the prepared labels, and applies that update before the inverse \CSWAP{} reads it.} The adjacent-\AND{} merge of Ref.~\cite{Babbush2018} gives the same saving when the two openings can be brought next to each other. \Cref{eq:repair-bit} removes that requirement: the forward work bit is held across the target action, which does not use it, and it serves only the inverse \CSWAP{} that undoes the permutation. \Cref{app:and-merge} makes the relation explicit, since the merged pair leaves exactly the repair of \Cref{eq:repair-bit}.

Bridging trades $T$ gates for ancillae. Let $\Nbr$ matched pairs be bridged, and let their repairs need $\Nrep$ additional openings. Then
\begin{equation}
\begin{split}
\Tsel &= 8\,\Sw N - 4\bigl(\Nbr-\Nrep\bigr)\\
&\quad+ 4\Npred + \newedit{o(N)},\\
\Apeak &= \Nbr+\Nrep+\Npred+O(1).
\end{split}
\label{eq:pau-pareto}
\end{equation}
Each bridged pair saves $4\,T$ and keeps one work bit live across the action. Each repair opening gives $4\,T$ back, and its predicate stays live from $\mathsf G$ until it is applied. When every repair is Clifford, bridging all pairs halves the leading term to $4\Sw N$. The price is $\Sw N$ live work bits. Closed and bridged pairs can also be mixed to fit the available workspace. \newedit{\Cref{eq:pau-pareto} is thus a trade-off between $T$ gates and width, and the case studies of \Cref{sec:cases} sit near its low-width end.}
\section{Two Hamiltonian families instantiate the framework}
\label{sec:cases}

The construction of \Cref{sec:permutation} is instantiated for two models with complementary support geometries. In each case, we express the Hamiltonian as a structured family, identify its geometry, specify the information supplied by \PREPARE{}, determine which \CSWAP{}s can be bridged, and count the resulting resources.

The Heisenberg ring with even $N\geq4$ isolates the separation of operator class from support: its bonds have $\chi=2$, so two swap trees suffice, and no bridging is required. The Anderson impurity model combines bath chains with $\chi=2$ and an all-to-all impurity core in a single block encoding. Its one-body sector has both within-chain and across-chain structure, allowing repairs to be shared. It also exhibits the two useful bridging regimes of \Cref{sec:cswaps}: a shared repair and a diagonal action requiring no repair.

\subsection{Heisenberg ring}
\label{case:heisenberg}

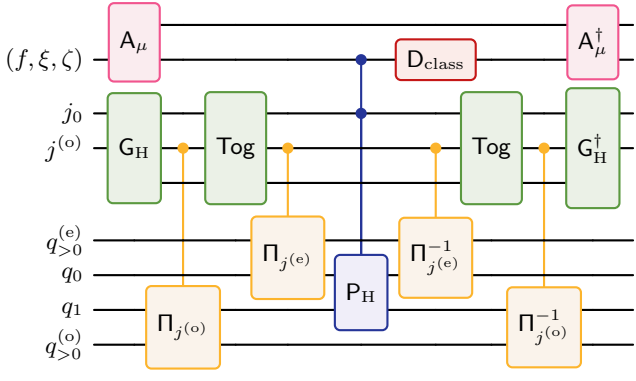
\begin{figure}[t]
\centering

\begin{quantikz}[row sep={0.46cm,between origins}, column sep=0.18cm]
& \gate[wires=2,style={draw=CrayonRose,fill=CrayonRoseFill,rounded corners=2pt}]{\mathsf A_\mu} \setwiretype{n} &[-0.375cm]\setwiretype{q}  &[-0.375cm]  &[-0.375cm]  &[-0.04cm] &[-0.07cm] &[-0.375cm] &[-0.375cm] &[-0.375cm] \gate[wires=2,style={draw=CrayonRose,fill=CrayonRoseFill,rounded corners=2pt, inner ysep=-2pt}]{\mathsf A_\mu^\dagger} & \\
\lstick{$(f,\xi,\zeta)$} & & & & & \control[style={draw=CrayonBlue,fill=CrayonBlue}]{}\wire[d][5][CrayonBlue]{q} & \gate[style={draw=CrayonRed,fill=CrayonRedFill,rounded corners=2pt}]{\mathsf D_\mathrm{class}} & & & & \\[0.25cm]
\lstick{$j_0$} & \gate[wires=3,style={draw=CrayonGreen,fill=CrayonGreenFill,rounded corners=2pt}]{\mathsf G_{\mathrm H}} & & \gate[wires=3,style={draw=CrayonGreen,fill=CrayonGreenFill,rounded corners=2pt}]{\Tog} & & \control[style={draw=CrayonBlue,fill=CrayonBlue}]{}\wire[d][4][CrayonBlue]{q} & & \gate[wires=3,style={draw=CrayonGreen,fill=CrayonGreenFill,rounded corners=2pt}]{\Tog} & & \gate[wires=3,style={draw=CrayonGreen,fill=CrayonGreenFill,rounded corners=2pt}]{\mathsf G_{\mathrm H}^\dagger} & \\
\lstick{$j^{(\mathrm o)}$} & & \control[style={draw=CrayonYellow,fill=CrayonYellow}]{}\wire[d][5][CrayonYellow]{q} & & \control[style={draw=CrayonYellow,fill=CrayonYellow}]{}\wire[d][3][CrayonYellow]{q} & & \control[style={draw=CrayonYellow,fill=CrayonYellow}]{}\wire[d][3][CrayonYellow]{q} & & \control[style={draw=CrayonYellow,fill=CrayonYellow}]{}\wire[d][5][CrayonYellow]{q} & & \\
& \setwiretype{n} & \setwiretype{q} & & & & & & & & \setwiretype{n} & \\[0.3cm]
\lstick{$q^{(\mathrm e)}_{>0}$} & & & & \gate[wires=2,style={draw=CrayonYellow,fill=CrayonYellowFill,rounded corners=2pt}]{\Perm_{\jev}} &  & \gate[wires=2,style={draw=CrayonYellow,fill=CrayonYellowFill,rounded corners=2pt, inner ysep=-2pt}]{\Perm_{\jev}^{-1}} & & & & \\
\lstick{$q_0$} & & & &  & \gate[wires=2,style={draw=CrayonBlue,fill=CrayonBlueFill,rounded corners=2pt}]{\PH} &  & & & & \\
\lstick{$q_1$} & & \gate[wires=2,style={draw=CrayonYellow,fill=CrayonYellowFill,rounded corners=2pt}]{\Perm_{\jod}} & & & & & & \gate[wires=2,style={draw=CrayonYellow,fill=CrayonYellowFill,rounded corners=2pt, inner ysep=-2pt}]{\Perm_{\jod}^{-1}} & & \\
\lstick{$q^{(\mathrm o)}_{>0}$} & & & & & & & & & &
\end{quantikz}
\vspace{-.3cm}
\caption{\textbf{Permute--act--unpermute block encoding of the Heisenberg ring for even $N$.}
The alias sampler $\mathsf A_\mu$ prepares the operator label $(f,\xi,\zeta)$, while the control generator $\mathsf G_{\mathrm H}$ prepares the position $j$ and the auxiliary controls consumed by the permutation. The odd and even swap trees, connected by the address toggle $\Tog$, bring the selected nearest-neighbour pair to the fixed target register $\mathcal T=(q_0,q_1)$. The target action $\PH$ applies the operator specified by the label, after which the permutation is reversed. The Clifford diagonal $\mathsf D_{\mathrm{class}}$ supplies the remaining Pauli phases and coefficient signs. \Cref{fig:ring-alias,fig:ring-native-register,fig:address-toggle,fig:swap-tree,fig:ring-label-action,fig:ring-class-diagonal} give the circuit decompositions of the corresponding blocks. All circuits share one colour code: \protect\crayonswatch{CrayonRose}\ pink for coefficients, \protect\crayonswatch{CrayonGreen}\ green for labels, \protect\crayonswatch{CrayonYellow}\ yellow for permutations, \protect\crayonswatch{CrayonBlue}\ blue for target actions, \protect\crayonswatch{CrayonViolet}\ violet for parity frames, and \protect\crayonswatch{CrayonRed}\ red for signs and repairs.
}
\label{fig:heisenberg-circuit}
\end{figure}

The Heisenberg chain is a minimal model for central questions of non-equilibrium many-body physics. It spans integrable and chaotic regimes and is a reference system for thermalisation and anomalous spin transport, Refs.~\cite{DAlessio2016,PilatowskyCameo2025,GopalakrishnanVasseur2023,Valli2025,Summer2026}; with random fields it becomes the canonical setting of many-body localisation, Refs.~\cite{Abanin2019,SierantZakrzewski2022}. These questions make it a standard target for quantum simulation and a testbed for fault-tolerant circuit optimisation, Refs.~\cite{Keenan2023,Childs2018,NamMaslov2019}. They concern highly excited states and long times, which classical methods reach poorly: exact diagonalisation is limited to modest sizes, Ref.~\cite{DAlessio2016}, and the entanglement generated after a quench confines matrix-product states to short times, Refs.~\cite{Schollwock2011,Paeckel2019}. For the present construction it also provides published block encodings for comparison, while its many terms per site occupy only two support shapes.

\begin{figure}
\centering
\includegraphics[width=\linewidth]{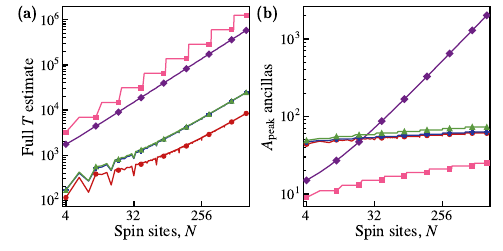}\par\smallskip
{\footnotesize\setlength{\tabcolsep}{3pt}%
\begin{tabular}{@{}l@{}rrr@{}}
\toprule
\textbf{(c)} & $\Ttot$ & $\Nrot$ & $\Apeak$\\
\midrule
\plotkey{CrayonRed}{\plotmarkercircle}\hspace{0.45em}This work & $8N+4\mu$ & $O(1)$ & $3\mu+2\log_2N$\\
\plotkey{CrayonBlue}{\plotmarkerplus}\hspace{0.45em}Flat walk & $24N+4\mu$ & $O(1)$ & $3\mu+2\log_2N$\\
\plotkey{CrayonGreen}{\plotmarkertriangle}\hspace{0.45em}Product-wise & $24N+4\mu$ & $O(1)$ & $3\mu+3\log_2N$\\
\plotkey{CrayonViolet}{\plotmarkerdiamond}\hspace{0.45em}Register mask$^\ast$ & $24N$ & $8N$ & $2N$\\
\plotkey{CrayonRose}{\plotmarkersquare}\hspace{0.45em}Tree \SELECT{} & $\tfrac{15}{2}\,2^{\lceil\log_2 6N\rceil}$ & $2^{\lceil\log_2 6N\rceil+1}$ & $2\log_2N$\\
\bottomrule
\end{tabular}}
\caption{\textbf{Resources for one uncontrolled block-encoding query of the Heisenberg ring at even $\mathbf N$.}
(a)~Total $T$ count, with the $\Nrot$ arbitrary rotations synthesised at $\epssynth=10^{-3}$ and alias precision $\mu=10$.
(b)~Peak ancilla count $\Apeak$, excluding the $N$ system qubits, with $\mu=10$.
(c)~Leading resource expressions, up to $O(\log N)$ corrections in $\Ttot$ and $O(1)$ corrections in $\Apeak$; the key beside each row identifies its curves in (a) and (b).
Curves in (a) and (b) evaluate the corresponding resource expressions at every even $N$; markers show exact gate counts for this work, the flat walk, and the product-wise walk at the sizes indicated, while the remaining constructions use the gate inventories of their sources.
The comparators are the flat walk of Ref.~\cite{Babbush2018}, the product-wise walk of Ref.~\cite{YOSKM2024}, the register mask of Ref.~\cite{DellaChiara2025}, and the tree \SELECT{} of Ref.~\cite{Childs2018}. The starred register-mask row retains the published open-chain Hamiltonian with $6N-3$ terms.
For the first three constructions, the dips at powers of two occur because preparing the uniform position state requires four arbitrary rotations away from powers of two and none at powers of two. The steps of the tree result from padding its $6N$ leaves to the next power of two.
}
\label{fig:heisenberg-comparison}
\end{figure}

We consider a ring of an even number $N\geq4$ of spins, with the Hamiltonian

\begin{equation}
\begin{split}
H &= \sum_{j=0}^{N-1} \bigl(g_X X_j+g_Y Y_j+g_Z Z_j \\
&\qquad\qquad +J_X X_jX_{j+1}+J_Y Y_jY_{j+1}+J_Z Z_jZ_{j+1}\bigr),
\end{split}
\label{eq:heisenberg-ring-hamiltonian}
\end{equation}
with site indices taken modulo $N$. There are six operator classes but only two support shapes. A field term sits on a single site, and a bond term on a nearest-neighbour pair.

We first cast the ring in the structured form of \Cref{eq:structured-family}, which separates the operator class, the target register and the permutation that transports the support. The label $b=(f,\xi,\zeta)$ specifies the operator class: $f=0$ denotes a field and $f=1$ a bond. The exponents $(\xi,\zeta)$ of \Cref{eq:target-xz} pick the axis, with $(0,1)$, $(1,0)$ and $(1,1)$ giving $Z$, $X$ and $Y$. The target register is $\mathcal T=(q_0,q_1)$. The permutation $\Perm_j$ brings site $j$ to $q_0$ and site $j+1$ to $q_1$. On the target, each class is then one fixed operator,

\begin{equation}
P^{(f\xi\zeta)}_{\mathcal T}=i^{(1+f)\xi\zeta}(X^{\xi}Z^{\zeta})_{q_0}(X^{\xi}Z^{\zeta})_{q_1}^{f}.
\end{equation}
The phase is there to turn $XZ$ into $Y$ on a site and $(XZ)(XZ)$ into $YY$ on a bond. With these pieces, \Cref{eq:heisenberg-ring-hamiltonian} becomes

\begin{equation}
H=\sum_{j=0}^{N-1}\sum_{f,\xi,\zeta}(-1)^{\sgn(f,\xi,\zeta)}\,W_{f\xi\zeta}\,\Perm_j^{-1}P^{(f\xi\zeta)}_{\mathcal T}\Perm_j.
\label{eq:H-heisenberg-block}
\end{equation}
Here $\sgn(f,\xi,\zeta)$ is the sign bit of the coupling of class $(f,\xi,\zeta)$, and $W_{f\xi\zeta}$ is its magnitude. The block encoding is normalised by $\Lnorm=N\sum_{f\xi\zeta}W_{f\xi\zeta}$. The permutation depends only on position, whereas the label determines the operator. A single $\Perm_j$ therefore serves all six classes.

The bond geometry then fixes how many \CSWAP{}s $\Perm_j$ needs. For even $N\geq4$, every bond joins an even site to an odd site, including the wraparound bond. The ring is therefore bipartite with $\chi=2$, with the even and the odd sites as its two groups. As in \Cref{sec:pau-resources}, each group gets one swap tree. The odd tree $\Perm_{\jod}$ brings the odd end of the bond to its root $q_1$. The even tree $\Perm_{\jev}$ brings the even end to its root $q_0$. Together they hold $N-2$ \CSWAP{}s. This is the minimum of \Cref{eq:pau-chi-switches} and the first row of \Cref{tab:pau-geometries}. The endpoint placed on $q_0$ depends on the parity of $j$. A $j_0$-controlled exchange of $q_0$ and $q_1$ puts the pair back in order. \Cref{fig:heisenberg-circuit} draws it inside $\PH$, but it completes $\Perm_j$, \Cref{sec:where-what}.

The two $j_0$-controlled exchanges that complete $\Perm_j$ cost $8\,T$.

What \PREPARE{} must supply follows from the coefficients. Each class has the same coefficient at every ring site, so \PREPARE{} splits into a class part and a position part. The class part is the coherent alias sampler $\mathsf A_\mu$ of Ref.~\cite{Babbush2018}. It loads only the six magnitudes $W_{f\xi\zeta}$ into the label, at precision $\mu$. The position part needs no table at all, since the position is uniform over the $N$ sites. \Cref{app:heisenberg-ring-complete-query} gives the alias circuit and padded magnitude table, and specifies the uniform position preparation $\Unif{N}$; \Cref{app:heisenberg-resource-comparison} counts its cost. The uniform-position block is drawn but not decomposed into a circuit there. The Hubbard \PREPARE{} of Ref.~\cite{Babbush2018} uses the same split, with three distinct coefficients spread uniformly over sites and spins.

The two trees want the address in two different forms. Storing both would cost a second register. Instead, we keep one register and switch it between the two forms with the \emph{toggle} $\Tog$, which \Cref{fig:heisenberg-circuit} places between the odd and the even tree. The control generator $\mathsf G_{\mathrm H}$ prepares the position $j$ and a word of borrows, which the toggle reads. These borrows are the nonlinear controls of the ring. The position is $j=(j_m\cdots j_1j_0)$, with $m=\lceil\log_2(N/2)\rceil$. The odd tree reads its high bits $\jod=\lfloor j/2\rfloor$ directly. The even tree holds site $2v+2$ at its leaf $v$, so that the wrap is the last link of the chain. It therefore needs $\jev=(\jod-\bar j_0)\bmod(N/2)=(\lceil j/2\rceil-1)\bmod(N/2)$. The toggle decrements the address only when $j_0=0$ and leaves it unchanged when $j_0=1$.

The toggle adds the borrow word to the register with \CNOT{}s. $\mathsf G_{\mathrm H}$ computes the borrows once, at $m-1$ openings, plus one when $N$ is not a power of two. They stay frozen while the register changes, so $\Tog$ is its own inverse.

\Cref{fig:heisenberg-circuit} puts these pieces together into \SELECT{} and the complete query, for even $N$. It reads from left to right, in the order the query runs. First, $\mathsf A_\mu$ loads the label, and $\mathsf G_{\mathrm H}$ prepares the position and the borrows. Then the odd tree, the toggle and the even tree bring the selected pair to $\mathcal T$. On the target, $\PH$ first orients the pair with the $j_0$-controlled exchange. It then applies $X^{\xi}Z^{\zeta}$ to site $j$ and, for a bond term, to site $j+1$. The compressed label of \Cref{sec:interface} uses three bits rather than an explicit four-bit mask. Accordingly, $\PH$ forms the bond exponents $f\xi$ and $f\zeta$ itself, as products of label bits. This target cost is independent of $N$.

The two products cost $8\,T$ and are opened sequentially on one recycled work qubit. An $S$ gate acts on each product while it is live inside $\PH$, supplying the phase $i^{f\xi+f\zeta}$ of \Cref{eq:ring-sequential-target-action}. The remaining phase and the coefficient signs form the label-only Clifford diagonal
\begin{equation}
\mathsf D_{\mathrm{class}}=\mathsf D_{\sgn}\bigl(-iS_\xi S_\zeta S_f^\dagger\bigr),
\end{equation}
where $\mathsf D_{\sgn}$ is the sign diagonal of \Cref{eq:ring-direct-sign-diagonal}. On the six populated labels,
\begin{equation}
i^{f(\xi+\zeta)}\bigl(-i\,i^{\xi+\zeta-f}\bigr)=i^{(1+f)\xi\zeta}.
\end{equation}
Thus the full Pauli-phase correction uses both the live products inside $\PH$ and the residual label diagonal drawn after it. The latter alone does not supply the full correction. \Cref{fig:ring-class-diagonal} realises this residual diagonal, including the coefficient signs, with $X$, $Z$, $S$, $S^\dagger$ and \CZ{} gates, without additional ancillae. The circuit reverses the trees and the toggle, after which $\mathsf G_{\mathrm H}^\dagger$ closes the stored products and the inverse loaders reverse preparation. 

We keep every \CSWAP{} pair of the ring trees closed. Each tree swap exchanges a single qubit, so an $X$ or $Y$ action can change its retained product and require a repair predicate. A tree \CSWAP{} sees that change only when it lies on the path of the selected leaf, \Cref{sec:pau-resources}, so its predicate is a different product of address bits for nearly every tree \CSWAP{}. Bridging the trees would therefore trade reopenings for repairs almost one for one, the same argument that keeps the AIM trees closed in \Cref{case:aim}. The resource counts below use this closed schedule, \newedit{so the reduction relative to the walks is obtained without bridging}.

We can now compare the complete query with published constructions. It costs $\Ttot=8N+4\mu+O(\log N)$. That is nearly three times fewer $T$ gates than \newedit{the flat and product-wise Pauli walks}, \newedit{with a smaller peak width than either walk at every plotted size}, as shown in \Cref{fig:heisenberg-comparison}. The ring comparators are evaluated as one uncontrolled call at the same normalisation, except that the register-mask row retains its published open-chain Hamiltonian. \Cref{app:heisenberg-resource-comparison} describes how each construction is adapted to \Cref{eq:heisenberg-ring-hamiltonian} and how exact $T$ gates and rotations are counted. \newedit{The two walks} are the flat Pauli walk of Ref.~\cite{Babbush2018} and the product-wise walk of Ref.~\cite{YOSKM2024}, both at \newedit{$24N+4\mu+O(\log N)$}. They use our sampler $\mathsf A_\mu$, so they differ from our query only in \SELECT{}.

The leading factor of three is explained by the per-term cost of a flat walk. A walk over the Pauli list pays $4\,T$ per term, and the ring has six terms per site. The two swap trees pay $8\,T$ per site, whatever acts there. Our permutation pays once per position, independently of the number of classes. The other two comparators are the tree \SELECT{} of Ref.~\cite{Childs2018} and the register mask of Ref.~\cite{DellaChiara2025}. Both move the non-Clifford cost into state preparation, where they pay $\Theta(N)$ arbitrary rotations. Once synthesised, their totals lie one to two orders of magnitude higher from $N=8$ on. The register mask is evaluated on the open chain for which it was published. It also needs $2N$ mask qubits, the only width in \Cref{fig:heisenberg-comparison}(b) that grows linearly.

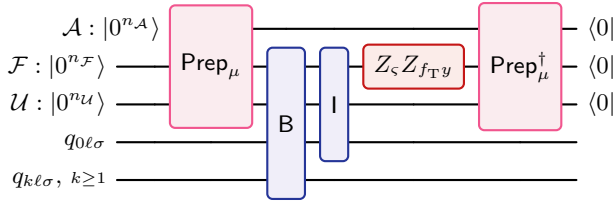
\begin{figure}[t]
\centering
{\small
\begin{quantikz}[row sep={0.5cm,between origins},column sep=0.2cm]
&[0.3cm] \setwiretype{n} & \gate[wires=3,style={draw=CrayonRose,fill=CrayonRoseFill,rounded corners=2pt}]{\mathsf {Prep}_\mu}\lstick{$\mathcal A:\ket{0^{n_{\mathcal A}}}$} & \setwiretype{q} &  &  & \gate[wires=3,style={draw=CrayonRose,fill=CrayonRoseFill,rounded corners=2pt}]{\mathsf {Prep}_\mu^\dagger} & \rstick{$\bra0$}  \\
\lstick{$\mathcal F:\ket{0^{n_{\mathcal F}}}$} &  &  & \gate[wires=4,style={draw=CrayonBlue,fill=CrayonBlueFill,rounded corners=2pt}]{\mathsf B} & \gate[wires=3,style={draw=CrayonBlue,fill=CrayonBlueFill,rounded corners=2pt}]{\mathsf I} & \gate[style={draw=CrayonRed,fill=CrayonRedFill,rounded corners=2pt}]{Z_{\sgn}Z_{f_\mathrm{T}y}} &  & \rstick{$\bra0$}  \\
\lstick{$\mathcal U:\ket{0^{n_{\mathcal U}}}$} &  &  &  &  &  &  & \rstick{$\bra0$}  \\
\lstick{$q_{0\ell\sigma}$} & &  &  &  &  &  &   \\
\lstick{$q_{k\ell\sigma},\,{\scriptstyle k\geq1}$} & &  &  &  &  &  &
\end{quantikz}}\par
\vspace{-.3cm}
\caption{\textbf{Complete block-encoding query for the AIM.}
The common preparation $\mathsf{Prep}_\mu$ prepares the term label $\mathcal F$ and its uniform-refinement register $\mathcal U$. Family flags in $\mathcal F$ select either the one-body circuit $\mathsf B$, expanded in \Cref{fig:aim-onebody-main-circuit}, or the interaction circuit $\mathsf I$, expanded in \Cref{fig:aim-interaction-main-circuit}. Each acts as the identity outside its own family, so the two run sequentially. Only $\mathsf B$ addresses the bath qubits. The Clifford sign corrections $Z_{\sgn}Z_{f_\mathrm{T}y}$ follow $\mathsf I$, and $\mathsf{Prep}_\mu^\dagger$ closes the query. \Cref{fig:aim-preparation-blocks} expands $\mathsf{Prep}_\mu$, whose alias sampler is the standard construction of Ref.~\cite{Babbush2018}.
}
\label{fig:aim-complete-query}
\end{figure}

\subsection{Anderson impurity model}
\label{case:aim}

The Anderson impurity model (AIM) describes interacting orbitals coupled to a non-interacting bath, Ref.~\cite{Anderson1961}. It is the auxiliary problem of dynamical mean-field theory, which maps a correlated lattice onto an impurity with a self-consistent bath, Refs.~\cite{Georges1996,Pavarini2025}. For real materials, the lattice model is itself derived from first principles, with localised orbitals and screened interactions, Refs.~\cite{Marzari2012,Aryasetiawan2004}. Accurate bath representations require many sites, beyond the reach of exact diagonalisation. The AIM has therefore been proposed as the quantum step of a classical embedding loop, Refs.~\cite{Bauer2016,Jamet2021}. Tensor-network solvers can also treat it efficiently, Ref.~\cite{Bauernfeind2017}, but rapid entanglement growth during real time evolution, particularly in multi-orbital systems, makes these problems promising candidates for quantum computing, Refs.~\cite{Bertrand_2025,JametAPLQuantum2025,Hogan_2026}.

The AIM combines the chain and all-to-all geometries of \Cref{tab:pau-geometries} in one Hamiltonian. Its one-body terms live on long nearest-neighbour chains. Its interactions, by contrast, couple every pair of impurity orbitals. \Cref{tab:cases-summary} sets the AIM beside the ring, step by step. Its one-body family is the ring on every chain at once, plus one stage that picks the chain. Its interaction family is an all-to-all network on the impurity orbitals. We first describe the common query, then treat its two operator families.

\begin{table*}[t]
\caption{The steps of \Cref{sec:permutation} in the two case studies. Each column is one structured family. The Heisenberg column assumes even $N\geq4$; its $N-2$ count is for the two trees, with the target-orientation exchange counted separately.}
\label{tab:cases-summary}
{\footnotesize\setlength{\tabcolsep}{5pt}%
\begin{tabular}{@{}llll@{}}
\toprule
& Heisenberg ring & AIM one-body & AIM interaction\\
\midrule
Address & site $j$ & site $k$, chain $(\ell,\sigma)$, flag $f_\mathrm{T}$ & orbital pair $(\ell,\ell')$\\
Label & field or bond, axis $(\xi,\zeta)$ & onsite or hopping, axis $y$ & density term or Hund word\\
Target register & $\mathcal T=(q_0,q_1)$ & $\mathcal T_\mathsf B$, two impurity qubits & $\mathcal T_\mathsf I$, four impurity qubits\\
Permutation & two swap trees & swap trees, then chain front $\PermC_{\ell\sigma}$ & all-to-all network of block \CSWAP{}s\\
\CSWAP{} count & $N-2$ & $N-2$ & all-to-all row of \Cref{tab:pau-geometries} at $N=n_\mathrm{imp}$\\
Bridging & none & chain front, shared repair & spin-down stage, no repair\\
Leading \SELECT{} cost & $8N$ & $8N$ & $O(n_\mathrm{imp})$\\
\bottomrule
\end{tabular}}
\end{table*}

We take the multi-orbital model of Ref.~\cite{JametAPLQuantum2025}. It has $n_\mathrm{imp}$ impurity orbitals $\ell=0,\ldots,n_\mathrm{imp}-1$, each with spin $\sigma\in\{\uparrow,\downarrow\}\equiv\{0,1\}$. Each spin orbital is coupled to its own non-interacting bath. A basis rotation turns each bath into a chain of sites $k=0,\ldots,n_\mathrm{bath}$, with the impurity at $k=0$. We write $c_{k\ell\sigma}$ for the fermionic mode of site $k$ on the chain of $(\ell,\sigma)$, and $n_{k\ell\sigma}=c^\dagger_{k\ell\sigma}c_{k\ell\sigma}$. The Hamiltonian reads

\begin{equation}
\begin{split}
H_\mathrm{AIM}={}&\sum_{\ell,\sigma}\sum_{k=0}^{n_\mathrm{bath}}a_{k\ell}\,n_{k\ell\sigma}\\ &+\sum_{\ell,\sigma}\sum_{k=0}^{n_\mathrm{bath}-1}t_{k\ell}\left(c^\dagger_{k\ell\sigma}c_{k+1,\ell\sigma}+\mathrm{h.c.}\right)\\
&+U\sum_\ell n_{0\ell\uparrow}n_{0\ell\downarrow}+(U-2J)\sum_{\ell\neq\ell'}n_{0\ell\uparrow}n_{0\ell'\downarrow}\\ & +(U-3J)\!\!\sum_{\substack{\ell<\ell'\\\sigma}}\!\!n_{0\ell\sigma}n_{0\ell'\sigma}\\
&-J\sum_{\ell\neq\ell'}c^\dagger_{0\ell\uparrow}c_{0\ell\downarrow}c^\dagger_{0\ell'\downarrow}c_{0\ell'\uparrow}\\ &+J\sum_{\ell\neq\ell'}c^\dagger_{0\ell\uparrow}c^\dagger_{0\ell\downarrow}c_{0\ell'\downarrow}c_{0\ell'\uparrow},
\end{split}
\label{eq:aim-fermionic-hamiltonian}
\end{equation}
with $U$ and $J$ the Coulomb and Hund couplings of Ref.~\cite{Kanamori1963}. The first two lines hold the onsite energies $a_{k\ell}$ and the hoppings $t_{k\ell}$ of each chain. The other lines are the Kanamori interaction, and they act on the impurity sites alone. The hopping $t_{0\ell}$ is the hybridisation between bath and impurity. The third and fourth lines are the Kanamori density repulsion. It is $U$ between the two electrons of one orbital. Between different orbitals, it is $U-2J$ for opposite spins and $U-3J$ for equal spins. The last two lines hold the spin-flip and pair-hopping terms. They make the interaction invariant under spin rotations.

\Cref{sec:permutation} acts on a Pauli LCU, so the modes must first become qubits. Under Jordan--Wigner, the mode $c_{k\ell\sigma}$ becomes the qubit $q_{k\ell\sigma}$, and we keep this labelling throughout. We order the Jordan--Wigner string as a snake, which leaves no parity string on any term. The one-body terms become $Z$ on every site and $XX+YY$ on every chain bond. The interaction becomes $ZZ$ on pairs of impurity qubits, together with four \emph{Hund words}. A Hund word is a Pauli string of $X$ and $Y$ factors on the four impurity qubits of two orbitals. The stated Pauli expressions use this fixed mode ordering. Each spin-up chain runs from its bath end to the impurity, and the spin-down chain of the same orbital runs back out.

Neighbouring sites of a chain are then adjacent in the string, and so are the two impurity modes of each orbital. Each Hund term is a product of two bilinears, each on the two adjacent spin modes of one orbital, so it carries no parity string either.

Before treating either family, we fix the query that holds them both. The mapped Hamiltonian contains two support shapes, so we write it as two structured families. Up to a constant $\eta$, and with the density-generated shift of the impurity energies folded into $a_{0\ell}$, it becomes $H_{\mathrm{1b}}+H_{\mathrm{int}}$, and we write out the two parts below, family by family. In their Pauli form, the reused symbol $a_{0\ell}$ denotes the folded coefficient $a_{0\ell}^{\mathrm{bare}}+[U+(n_\mathrm{imp}-1)(2U-5J)]/2$, where $a_{0\ell}^{\mathrm{bare}}$ is the parameter in \Cref{eq:aim-fermionic-hamiltonian}; the constant terms enter $\eta$. Both families have the shape $\Perm^{-1}P_{\mathcal T}\Perm$ of \Cref{eq:H-heisenberg-block}. Each support is transported to a fixed target, acted upon, and restored. Unlike the ring, the AIM requires a common preparation for two families. The composition of \Cref{sec:interface} allows this.

\Cref{fig:aim-complete-query} shows how, from left to right. The common preparation $\mathsf{Prep}_\mu$ writes the term label $\mathcal F$ and its uniform refinement bits $\mathcal U$. Family flags in the label then select the one-body circuit $\mathsf B$ of \SELECT{} or its interaction circuit $\mathsf I$. Each acts as the identity outside its own family, so the two simply run one after the other. The Clifford sign factors $Z_{\sgn}Z_{f_\mathrm{T}y}$ follow, and $\mathsf{Prep}_\mu^\dagger$ closes the query. Otherwise the interface is that of the ring. The label is compressed, and each circuit computes its few nonlinear controls in its own control generator, $\mathsf G_\mathsf{B}$ or $\mathsf G_\mathsf{I}$.

\Cref{tab:aim-row-map} lists what each generator writes for each row. Coefficients are loaded as in the ring and in Ref.~\cite{Babbush2018}. The alias table holds one row per coefficient row, about $N$ rows in all; equal numerical magnitudes need not be merged. Uniform choices then pick one of the Pauli terms represented by that row. 

We start with the one-body family. Because its terms address all $N$ qubits, $\mathsf B$ determines the leading cost. The address is a site $k$ on the chain of $(\ell,\sigma)$, together with the hopping flag $f_\mathrm{T}$. The flag belongs to the address because an onsite term and a hopping have different supports, one site or a bond. The target register is $\mathcal T_\mathsf B=(q_{00\uparrow},q_{00\downarrow})$, the two impurity qubits of orbital $0$. The permutation $\Perm_{k\ell\sigma}=\PermC_{\ell\sigma}\Perm_k$ brings site $k$ to qubit $k_0$ of $\mathcal T_\mathsf B$ and, for a hopping, site $k+1$ to the other qubit. The one-body part then reads

\begin{equation}
\begin{split}
H_{\mathrm{1b}}={}&-\sum_{\ell,\sigma}\sum_{k=0}^{n_\mathrm{bath}}\tfrac12a_{k\ell}\,\Perm_k^{-1}\bigl[\PermC_{\ell\sigma}^{-1}Z_{00\uparrow}^{1-k_0}Z_{00\downarrow}^{k_0}\PermC_{\ell\sigma}\bigr]\Perm_k\\
&+\sum_{\ell,\sigma}\sum_{k=0}^{n_\mathrm{bath}-1}\sum_{y=0}^{1}\tfrac12t_{k\ell}\,\Perm_k^{-1}\bigl[\PermC_{\ell\sigma}^{-1}P^{(y)}_{00\uparrow}P^{(y)}_{00\downarrow}\PermC_{\ell\sigma}\bigr]\Perm_k.
\end{split}
\label{eq:aim-swap-form}
\end{equation}
Here $P^{(0)}=X$ and $P^{(1)}=Y$, so the sum over $y$ gives the $XX+YY$ of a hopping. The trees $\Perm_k$ also depend on $f_\mathrm{T}$, since onsite rows skip the toggle, so the $\Perm_k$ of the onsite line and of the hopping line are different permutations. An onsite $Z$ acts on qubit $k_0$ of $\mathcal T_\mathsf B$, where the permutation puts site $k$. Equivalently, a virtual exchange $\SWAP^{k_0}$ of the two qubits of $\mathcal T_\mathsf B$ would always bring site $k$ to $q_{00\uparrow}$, as the $j_0$-controlled exchange of the ring does. Reading $k_0$ in the target action instead is the alternative of \Cref{sec:where-what}. $\SWAP^{k_0}$ drops out of a hopping, whose target is symmetric in its two qubits, and for an onsite term it is never built.

The support reaches $\mathcal T_\mathsf B$ in two stages, which one example makes concrete, a hopping between sites $4$ and $5$ on the chain of $(1,\downarrow)$. First, the swap trees $\Perm_k$ choose the site. They act inside every chain at once and depend on $k$, but not on the chain. For this hopping, they bring sites $4$ and $5$ of every chain to the roots of its two trees, its front pair $k=0,1$. Then the chain-front permutation $\PermC_{\ell\sigma}$ chooses the chain. It depends on $(\ell,\sigma)$ alone and exchanges whole front pairs. Here it brings the front pair of the chain of $(1,\downarrow)$ to $\mathcal T_\mathsf B$, so both ends of the hopping travel together. The two stages rest on two different structures. The trees rest on the \emph{within-chain geometry}, which is that of the ring. The bond graph is a disjoint union of chains, each bipartite, with the even and the odd positions $k$ as its two groups. Consequently $\chi=2$, as in the ring.

The chain-front permutation rests on the \emph{across-chain structure}, which the ring lacks. The $2n_\mathrm{imp}$ chains, one per spin orbital, share one geometry, and no one-body term leaves its chain. One address $k$ therefore serves every chain, and $\PermC_{\ell\sigma}$ moves the whole support of a term under one control. It does so with \emph{block} \CSWAP{}s, which exchange several qubits under one control. $\PermC_{\ell\sigma}$ is a swap tree of block \CSWAP{}s whose leaves are the $2n_\mathrm{imp}$ front pairs, so each block exchanges the front pairs of two chains. Counting each block \CSWAP{} as two \CSWAP{}s, one per exchanged qubit pair, $\mathsf B$ holds $N-2$ \CSWAP{}s in all, trees included, at $N=2n_\mathrm{imp}(n_\mathrm{bath}+1)$. This is the minimum of \Cref{eq:pau-chi-switches} at $\chi=2$. The $2n_\mathrm{imp}$ chains thus cost as much as a single chain of $N$ qubits.

\begin{figure}[t]
\centering
\crayontitle{CrayonBlue}{\textbf{One-body circuit} $\mathsf B$}\par
\resizebox{\columnwidth}{!}{%
\begin{quantikz}[row sep={.46cm,between origins},column sep=.1cm]
\lstick{$\mathcal F$} & \gate[wires=6,disable auto height,style={draw=CrayonGreen,fill=CrayonGreenFill,rounded corners=2pt}]{\mkern-3mu\mathsf G_\mathsf{B}\mkern-3mu} &[-.3cm]  &[-.3cm]  &[-.3cm]  & \control[style={draw=CrayonYellow,fill=CrayonYellow}]{}\wire[d][7][CrayonYellow]{q} & \control[style={draw=CrayonBlue,fill=CrayonBlue}]{}\wire[d][7][CrayonBlue]{q} &[-.4cm] \control[style={draw=CrayonRed,fill=CrayonRed}]{}\wire[d][10][CrayonRed]{q} &[-.4cm] \control[style={draw=CrayonYellow,fill=CrayonYellow}]{}\wire[d][7][CrayonYellow]{q} &  &[-.3cm]  &[-.3cm]  &[-.3cm] \gate[wires=6,disable auto height,style={draw=CrayonGreen,fill=CrayonGreenFill,rounded corners=2pt}]{\mkern-3mu\mathsf G_\mathsf{B}^\dagger\mkern-4.5mu} &   \\
\lstick{$\sigma$} &  &  &  &  & \control[style={draw=CrayonYellow,fill=CrayonYellow}]{} &  &  & \control[style={draw=CrayonYellow,fill=CrayonYellow}]{} &  &  &  &  &   \\
\lstick{$j{:}\ket{0^{m}}$} &  & \control[style={draw=CrayonYellow,fill=CrayonYellow}]{}\wire[d][6][CrayonYellow]{q} & \gate[wires=2,disable auto height,style={draw=CrayonGreen,fill=CrayonGreenFill,rounded corners=2pt}]{\Tog} & \control[style={draw=CrayonYellow,fill=CrayonYellow}]{}\wire[d][4][CrayonYellow]{q} &  &  &  &  & \control[style={draw=CrayonYellow,fill=CrayonYellow}]{}\wire[d][4][CrayonYellow]{q} & \gate[wires=2,disable auto height,style={draw=CrayonGreen,fill=CrayonGreenFill,rounded corners=2pt}]{\Tog} & \control[style={draw=CrayonYellow,fill=CrayonYellow}]{}\wire[d][6][CrayonYellow]{q} &  & \rstick{$\ket0$}  \\
\lstick{$\gamma{:}\ket{0^{m}}$} &  &  &  &  &  &  &  &  &  &  &  &  & \rstick{$\ket0$}  \\
\lstick{$f_\mathrm{A}k_0{:}\ket0$} &  &  &  &  &  & \control[style={draw=CrayonBlue,fill=CrayonBlue}]{} &  &  &  &  &  &  & \rstick{$\ket0$}  \\
\lstick{$\beta{:}\ket0$} &  &  &  &  &  &  & \control[style={draw=CrayonRed,fill=CrayonRed}]{} &  &  &  &  &  & \rstick{$\ket0$}  \\
\lstick{$q_{k\ell\sigma},\,{\scriptstyle k=2,4,\dots}$} &  &  &  & \gate[wires=2,disable auto height,style={draw=CrayonYellow,fill=CrayonYellowFill,rounded corners=2pt}]{\makebox[\widthof{$\Perm_{j}$}][c]{\ensuremath{\Perm_{\jev}}}} &  &  &  &  & \gate[wires=2,disable auto height,style={draw=CrayonYellow,fill=CrayonYellowFill,rounded corners=2pt}]{\makebox[\widthof{$\Perm_{j}$}][c]{\ensuremath{\Perm_{\jev}^{-1}}}} &  &  &  &   \\
\lstick{$q_{0\ell\sigma}$} &  &  &  &  & \gate[wires=2,disable auto height,style={draw=CrayonYellow,fill=CrayonYellowFill,rounded corners=2pt}]{\makebox[\widthof{$\PermC_{\ell\sigma}$}][c]{\ensuremath{\PermC_{\ell\sigma}}}} & \gate[style={draw=CrayonBlue,fill=CrayonBlueFill,rounded corners=2pt}]{\Pb} &  & \gate[wires=2,disable auto height,style={draw=CrayonYellow,fill=CrayonYellowFill,rounded corners=2pt}]{\makebox[\widthof{$\PermC_{\ell\sigma}$}][c]{\ensuremath{\PermC_{\ell\sigma}^{-1}}}} &  &  &  &  &   \\
\lstick{$q_{1\ell\sigma}$} &  & \gate[wires=2,disable auto height,style={draw=CrayonYellow,fill=CrayonYellowFill,rounded corners=2pt}]{\makebox[\widthof{$\Perm_j$}][c]{\ensuremath{\Perm_{\jod}}}} &  &  & \wire[d][2][CrayonYellow]{q} &  &  & \wire[d][2][CrayonYellow]{q} &  &  & \gate[wires=2,disable auto height,style={draw=CrayonYellow,fill=CrayonYellowFill,rounded corners=2pt}]{\makebox[\widthof{$\Perm_j$}][c]{\ensuremath{\Perm_{\jod}^{-1}}}} &  &   \\
\lstick{$q_{k\ell\sigma},\,{\scriptstyle k=3,5,\dots}$} &  &  &  &  &  &  &  &  &  &  &  &  &   \\
\lstick{$\wb_{\mathrm{imp}}{:}\ket0$} & \setwiretype{n} &  &  &  &  &  & \gate[style={draw=CrayonRed,fill=CrayonRedFill,rounded corners=2pt}]{\mathsf{Rep}_\mathsf{B}} & \wire[l][3][draw=CrayonYellow]{q} &  &  &  &  &
\end{quantikz}%
}\par
\vspace{-.3cm}
\caption{\textbf{One-body circuit $\mathsf B$ of \SELECT{}.}
The control generator $\mathsf G_\mathsf{B}$ derives the address and repair predicates from the label $\mathcal F$. The swap trees $\Perm_{\jev}$ and $\Perm_{\jod}$, connected by the address toggle $\Tog$, form $\Perm_k$ and select the site or hopping endpoints within each chain. The chain-front permutation $\PermC_{\ell\sigma}$ selects the chain and brings its front to the fixed target register $\mathcal T_\mathsf{B}$. The target action $\Pb$ then applies the onsite or hopping Pauli. Only $\PermC_{\ell\sigma}$ is bridged across the target action. The repair $\mathsf{Rep}_\mathsf{B}$ updates its live work bits before the inverse permutation. \Cref{fig:aim-onebody-generator,fig:address-toggle,fig:swap-tree,fig:aim-onebody-action,fig:aim-onebody-repair-bridge} give the circuit decompositions of $\mathsf G_\mathsf{B}$, $\Tog$, the swap trees with $\PermC_{\ell\sigma}$, $\Pb$, and $\mathsf{Rep}_\mathsf{B}$, respectively. }
\label{fig:aim-onebody-main-circuit}
\end{figure}
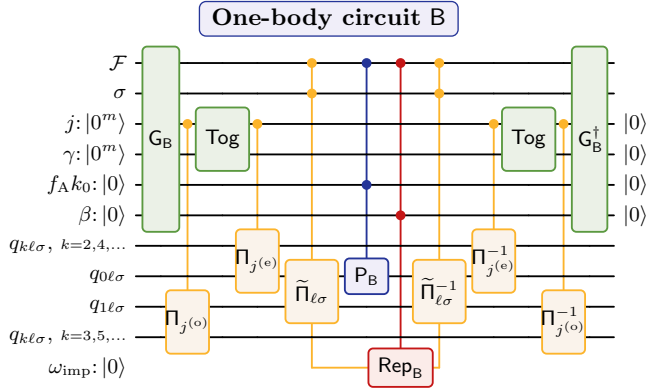

The one-body circuit, \Cref{fig:aim-onebody-main-circuit}, runs these stages in order, from left to right. $\mathsf G_\mathsf{B}$ derives the address and the repair predicates from the label $\mathcal F$. The odd tree, the toggle $\Tog$ and the even tree bring the selected sites to the chain fronts, as in the ring. $\PermC_{\ell\sigma}$ then brings the chosen front to $\mathcal T_\mathsf B$. The target action $\Pb$ applies $Z$, $XX$ or $YY$ with Clifford gates controlled by label bits. $\mathsf{Rep}_\mathsf{B}$ then repairs the work bits of the bridged $\PermC_{\ell\sigma}$ before its inverse, and the rest of the circuit undoes the permutations and $\mathsf G_\mathsf{B}$.

Only the chain-front permutation is bridged. One control of $\Pb$, $f_\mathrm{A}k_0$ with $f_\mathrm{A}$ the onsite flag, puts an onsite $Z$ on qubit $k_0$ of $\mathcal T_\mathsf B$. It stands in for the virtual exchange $\SWAP^{k_0}$. $\mathsf G_\mathsf{B}$ writes it. The odd tree reads $\jod=\lfloor k/2\rfloor$, and the toggle $\Tog$ moves the same register to the address of the even tree. An onsite term leaves the address untouched; only the site that the tree of its own parity group brings to $\mathcal T_\mathsf B$ receives a nonidentity target action, although both trees are present in the circuit. 

Why bridge the chain-front permutation and not the trees? Under the criterion of \Cref{sec:cswaps}, bridging $\PermC_{\ell\sigma}$ is beneficial because several work bits share one repair opening. Here the block \CSWAP{} supplies this sharing by carrying both ends of a hopping at once, as for the hopping between sites $4$ and $5$ above. The repair accounts for the change in data parity induced by $\Pb$. An onsite $Z$ flips no qubit, and a hopping flips both front qubits of the chain of $(\ell,\sigma)$. The hopping flag $f_\mathrm{T}$ thus sets the parity change. Only the blocks on the path of the leaf of chain $(\ell,\sigma)$ see that change. The indicator $\rep_e$ equals one exactly when block $e$ does, so the repair predicate $c\,\delta$ of \Cref{eq:repair-bit} is $f_\mathrm{T}\rep_e$.

The repair is shared because a block \CSWAP{} contains two \CSWAP{}s on one control, one for the sites $k=0$ of two chain fronts and one for their sites $k=1$. A hopping flips both, so one predicate repairs both work bits,

\begin{equation}
\wb_{e,k}'=\wb_{e,k}\oplus f_\mathrm{T}\rep_e,\qquad k=0,1.
\label{eq:aim-shared-repair}
\end{equation}
A closed block pays two openings per \CSWAP{}, four in all. A bridged block pays one per \CSWAP{} plus the shared repair, three in all. In \Cref{fig:aim-repair-application}(b), the yellow panels are the two bridged \CSWAP{}s and their inverses, the blue panel is $\Pb$, and the red panel holds the shared repair. A tree \CSWAP{}, by contrast, moves a single site, so it would need a repair of its own. That repair is a different product of address bits for nearly every tree \CSWAP{}. The trees therefore stay closed, and they carry the leading $8N$ of $\mathsf B$. The indicator is defined in \Cref{eq:aim-transport-indicator}. It already contains the control of $e$, and it is a product of address bits.

The predicate is a parity of products of $f_\mathrm{T}$ with address bits. $\mathsf G_\mathsf{B}$ holds one such product per block and builds each from $f_\mathrm{T}$ or from a product it already holds by one further factor, so each block pays one opening, \Cref{app:repair-count}. The \CNOT{}s of $\mathsf{Rep}_\mathsf{B}$ then apply the predicate to both work bits. Bridging thus removes a quarter of the cost of the chain-front permutation; of the rest, two thirds are the bridged \CSWAP{}s and one third their shared repairs. The parity of a tree \CSWAP{} changes only when it acts on the chain of $(\ell,\sigma)$ and lies on the path of the selected leaf. Bridging the trees would therefore trade reopenings for repairs almost one for one, while holding one live work bit per tree \CSWAP{}, \Cref{app:repair-count}.

At five orbitals, the chain-front stage holds eighteen bridged work bits and nine repair predicates. This stage sets the reported peak ancilla count over the plotted sizes.

\begin{figure}[t]
\centering
\makebox[\columnwidth][l]{\textbf{(a)}\hfill$\vcenter{\hbox{%
\begin{quantikz}[row sep={0.43cm,between origins},column sep=0.08cm]
\lstick{$c$} &  & \control{} \wire[d][3]{q} &  &  & \\
\lstick{$q_0$} & \targ{} & \control{} &  & \targ{} & \\
\lstick{$q_1$} & \ctrl{-1} &  & \targ{} & \ctrl{-1} & \\
\vphantom{\ket{0}} & \lstick{$\wb$}\setwiretype{n} &  & \ctrl{-1}\setwiretype{q} &  &
\end{quantikz}%
}}\hspace{0.01cm}=\hspace{0.01cm}\vcenter{\hbox{%
\begin{quantikz}[row sep={0.43cm,between origins},column sep=0.12cm]
 & \ctrl{3} & \\
 & \swap{1} & \\
 & \targX{} & \\
\vphantom{\ket{0}} & \setwiretype{n} & \setwiretype{q}
\end{quantikz}%
}}$\hfill\hphantom{\textbf{(a)}}}\par\smallskip
\makebox[\columnwidth][l]{\raisebox{-\height}{\textbf{(b)}}\hfill\raisebox{-\height}{%
\begin{quantikz}[row sep={0.35cm,between origins}, column sep=0.2cm]
\lstick{$g_e$} & \ctrl{9} \gategroup[11,steps=2,style={draw=none,fill=CrayonYellowWash,rounded corners=2pt,inner xsep=1pt,inner ysep=-1pt},background]{} & \ctrl{10} &  &  &  &  &  &  & \control{} \wire[d][11]{q} \gategroup[12,steps=3,style={draw=none,fill=CrayonRedWash,rounded corners=2pt,inner xsep=1pt,inner ysep=-1pt},background]{} &  & \control{} \wire[d][11]{q} &  & \ctrl{10} \gategroup[11,steps=2,style={draw=none,fill=CrayonYellowWash,rounded corners=2pt,inner xsep=1pt,inner ysep=-1pt},background]{} & \ctrl{9} &  \\
\lstick{$f_\mathrm{T}$} &  &  &  & \gategroup[7,steps=4,style={draw=none,fill=CrayonBlueWash,rounded corners=2pt,inner xsep=1pt,inner ysep=-1pt},background]{} &  &  & \control{} \wire[d][4]{q} &  & \control{} &  & \control{} &  &  &  &  \\
\lstick{$f_\mathrm{A}$} &  &  &  & \ctrl{3} &  &  &  &  &  &  &  &  &  &  &  \\
\lstick{$f_\mathrm{A}k_0$} &  &  &  &  & \control{} \wire[d][2]{q} &  &  &  &  &  &  &  &  &  &  \\
\lstick{$f_\mathrm{T}y$} &  &  &  &  &  & \control{} \wire[d][1]{q} &  &  &  &  &  &  &  &  &  \\
\lstick{$q_{00\uparrow}$} & \swap{1} &  &  & \gate{Z} & \gate{Z} \wire[d][2]{q} & \gate{Z} \wire[d][2]{q} & \gate{X} \wire[d][2]{q} &  &  &  &  &  &  & \swap{1} &  \\
\lstick{$q_{0\ell\sigma}$} & \targX{} &  &  &  &  &  &  &  &  &  &  &  &  & \targX{} &  \\
\lstick{$q_{10\uparrow}$} &  & \swap{1} &  &  & \gate{Z} & \gate{Z} & \gate{X} &  &  &  &  &  & \swap{1} &  &  \\
\lstick{$q_{1\ell\sigma}$} &  & \targX{} &  &  &  &  &  &  &  &  &  &  & \targX{} &  &  \\
& \setwiretype{n} & \setwiretype{q} &  &  &  &  &  &  &  & \targ{} \wire[d][1]{q} &  &  &  &  & \setwiretype{n} \\
& \setwiretype{n} &  & \setwiretype{q} &  &  &  &  &  &  & \targ{} &  &  &  & \setwiretype{n} &  \\
& \setwiretype{n} &  &  &  &  &  &  &  &  & \ctrl{-2} \setwiretype{q} &  & \setwiretype{n} &  &  &
\end{quantikz}%
}\hfill\hphantom{\textbf{(b)}}}
\caption{\textbf{Bridged \CSWAP{}s.} (a)~The \CSWAP{} of \Cref{eq:work-bit} with its work bit $\wb$ left open is drawn as a \CSWAP{} with a corner on the work bit. The matching corner of the inverse \CSWAP{} consumes it. Every circuit draws bridged \CSWAP{}s this way, \Cref{sec:cswaps}. (b)~Bridged chain-front block. The two yellow panels exchange the front pairs of the chains $(0,\uparrow)$ and $(\ell,\sigma)$ and later undo the exchange. Each \CSWAP{} is drawn as in (a), with its work bit opened from $g_e$ below it and retained across the target action instead of being closed and reopened. The blue panel applies $\Pb$ to the front pair $(q_{00\uparrow},q_{10\uparrow})$, which the terminal fixed swap of $\PermC_{\ell\sigma}$ maps onto the target register $\mathcal T_\mathsf B$. A hopping changes the parity of both endpoints by $f_\mathrm{T}$. The red panel therefore opens the shared repair predicate $g_e f_\mathrm{T}$ once and uses it to repair both retained work bits before the inverse \CSWAP{}s consume them. The block requires three openings, for a cost of $12\,T$, instead of the four openings and $16\,T$ of two closed \CSWAP{} pairs.}
\label{fig:aim-repair-application}
\end{figure}
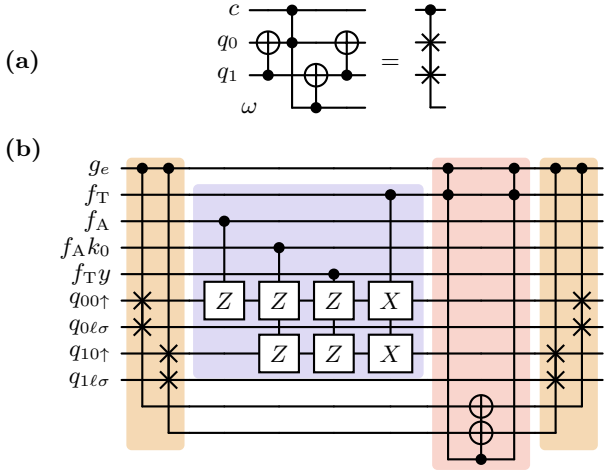

\begin{figure}[t]
\centering
\crayontitle{CrayonBlue}{\textbf{Interaction circuit} $\mathsf I$}\par
{\small
\begin{quantikz}[row sep={.42cm,between origins},column sep=.10cm]
\lstick{$\mathcal F$}
& \gate[wires=4,style={draw=CrayonGreen,fill=CrayonGreenFill,
rounded corners=2pt,inner ysep=-3.0pt}]{\mathsf G_\mathsf{I}}
& \control[style={draw=CrayonYellow,fill=CrayonYellow}]{}
\wire[d][4][CrayonYellow]{q}
&[-.3cm] \control[style={draw=CrayonYellow,fill=CrayonYellow}]{}
\wire[d][5][CrayonYellow]{q}
&[-.3cm] \control[style={draw=CrayonBlue,fill=CrayonBlue}]{}
\wire[d][4][CrayonBlue]{q}
&[-.3cm] \control[style={draw=CrayonBlue,fill=CrayonBlue}]{}
\wire[d][5][CrayonBlue]{q}
&[-.3cm] \control[style={draw=CrayonYellow,fill=CrayonYellow}]{}
\wire[d][4][CrayonYellow]{q}
&[-.3cm] \control[style={draw=CrayonYellow,fill=CrayonYellow}]{}
\wire[d][5][CrayonYellow]{q}
&[-.3cm] \gate[style={inner ysep=-.5pt}]{Z_{f_\mathrm{J}y}}
\gategroup[
wires=4,
steps=1,
style={
draw=none,
fill=CrayonRedWash,
rounded corners=2pt,
inner xsep=-1.8pt,
inner ysep=-1.8pt
},
background
]{}
& [.1cm]\gate[wires=4,style={draw=CrayonGreen,fill=CrayonGreenFill,
rounded corners=2pt,inner ysep=-5.0pt}]{\mathsf G_\mathsf{I}^\dagger}
& \\
\lstick{$\sigma$}
& & & & & & & & & & \\
\lstick{$f_\mathrm{D}\sigma:\ket0$}
& & & & & \control[style={draw=CrayonBlue,fill=CrayonBlue}]{}
& & & & & \rstick{$\ket0$} \\
\lstick{$f_\mathrm{J}\sigma:\ket0$}
& & & &
\control[style={draw=CrayonBlue,fill=CrayonBlue}]{}
& \control[style={draw=CrayonBlue,fill=CrayonBlue}]{}
& & & \gate[style={inner ysep=0pt}]{Z}
& & \rstick{$\ket0$} \\[.2cm]
\lstick{$q_{0\ell\uparrow}$}
& \gate[wires=2,style={draw=CrayonViolet,fill=CrayonVioletFill,
rounded corners=2pt,inner ysep=-3.0pt}]{\mathsf E_\mathsf{I}}
& \gate[style={draw=CrayonYellow,fill=CrayonYellowFill,
rounded corners=2pt,inner ysep=0pt}]{\PermC_\uparrow}
&
& \gate[style={draw=CrayonBlue,fill=CrayonBlueFill,
rounded corners=2pt,inner ysep=0pt}]{\Pup}
&
& \gate[style={draw=CrayonYellow,fill=CrayonYellowFill,
rounded corners=2pt,inner ysep=0pt}]{\PermC_\uparrow^{-1}}
&
&
& \gate[wires=2,style={draw=CrayonViolet,fill=CrayonVioletFill,
rounded corners=2pt,inner ysep=-3.0pt}]{\mathsf E_\mathsf{I}}
& \\[.42cm]
\lstick{$q_{0\ell\downarrow}$}
&
&
& \gate[style={draw=CrayonYellow,fill=CrayonYellowFill,
rounded corners=2pt,inner ysep=0pt}]
{\PermC_\downarrow}
\wire[d][1][draw=CrayonYellow]{q}
&
& \gate[style={draw=CrayonBlue,fill=CrayonBlueFill,
rounded corners=2pt,inner ysep=0pt}]{\Pdn}
&
& \gate[style={draw=CrayonYellow,fill=CrayonYellowFill,
rounded corners=2pt,inner ysep=0pt}]
{\PermC_\downarrow^{-1}}
\wire[d][1][draw=CrayonYellow]{q}
&
&
& \\[.1cm]
\lstick{$\wb_{\mathrm c}:\ket0$}
& \setwiretype{n}
&
&
&
&
&
& \wire[l][4][draw=CrayonYellow]{q}
&
&
&
\end{quantikz}}\par
\vspace{-.3cm}
\caption{\textbf{Interaction circuit $\mathsf I$ of \SELECT{}.}
The control generator $\mathsf G_\mathsf{I}$ opens the products $f_\mathrm{D}\sigma$ and $f_\mathrm{J}\sigma$ used by the density and Hund rows. The basis change $\mathsf E_\mathsf{I}$ encloses two stages acting on disjoint sets of impurity qubits. In the spin-up stage, $\PermC_\uparrow$ brings the selected orbital pair to the fixed target register $\mathcal T_\mathsf{I}$, where $\Pup$ acts. In the spin-down stage, $\Pdn$ acts between the forward and inverse bridged permutations $\PermC_\downarrow$ of the complementary orbital pairs $\{1,2\},\{3,4\},\ldots$. The register $\wb_\mathrm{c}$ retains one work bit per complementary pair across the spin-down target action. The red panel applies the Clifford correction to the Hund phase. The bath is a spectator throughout. \Cref{fig:aim-interaction-generator,fig:aim-spin-up-action,fig:aa-network,fig:aim-interaction-walks} give the circuit decompositions of $\mathsf G_\mathsf{I}$, $\mathsf E_\mathsf{I}$ with $\Pup$, the all-to-all network used by $\PermC_\uparrow$, and the spin-down stage with $\PermC_\downarrow$ and $\Pdn$, respectively. 
}
\label{fig:aim-interaction-main-circuit}
\end{figure}
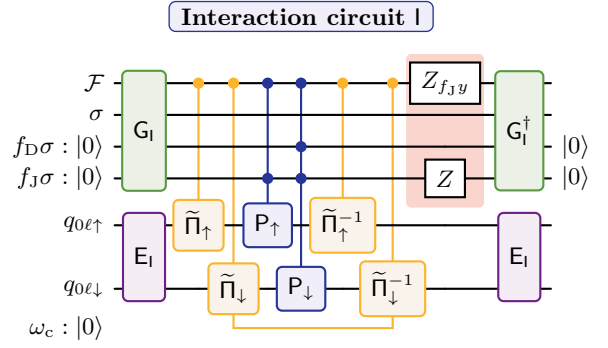

We now turn to the interaction family. Its terms act on the impurity alone, so $\mathsf I$ costs only $O(n_\mathrm{imp})$. They also exhibit bridging without a repair. The address is an orbital pair and the label specifies the interaction word. Specifically, the address is $(\ell,\ell')$, while $b$ selects a density term or one of the four Hund words. A Hund word acts on all four impurity qubits of the pair, a density term only on two. We take all four as the support. A density term then acts as the identity on the other two, which \Cref{sec:where-what} allows. The permutation $\PermC_{\ell\ell'}$ brings these four qubits to $\mathcal T_\mathsf I$, the impurity qubits of orbitals $0$ and $1$. This register contains $\mathcal T_\mathsf B$. With $\alpha_{\ell\ell',b}$ the coefficient of word $b$ on the pair, the interaction part reads

\begin{equation}
H_{\mathrm{int}}=\sum_{\ell<\ell'}\sum_{b}\alpha_{\ell\ell',b}\,\PermC_{\ell\ell'}^{-1}P^{(b)}_{\mathcal T_\mathsf I}\PermC_{\ell\ell'}.
\label{eq:aim-swap-form-int}
\end{equation}
Explicitly, $\mathcal T_\mathsf I=(q_{00\uparrow},q_{00\downarrow},q_{01\uparrow},q_{01\downarrow})$, and the formal structured permutation orders the four qubits as $(\ell\!\uparrow,\ell\!\downarrow,\ell'\!\uparrow,\ell'\!\downarrow)$.  \Cref{tab:aim-row-map} lists the Pauli string that each row and refinement select.  For interorbital density rows, $\alpha_{\ell\ell',b}$ is one quarter of the corresponding density coupling in \Cref{eq:aim-fermionic-hamiltonian}; for a Hund row, the appendix establishes $|\alpha_{\ell\ell',b}|=|J|/4$. The selected Pauli word is unsigned, and its term-dependent sign is supplied by the coefficient-sign and Clifford phase controls. A $U$ term acts on a single orbital. In the five-orbital construction, its coefficient $U/4$ is shared over the two oriented-pair ranks whose head is that orbital, giving $U/8$ per rank in the structured sum. 

Any two orbitals may couple, so the network of $\PermC_{\ell\ell'}$ is the all-to-all network of \Cref{tab:pau-geometries} at $N=n_\mathrm{imp}$. Each of its block \CSWAP{}s exchanges the impurity qubits of two orbitals. \Cref{eq:aim-swap-form-int} is the structured form, not the circuit. The circuit of \Cref{fig:aim-interaction-main-circuit} implements this operator without permuting the selected spin-down pair onto fixed qubits of $\mathcal T_\mathsf{I}$, and it replaces each two-qubit block by a cheaper equivalent. Only the $X$ part of a factor flips bits, since $Y=iXZ$, so only that part needs a permutation to reach $\mathcal T_\mathsf{I}$. A $Z$ factor can be applied where its qubit sits, once its position is known.

The interaction circuit therefore starts with a basis change $\mathsf E_\mathsf{I}$ that gathers all the $X$ factors on the spin-up qubits. It is one \CNOT{} per orbital, from its spin-up to its spin-down impurity qubit, and it maps the occupations $(n_\uparrow,n_\downarrow)$ of each orbital to $(n_\uparrow,n_\uparrow\oplus n_\downarrow)$. This is allowed by a $\mathbb Z_2$ symmetry of the Kanamori interaction. Spin flip and pair hopping change both occupations of every orbital they touch, and density terms change none, so no interaction word changes the parity $n_\uparrow\oplus n_\downarrow$ of any orbital. In the new basis, every word on the pair $(\ell,\ell')$ becomes
\begin{equation}
\begin{split}
\mathsf E_\mathsf{I}P^{(b)}\mathsf E_\mathsf{I}={}&(-1)^{f_\mathrm{J}}\underbrace{\bigl(X^{f_\mathrm{J}}Z^{\zeta_\uparrow}\bigr)_{0\ell\uparrow}\bigl(X^{f_\mathrm{J}}Z^{\zeta_\uparrow}\bigr)_{0\ell'\uparrow}}_{\Pup}\\
&\times\underbrace{Z_{0\ell\downarrow}^{\zeta_{\ell\downarrow}}Z_{0\ell'\downarrow}^{\zeta_{\ell'\downarrow}}}_{\Pdn},
\end{split}
\label{eq:aim-interaction-frame}
\end{equation}
where the four exponents record what the word does in the new basis. The flag $f_\mathrm{J}$ equals $1$ for a Hund word and $0$ for a density word. It decides whether the word flips both spin-up qubits, which only spin flip and pair hopping do, since they change the spin-up occupation of both orbitals. The exponent $\zeta_\uparrow$ puts a common $Z$ on the two spin-up qubits. For a Hund word, it selects the $XX$ or the $YY$ spin-up part, since $(XZ)(XZ)=-YY$. For a density word, it equals $1$ when the word couples the two orbitals and $0$ for the single-orbital $U$ word. The exponents $\zeta_{\ell\downarrow}$ and $\zeta_{\ell'\downarrow}$ put a $Z$ on the spin-down qubit of $\ell$ or $\ell'$, which now holds the parity $n_\uparrow\oplus n_\downarrow$ of that orbital. A density word has such a $Z$ wherever it had one on that spin-down qubit, and a Hund word has it on exactly one of the two. The Hund words are unsigned $X/Y$ strings. Their parity-frame image carries the intrinsic factor $(-1)^{f_\mathrm{J}}$ in this ordered-$XZ$ convention; their signed coefficients are separate. The spin-up mask $\Pup$ carries all $X$ gates and the spin-down mask $\Pdn$ is diagonal.

\Cref{fig:aim-interaction-main-circuit} runs this frame in order. $\mathsf G_\mathsf{I}$ first opens the two products $f_\mathrm{D}\sigma$ and $f_\mathrm{J}\sigma$, where $f_\mathrm{D}$ flags the density rows, and $\mathsf E_\mathsf{I}$ changes the basis. The all-to-all network $\PermC_\uparrow$ then brings the spin-up pair to the two tree roots $q_{00\uparrow}$ and $q_{01\uparrow}$ in $\mathcal T_\mathsf{I}$, where $\Pup$ acts. The spin-down stage uses $\PermC_\downarrow$, one bridged \CSWAP{} per complementary pair of orbitals, to orient the fronts within each pair. Two short walks addressed by the pair then find the selected spin-down qubits, and $\Pdn$ phases them where they sit. Such an exchange can move a spin-down site onto a qubit of $\mathcal T_\mathsf I$, but the spin-down pair is never permuted onto fixed qubits of $\mathcal T_\mathsf I$. The register $\wb_\mathrm{c}$ holds the bridged work bits. Only phases lie between the two occurrences of $\PermC_\downarrow$, so its bridged \CSWAP{}s close without repair. This is the diagonal case of \Cref{sec:cswaps}. The final $\mathsf E_\mathsf{I}$ copies each flip onto the spin-down qubit of the same orbital. Before $\mathsf G_\mathsf{I}^\dagger$ closes the two products, Clifford label gates supply the row sign and correct the raw and relative Hund phases.

The closed spin-up network costs two openings per \CSWAP{}. Across the whole network, the bridged spin-down pairs and their phase walks together cost one further opening per spin-up \CSWAP{} on average. The network thus costs three openings per spin-up \CSWAP{} on average, giving the base count $3\lceil3(n_\mathrm{imp}-2)/2\rceil+\Npre$ in \Cref{eq:aim-interaction-openings}. These counts cover the permutations and the phase walks; the two products opened by $\mathsf G_\mathsf{I}$ are additional. At five orbitals, the permutations and walks cost fifteen openings ($60\,T$), and those two products add $8\,T$. This aggregate accounting does not assign a separate third opening to every individual block.

Putting the two families together gives the cost of the AIM query. The swap trees touch nearly every system qubit, so the one-body sector sets the leading cost of \SELECT{},

\begin{equation}
\begin{split}
\Tsel=&16n_\mathrm{imp}(n_\mathrm{bath}-1)
+O\!\left(n_\mathrm{imp}+\log n_\mathrm{bath}\right)\\
=8&N+O\!\left(n_\mathrm{imp}+\log n_\mathrm{bath}\right).
\end{split}
\label{eq:aim-main-select-scaling}
\end{equation}

This saving has the same origin as in the ring. The Jordan--Wigner list holds three strings per qubit, $Z$, $XX$ and $YY$. A walk over it pays $4\,T$ per string, or $12N$. The trees pay $8\,T$ per qubit, whatever acts there. The saving is thus a constant factor, set by the number of classes per site. $12N$ is also the leading \SELECT{} cost of the linear-$T$ construction of Ref.~\cite{Babbush2018}. Bridging changes only the subleading chain-front and interaction terms. Preparation adds a second linear term, about $4N$ for the alias lookup over the $N$ rows of \Cref{eq:aim-factored-lambda}. For fixed $n_\mathrm{imp}$ and $\mu$ as the bath length grows, the complete query therefore costs about $12N$ at leading order. If the impurity count also grows, the rank decoder contributes an additional $O(n_\mathrm{imp}^2)$ term, \Cref{eq:aim-query-T-scaling}.

\begin{figure}[t]
\centering
\includegraphics[width=\columnwidth]{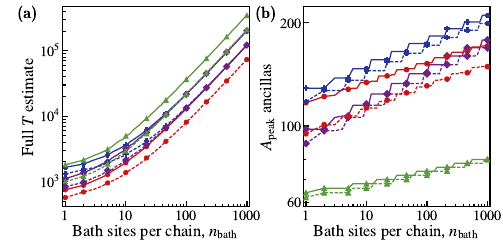}\par\smallskip
{\footnotesize\setlength{\tabcolsep}{3pt}%
\begin{tabular}{@{}llr@{}}
\toprule
\textbf{(c)} & & Leading order\\
\midrule
\multirow{2}{*}{\plotkey{CrayonRed}{\plotmarkercircle}\hspace{0.45em}This work} & $\Ttot$ & $24n_\mathrm{imp}n_\mathrm{bath}+4\mu$\\
& $\Apeak$ & $3\mu+6n_\mathrm{imp}+6\log_2n_\mathrm{bath}+10\log_2n_\mathrm{imp}$\\
\addlinespace[3pt]
\multirow{2}{*}{\plotkey{CrayonBlue}{\plotmarkerplus}\hspace{0.45em}Structured} & $\Ttot$ & $40n_\mathrm{imp}n_\mathrm{bath}+16\mu$\\
& $\Apeak$ & $5\mu+9\log_2n_\mathrm{bath}+11\log_2n_\mathrm{imp}$\\
\addlinespace[3pt]
\multirow{2}{*}{\plotkey{CrayonGreen}{\plotmarkertriangle}\hspace{0.45em}Flat} & $\Ttot$ & $n_\mathrm{imp}(72n_\mathrm{bath}+48n_\mathrm{imp})+16\mu$\\
& $\Apeak$ & $5\mu+2\log_2(n_\mathrm{imp}n_\mathrm{bath})$\\
\addlinespace[3pt]
\multirow{2}{*}{\plotkey{CrayonViolet}{\plotmarkerdiamond}\hspace{0.45em}Sparse} & $\Ttot$ & $n_\mathrm{imp}(40n_\mathrm{bath}+4n_\mathrm{imp})+16\mu$\\
& $\Apeak$ & $5\mu+9\log_2(n_\mathrm{imp}n_\mathrm{bath})$\\
\bottomrule
\end{tabular}}
\caption{\textbf{Resources for one uncontrolled complete AIM block-encoding query as a function of the number of bath sites per chain.}
(a)~Total $T$ count, with the $\Nrot$ arbitrary rotations synthesised at $\epssynth=10^{-3}$.
(b)~Peak ancilla count $\Apeak$, excluding the $N=2n_\mathrm{imp}(n_\mathrm{bath}+1)$ system qubits.
(c)~Bath-leading resource expressions for $n_\mathrm{bath}\gtrsim n_\mathrm{imp}$ at fixed impurity count. The flat row assumes that no two entries of its lookup coincide; for the plotted coefficients, the lookup merges repeated entries, so the flat curve in (a) grows \newedit{about 4\%} more slowly. Our $\Apeak$ row holds over the plotted range; beyond a few thousand bath sites, the lookup erasure sets our peak and adds the $\sqrt{n_\mathrm{imp}n_\mathrm{bath}}$ term of \Cref{eq:aim-query-space-scaling}. The key beside each row identifies its curves in (a) and (b).
Solid and dashed curves correspond to $n_\mathrm{imp}=5$ and $n_\mathrm{imp}=3$, respectively. The comparators are the structured and flat readings of Ref.~\cite{Babbush2018} and the sparse reading of Refs.~\cite{Berry2019,Lee2021THC}. Alias precision is $\mu=10$ in (a) and (b).
}
\label{fig:aim-query-resources}
\end{figure}

As the bath grows at a fixed number of impurity orbitals, the complete query asymptotically needs three fifths of the $T$ gates of the cheapest of three matched comparators, 40\% fewer, at comparable peak ancillae, \Cref{fig:aim-query-resources}(c). The comparators are the structured and flat readings of Ref.~\cite{Babbush2018} and the sparse reading of Refs.~\cite{Berry2019,Lee2021THC}, constructed as in \Cref{app:aim-babbush-comparison}. \newedit{Each comparator uses only the methods of the references it follows. If the structured comparator also adopted the measured lookup erasure of Ref.~\cite{Berry2019}, its leading cost would fall from $20N$ to $16N$, and our query would need three quarters of its $T$ gates.} All four calls are linear in the bath length, so the margin is a constant factor rather than a change of scaling. The sparse and structured readings are the cheapest, and at five orbitals they agree within two per cent at bath lengths $100$ and $1000$.

\section{Conclusion}
\label{sec:conclusion}
In this paper, we have shown how the repeated structure of a Hamiltonian can be compiled into a block encoding whose \newedit{\SELECT{} has a $T$ count} set by the geometry of its supports. \newedit{Writing each term as a local operator conjugated by a support-dependent permutation separates \emph{where} it acts from \emph{what} it applies, so that the two costs add instead of multiplying. This separation has the precedents discussed in \Cref{sec:intro}, and \SELECT{} circuits whose $T$ count is linear in $N$ already exist for spin terms, Ref.~\cite{YOSKM2024}, for fermionic terms, Ref.~\cite{Wan2021}, and for particular term types, Refs.~\cite{Babbush2018,Berry2019,Sanders2020Heuristics}. What we add is a systematic way to make the separation cheap. A \CSWAP{} network fixed by the support geometry brings the selected support to the target register, an $XZ$ mask on that register alone applies the operator, and \PREPARE{} writes the address in the form the network consumes. The geometry of the supports then sets how the $T$ count of \SELECT{} scales with $N$. The permutation depends on the address alone, so every operator class reuses it, and further operator classes change only the $XZ$ mask, at a cost that does not grow with $N$. Families with different geometries, such as the bath chains and the all-to-all core of the impurity model, combine in one query, and their costs add.} Against constructions that are already linear, the gain is at most a constant factor. \newedit{For the Heisenberg ring, the leading cost is $8N$ $T$ gates, against $24N$ for flat and product-wise Pauli walks. For all-to-all pairs, it is} the same $12N$ as the plane-wave chemistry construction of Ref.~\cite{Babbush2018} but for any Pauli operator on the pair, and, \newedit{since the Jordan--Wigner signs of a controlled fermionic transposition are Clifford,} four times fewer $T$ gates than the ancilla-free quadratic fermionic circuit of Ref.~\cite{Wan2021}.

The cost of selecting \emph{where} a term acts is determined by the connectivity of the Hamiltonian. This connectivity fixes the \CSWAP{} network that brings the selected support to a fixed target register. For two-qubit supports, the resulting \SELECT{} costs \newedit{$8\Sw N+o(N)$} $T$ gates with $O(\log N)$ ancillae, where $\Sw$ depends only on the connectivity. It ranges from $\Sw=1$ for nearest-neighbour and bounded-range connectivity to $\Sw=3/2$ for all-to-all connectivity\newedit{, and at both ends the correction is $O(\log N)$}. In particular, the leading cost \newedit{of \SELECT{}} is independent of the number and values of the coupling coefficients, so power-law couplings cost the same \newedit{there} as uniform ones\newedit{; the coefficients enter through \PREPARE{}}. We also proved that the \CSWAP{} count is minimal within this class of networks, \Cref{app:cswap-count}. \newedit{When every set of three sites is a support, the density is again $\Sw=3/2$, but the minimum exceeds $3N/2$ by a term of order $\sqrt N$, \Cref{app:all-triples}.} \newedit{For supports of any fixed size, any network that delivers all supports can be controlled with logarithmic control workspace when \PREPARE{} writes one address per support, but the $T$ count of its controls depends on the form of that address, \Cref{app:generic-controls}.} For bounded connectivity, part of the saving comes from how the address and operator label are represented, suggesting that the same principle may benefit other selection mechanisms.

The matched forward and inverse permutations provide a second saving. Instead of erasing a temporary AND and recomputing it for the inverse permutation, a bridged \CSWAP{} remembers its logical value across the target action and repairs it using the prepared labels. When the repair is Clifford, this halves the $T$ cost of a matched pair at the price of one additional ancilla. \newedit{The same applies to any temporary value whose inputs an intervening operation changes in a way fixed by registers that the circuit already holds.} In the examples studied here, however, bridging affects only subleading terms. \newedit{The dominant saving comes from a permutation that every operator class reuses, where a walk over the Pauli list pays for each site once per class.}

We also adapted \PREPARE{} to the information that \SELECT{} consumes and compared complete block-encoding queries. For the Heisenberg ring, the resulting query uses about three times fewer $T$ gates than \newedit{flat and product-wise Pauli walks}. For the Anderson impurity model, which combines long bath chains with an all-to-all interacting core, it uses about $1.7$ times fewer $T$ gates than the cheapest \newedit{of three comparators, each built from the methods of the references it follows}. These reductions are obtained at comparable peak workspace and at the same block-encoding normalisation\newedit{, to within $0.3\%$ for the sparse AIM comparator}, so an algorithm built on these block encodings makes \newedit{essentially} the same number of queries, each at the reduced cost.

Several questions remain open. For general families of supports of three or more sites, the minimum \CSWAP{} count is known \newedit{in closed form} only for $\chi=n_{\mathcal T}$, \Cref{app:cswap-count}\newedit{. Its leading order is $N$ whenever $\chi$ grows more slowly than $N$, and $3N/2$ for the complete family of three-site supports, \Cref{eq:triple-minimum}. For other families whose $\chi$ grows in proportion to $N$, including the complete families of four or more sites, the leading order remains open}. The nonlinear control cost remains $O(\log N)$ for supports of fixed size, \Cref{app:sequential-tree-controls}, \newedit{and a minimum-size network can always be controlled with $O(\log N)$ control qubits when \PREPARE{} writes one address per support, but whether some address encoding also keeps its control openings $O(\log N)$ is open}. \newedit{Since every operator class on a support reuses the same permutation, the margin over walks that pay once per class grows with the number of classes, so} \newedit{models with many operator classes per support, such as multi-orbital lattice models,} are natural next targets. More broadly, these results suggest \newedit{compiling the geometry of a Hamiltonian into the permutation that selects its supports, which then sets the leading cost, and its repeated local operators into a target action that every support shares}.
\section*{Code availability}
\newedit{The Qualtran~\cite{Qualtran2024} implementations used to obtain the resource counts, together with the verification scripts, are available from the authors upon request.}

\section{Acknowledgements}
\label{sec: acknowledgements}

We thank Élie Gouzien, Vivien Londe, \newedit{and Alice Moutenet} for fruitful discussions.

Generative artificial-intelligence tools assisted in this work. They helped to design some of the circuit subroutines, to develop parts of the proofs, and to edit the manuscript. The authors checked every AI-assisted construction and proof, and they take full responsibility for the content of this paper.
 
\hbadness=3500 %

\hbadness=1000

\appendix
\crefalias{section}{appendix}
\section{Permutations}
\label{app:permutations}

This appendix derives the counts that enter \Cref{eq:pau-pareto}: the \CSWAP{}s, the nonlinear controls $\Npred$ and the repair openings $\Nrep$. We first give the gate identities and prove the lower bound on the number of \CSWAP{}s, then construct the address controls and the nearest-neighbour and all-to-all networks. The final subsection derives the repairs needed to bridge gates inside these networks.

\subsection{Gate cost and work-bit reuse}
\label{app:bridges}
\label{app:cswap-opening}

A temporary \AND{} closes for free when the stored bit still equals the product of its current controls. In the circuits, a stored product begins one column after its opening. Whether the work bit of a \CSWAP{} can be reused for its inverse depends on how the action between them changes its inputs.

\subsubsection{Keeping the work bit across the action}
\label{app:cswap-bridge}

For an isolated matched pair, suppose the target action changes the pair parity by $\delta$. The inverse gate needs $\wb\oplus c\delta$, so applying $X_\wb^{c\delta}$ repairs the stored bit and allows a free close, \Cref{eq:repair-bit}. This requires $\delta$ to depend only on the prepared address and label. Pauli actions satisfy this condition: their $X$ exponents determine the bit flips, while their $Z$ factors leave the parity unchanged. An action that superposes parity sectors has no such fixed $\delta$ and uses the closed schedule.

A diagonal action needs no repair. An affine repair predicate $c\delta$ is applied by \CNOT{}s, so the pair again pays only one opening. A product predicate needs an opening of its own, so an isolated pair then costs two openings, as a closed one does; bridging saves only when several work bits share that product. A bridge that saves an opening saves $4\,T$ and retains one work bit, as counted in \Cref{eq:pau-pareto}. \Cref{app:repair-count} extends this rule to a network.

\subsubsection{Relation to the adjacent-\texorpdfstring{\AND{}}{AND} merge}
\label{app:and-merge}

The adjacent-\AND{} merge of Ref.~\cite{Babbush2018} also removes a reopening, by commuting it next to the preceding close.

\begin{figure}[htbp]
\centering

\begin{tabular}{@{}c@{\hspace{0.1em}}c@{\hspace{0.1em}}c@{}}

\multicolumn{3}{c}{
\textbf{(a) $\delta=0$: the \AND{} pair cancels}
}
\\[0.4em]
\begin{quantikz}[row sep={0.45cm,between origins}, column sep=0.2cm,align equals at=2.5]
\lstick{$c$} & \control{} \wire[d][3]{q} &  &  &  &  &  & \control{} \wire[d][3]{q} &  \\
\lstick{$q_0$} & \control{} &  & \targ{} \gategroup[2,steps=3,style={draw=none,fill=gray!25!white,rounded corners=2pt,inner xsep=1pt,inner ysep=-1pt},background]{} & \gate[wires=2,style={draw=CrayonBlue,fill=CrayonBlueFill,
    rounded corners=2pt,inner ysep=0pt}]{\Pact} & \targ{} &  & \control{} &  \\
\lstick{$q_1$} &  &  & \ctrl{-1} &  & \ctrl{-1} &  &  &  \\
\lstick{$\wb$} &  & \setwiretype{n} &  &  &  &  &  & \setwiretype{q}
\end{quantikz}
&
$\overset{\scalebox{0.7}{
\begin{quantikz}[row sep={0.45cm,between origins}, column sep=0.2cm]
& \control{} \wire[d][2]{q} & \control{} \wire[d][2]{q} &  \\
& \control{} & \control{} &  \\
&  & \setwiretype{n} & \setwiretype{q}
\end{quantikz}}}{=}$
&
\begin{quantikz}[row sep={0.45cm,between origins}, column sep=0.2cm,align equals at=2.5]
&[1.79768pt] &  &  &  & &[1.79768pt] \\
& & \targ{} \gategroup[2,steps=3,style={draw=none,fill=gray!25!white,rounded corners=2pt,inner xsep=1pt,inner ysep=-1pt},background]{} & \gate[wires=2,style={draw=CrayonBlue,fill=CrayonBlueFill,
    rounded corners=2pt,inner ysep=0pt}]{\Pact} & \targ{} & &  \\
& & \ctrl{-1} &  & \ctrl{-1} & &  \\
& &  &  &  & &
\end{quantikz}
\\[2.4em]
\multicolumn{3}{c}{
\textbf{(b) $\delta=1$: the \AND{} pair leaves $\CNOT_{c\to\wb}$}
}
\\[0.em]
\begin{quantikz}[row sep={0.45cm,between origins}, column sep=0.2cm,align equals at=2.5]
\lstick{$c$} & \control{} \wire[d][3]{q} &  &  &  &  &  & \control{} \wire[d][3]{q} &  \\
\lstick{$q_0$} & \control{} &  & \targ{} \gategroup[2,steps=3,style={draw=none,fill=gray!25!white,rounded corners=2pt,inner xsep=1pt,inner ysep=-1pt},background]{} & \gate[wires=2,style={draw=CrayonBlue,fill=CrayonBlueFill,
    rounded corners=2pt,inner ysep=0pt}]{\Pact} & \targ{} &  & \control{} &  \\
\lstick{$q_1$} &  &  & \ctrl{-1} &  & \ctrl{-1} &  &  &  \\
\lstick{$\wb$} &  & \setwiretype{n} &  &  &  &  &  & \setwiretype{q}
\end{quantikz}
&
$\overset{\scalebox{0.7}{
\begin{quantikz}[row sep={0.45cm,between origins}, column sep=0.2cm]
& \control{} \wire[d][2]{q} & \control{} \wire[d][2]{q} &  \\
& \control{} & \ocontrol{} &  \\
&  & \setwiretype{n} & \setwiretype{q}
\end{quantikz}}}{=}$
&
\begin{quantikz}[row sep={0.45cm,between origins}, column sep=0.2cm,align equals at=2.5]
& &  &  & \ctrl{3} \gategroup[4,steps=1,style={draw=none,fill=CrayonRedWash,rounded corners=2pt,inner xsep=-1pt,inner ysep=-1pt},background]{} &   \\
& \targ{} \gategroup[2,steps=3,style={draw=none,fill=gray!25!white,rounded corners=2pt,inner xsep=-1pt,inner ysep=-1pt},background]{} & \gate[wires=2,style={draw=CrayonBlue,fill=CrayonBlueFill,
    rounded corners=2pt,inner ysep=0pt}]{\Pact} & \targ{} &  &  \\
& \ctrl{-1} &  & \ctrl{-1} &  &  \\
& &  &  & \targ{}  &  
\end{quantikz}
\end{tabular}
\caption{The two cases of the \AND{}-merge identity, \Cref{eq:and-merge}. In each left-hand circuit, the first blank corner is the free close of the \AND{} opened in the forward permutation. The second corner is its $4\,T$ reopening for the inverse permutation. The grey block between them is the target action $\Pact$, conjugated by the two $\CNOT_{q_1\to q_0}$ of the \CSWAP{}s. This block leaves $c$ unchanged and sends $q_0\mapsto q_0\oplus\delta$. Commuting the reopening to its left therefore shifts its data control by $\delta$. The small circuit above each equals sign is the resulting pair of corners. (a)~For $\delta=0$ the corners cancel. (b)~For $\delta=1$ the shifted control is reversed, drawn as an open circle. The pair then leaves the Clifford $\CNOT_{c\to\wb}$, which commutes with the grey block. Neither right-hand circuit contains a paid corner, and in both $\wb$ stays live across $\Pact$.}
\label{fig:and-merge}
\end{figure}
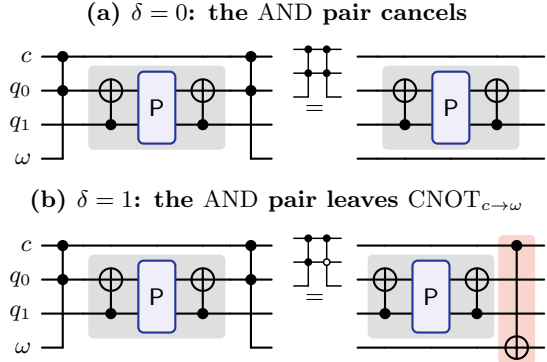

Consider a closed pair whose target action flips the pair parity by $\delta$. Commuting the reopening backwards through the central $\CNOT$--$\Pact$--$\CNOT$ block sends $q_0\mapsto q_0\oplus\delta$. The reopening then sits next to the deferred close, where
\begin{equation}
\AND_{c,\,q_0\oplus\delta\to\wb}\;
\AND_{c,\,q_0\gets\wb}
=
X_{\wb}^{c\delta}.
\label{eq:and-merge}
\end{equation}
For $\delta=0$ the two corners cancel. For $\delta=1$ they leave the Clifford $\CNOT_{c\to\wb}$, \Cref{fig:and-merge}. In both cases the pair pays a single opening, and $\wb$ stays live across $\Pact$.

The remaining gate is precisely the repair of \Cref{eq:repair-bit}. Bridging applies this relation directly, even when intervening \CSWAP{}s prevent the corners from becoming adjacent.

\subsubsection{The fermionic \texorpdfstring{\CSWAP{}}{CSWAP}}
\label{app:fermionic-cswap}

For fermionic modes the network keeps its layout, and each \CSWAP{} becomes a controlled fermionic transposition, the non-adjacent form of the fermionic swap of Refs.~\cite{Verstraete2009,Kivlichan2018}. It exchanges modes $q_i$ and $q_j$ with the Jordan--Wigner sign $(-1)^{q_iq_j+(q_i\oplus q_j)\sum_{i<k<j}q_k}$. The first term is the sign of exchanging two occupied modes. The second is the sign of carrying one fermion past the occupied modes between them. Consider the work bit $\wb=c\,(q_i\oplus q_j)$, once it is open and before the exchange writes $q_j$. The identities $c\,q_iq_j=c\,q_j\oplus q_j\wb$ and $c\,(q_i\oplus q_j)\,q_k=\wb\,q_k$ then turn the controlled sign into
\begin{equation}
\begin{split}
&(-1)^{c\,[q_iq_j+(q_i\oplus q_j)\sum_{i<k<j}q_k]}\\
&\qquad=\CZ_{c,q_j}\,\CZ_{q_j,\wb}\prod_{i<k<j}\CZ_{\wb,q_k}.
\end{split}
\label{eq:fermionic-transposition}
\end{equation}
A transposition thus costs the $4\,T$ of its \CSWAP{} plus up to $N$ \CZ{} gates, and it needs no further qubit. A network of \CSWAP{}s therefore keeps its $T$ count and its ancillae when it permutes fermionic modes. \Cref{app:lattice-addresses} applies the transposition to lattice fermions, and \Cref{app:aa-network} reduces its \CZ{} gates \newedit{to a number linear in $N$} for the all-to-all network.

\subsection{Swap trees and the \texorpdfstring{\CSWAP{}}{CSWAP} count}
\label{app:cswap-minimum}

A swap tree selects one site. Combining trees for different site groups gives the support-dependent count of \Cref{eq:pau-chi-switches}. We construct the tree before deriving the bound that it attains for pairs.

\subsubsection{The swap tree}
\label{app:swap-tree}

\begin{figure}[t]
\centering
\crayontitle{CrayonYellow}{\textbf{Swap tree at $M=10$}}\par\smallskip
\begin{quantikz}[row sep={0.36cm,between origins},column sep=0.30cm]
\lstick{$a_0$} & \ctrl{4} & \ctrl{6} & \ctrl{8} & \ctrl{10} & \ctrl{12} &  &  &  &  & \\
\lstick{$a_1$} &  &  &  &  &  & \ctrl{3} & \ctrl{7} &  &  & \\
\lstick{$a_2$} &  &  &  &  &  &  &  & \ctrl{2} &  & \\
\lstick{$a_3$} &  &  &  &  &  &  &  &  & \ctrl{1} & \\
\lstick{$0$} & \swap{1} &  &  &  &  & \swap{2} &  & \swap{4} & \swap{8} & \rstick{$a$} \\
\lstick{$1$} & \targX{} &  &  &  &  &  &  &  &  & \\
\lstick{$2$} &  & \swap{1} &  &  &  & \targX{} &  &  &  & \\
\lstick{$3$} &  & \targX{} &  &  &  &  &  &  &  & \\
\lstick{$4$} &  &  & \swap{1} &  &  &  & \swap{2} & \targX{} &  & \\
\lstick{$5$} &  &  & \targX{} &  &  &  &  &  &  & \\
\lstick{$6$} &  &  &  & \swap{1} &  &  & \targX{} &  &  & \\
\lstick{$7$} &  &  &  & \targX{} &  &  &  &  &  & \\
\lstick{$8$} &  &  &  &  & \swap{1} &  &  &  & \targX{} & \\
\lstick{$9$} &  &  &  &  & \targX{} &  &  &  &  &
\end{quantikz}\par
\caption{Swap tree with ten leaves and nine exchanges. The root, leaf $0$, receives the content of the selected leaf $a$. Each exchange reads one address bit $a_t$; exchanges that would reach beyond leaf $9$ are omitted. A leaf may hold one qubit, giving a \CSWAP{}, or a block of qubits, giving one \CSWAP{} per qubit under the same control. The latter is the chain-front permutation of \Cref{app:aim-onebody-bridge}. Neither case needs products of address bits. The inverse runs the columns in reverse order.}
\label{fig:swap-tree}
\label{fig:aim-set-tournament}
\label{fig:aim-impurity-tournament}
\label{fig:ring-permutation-fredkin}
\end{figure}
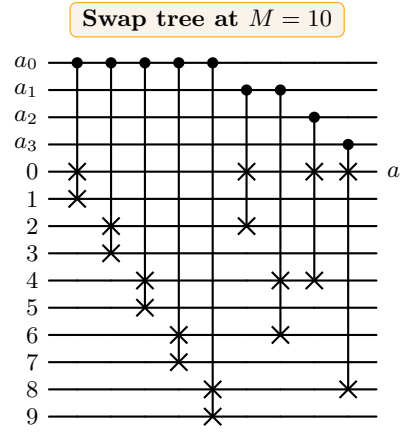

A swap tree acts on $M$ leaves, numbered by the address $a$ that selects them, and its root is leaf zero. Layer $t$ exchanges every leaf whose lowest $t+1$ bits are zero with the leaf $2^t$ above it, under the control of the address bit $a_t$. These two leaves are the roots of two \emph{subtrees} of up to $2^t$ leaves each, whose selected contents the earlier layers have already gathered there. The selected content moves down by $2^t$ exactly when $a_t=1$, so after $\lceil\log_2M\rceil$ layers it sits at the root, \Cref{fig:swap-tree}. It thus passes through the \emph{path} of leaf $a$, the \CSWAP{}s that act on the roots of the subtrees holding $a$, one per layer. For $M$ a power of two, the layers hold $M/2+M/4+\dots+1=M-1$ \CSWAP{}s. For other $M$, the \CSWAP{}s whose upper leaf lies beyond the $M$ leaves are omitted. The tree still permutes its $M$ leaves reversibly, and it still holds $M-1$ \CSWAP{}s, \Cref{fig:swap-tree}. Its inverse runs the layers in reverse order.

Every \CSWAP{} of the tree is enabled by one address bit, so a closed swap tree costs $M-1$ openings in each direction and needs no nonlinear control. A leaf may also hold several qubits. A \emph{block} \CSWAP{} then exchanges all of them under one control. It counts as one \CSWAP{} per qubit, since each qubit needs its own work bit, but a tree of block \CSWAP{}s moves whole supports at once, which matters for bridging, \Cref{app:repair-count}.

\subsubsection{Support geometry fixes the \texorpdfstring{\CSWAP{}}{CSWAP} count}
\label{app:cswap-count}

Let each support contain $n_{\mathcal T}$ of the $N$ sites, with every site in at least one support. Let $\chi$ be the least number of groups such that no support contains two sites from the same group, the strong chromatic number of the family.

Consider a fixed sequence of \CSWAP{}s, each enabled by a Boolean function of the prepared address, independently of the system state. For every support, some address must bring its sites to the target positions, in some order. Extra positions are allowed, and spectators are unconstrained. Unconditional exchanges can be absorbed into position labels. The bound counts conditional transpositions only; other primitives, such as controlled three-cycles, lie outside this model. It also excludes the cost of computing the controls. A network with the fewest \CSWAP{}s need not have the lowest total $T$ count, as \Cref{app:address-predicates} explains.

No such network can bring every support to the target register with fewer than $(N-n_{\mathcal T})+\lceil(\chi-n_{\mathcal T})/2\rceil$ \CSWAP{}s. To prove this, we run the network backwards from the target positions and relax it, letting every \CSWAP{} be enabled or disabled independently of the others. The relaxation can only enlarge the set of site pairs that reach the target register together, so a bound proved under it holds for the network itself. Let $G$ be the graph on positions whose edges are the pairs reachable in this way. Initially, $G$ is the complete graph on the $n_{\mathcal T}$ target positions. A \CSWAP{} that exchanges positions $u$ and $v$ carries $G$ to its union with its image under the transposition $(u\,v)$.

Under this map, $G$ remains a complete multipartite graph on the positions it has activated, with every other position isolated. The initial graph has this form, with singleton parts. Four cases exhaust a step, as follows.
\begin{itemize}
\item With both endpoints isolated, or both in one part, $G$ is unchanged.
\item With one endpoint isolated, that endpoint joins the part of the other. This activates exactly one position.
\item With the endpoints in different parts, each is split off as a singleton whenever its part held at least two positions. The step adds exactly the edges from each endpoint to the other members of its own old part, and it creates at most two singletons.
\end{itemize}

Write $a$ for the number of active positions and $p$ for the number of parts. We follow $D=a-p$, which starts at zero since $a=p=n_{\mathcal T}$. An activation raises $D$ by one, a split lowers it by at most two, and no other step changes it. A sequence with $n_{\mathrm{act}}$ activations and $n_{\mathrm{spl}}$ splits thus ends at $D\ge n_{\mathrm{act}}-2n_{\mathrm{spl}}$.

A network that serves every support must reach, under the relaxation, a graph that contains every pair of sites sharing a support. Every site is then active, so $a\ge N$. No part of a complete multipartite graph contains an edge, so no support holds two sites of one part. The parts that contain sites therefore form a grouping of the kind that defines $\chi$, and $p\ge\chi$. At the end, $n_{\mathrm{act}}=a-n_{\mathcal T}$ and $D=a-p$, so the inequality gives $n_{\mathrm{spl}}\ge(p-n_{\mathcal T})/2$. Activations and splits are distinct \CSWAP{}s, so the number of \CSWAP{}s obeys
\begin{equation}
\begin{aligned}
n_{\mathrm{act}}+n_{\mathrm{spl}}&\ \ge\ (a-n_{\mathcal T})+\left\lceil\frac{p-n_{\mathcal T}}2\right\rceil\\
&\ \ge\ (N-n_{\mathcal T})+\left\lceil\frac{\chi-n_{\mathcal T}}2\right\rceil .
\end{aligned}
\label{eq:cswap-minimum}
\end{equation}
Extra positions cannot help. Activating one raises the first term as fast as it could ever lower the second.

For pairs, the bound is attained for every bond graph. At $n_{\mathcal T}=2$ it reads $\lceil N+\chi/2-3\rceil$, and the network of \Cref{sec:pau-resources} spends exactly that many: it uses $N-\chi$ \CSWAP{}s in the swap trees of the groups and $\lceil3(\chi-2)/2\rceil$ in the all-to-all network on the $\chi$ tree roots, which is \Cref{eq:pau-chi-switches}. The network therefore attains the minimum \CSWAP{} count for pairs. In particular, the all-to-all network of \Cref{tab:pau-geometries,app:aim-query-interaction}, the case $\chi=N$, is \CSWAP{}-optimal at every $N$. Between the two extremes of \Cref{tab:pau-geometries}, the geometry enters only the constant. The triangular lattice, with $\chi=3$ under open boundaries or periodic sides divisible by three, needs $N-1$ \CSWAP{}s. Bonds of range $R$ on an open chain, with $\chi=R+1$, need $N-2+\lceil(R-1)/2\rceil$.

For supports of $n_{\mathcal T}\ge3$ sites, the bound is attained only in part. When $\chi=n_{\mathcal T}$, the network on the roots is empty, and the swap trees of the groups alone spend $N-n_{\mathcal T}$ \CSWAP{}s, which meets \Cref{eq:cswap-minimum}. The plaquettes of a square lattice, with the four groups of a $2\times2$ cell, and the triangles of a triangular lattice, with three, are of this kind when the boundaries allow these groups: open boundaries always do, and periodic ones do when every side is even for plaquettes and divisible by three for triangles. Other periodic sides need more groups, for instance nine on a $3\times3$ torus of plaquettes and five on a $5\times5$ one.

When $\chi>n_{\mathcal T}$, the minimum is bracketed\newedit{. The bracket fixes its leading order whenever $\chi$ grows more slowly than $N$, as shown below. When $\chi$ grows in proportion to $N$, the leading order is known only for all three-site supports, \Cref{app:all-triples}}. From below, \Cref{eq:cswap-minimum} reads $\lceil N+(\chi-3n_{\mathcal T})/2\rceil$. From above, the swap trees of the groups can be followed by $n_{\mathcal T}$ sequential swap trees on the roots, the $t$-th of which takes the $\chi-t+1$ roots still available as its leaves and brings the $t$-th member of the support to its root, the $t$-th qubit of $\mathcal T$. This spends exactly $N-\chi+\sum_{t=1}^{n_{\mathcal T}}(\chi-t)=N-\chi+n_{\mathcal T}\chi-n_{\mathcal T}(n_{\mathcal T}+1)/2$ \CSWAP{}s, and $\sum_{t=1}^{n_{\mathcal T}}(N-t)=n_{\mathcal T}N-n_{\mathcal T}(n_{\mathcal T}+1)/2$ at $\chi=N$. Its controls cost $O(\log N)$ openings, \Cref{eq:sequential-tree-controls}. The two ends differ by an amount that depends on $\chi$ and $n_{\mathcal T}$ alone, and both are $N+O(n_{\mathcal T}\chi)$. Hence $\Sw=1$ whenever $\chi$ grows more slowly than $N$, in particular whenever it stays bounded, and $\Sw>1$ exactly when $\chi$ grows in proportion to $N$.

\newedit{Whether a given network delivers a given support is a question of vertex-disjoint paths. Unfold the network in time, with one vertex of capacity one per position and stage, a straight arc for each position that a \CSWAP{} leaves untouched, and all four arcs through each \CSWAP{}. Disjoint paths from the sites of the support to $\mathcal T$ fix a partial matching of the ports of each \CSWAP{}, which extends to the exchange or the identity, so the support is delivered exactly when such paths exist. A maximum-flow computation therefore checks a network support by support, but it gives no closed form, and \Cref{eq:cswap-minimum} remains the closed-form bound.}

\subsubsection{All three-site supports}
\label{app:all-triples}

\newedit{When every set of three sites is a support, $n_{\mathcal T}=3$ and $\chi=N$. \Cref{eq:cswap-minimum} then gives $\lceil3(N-3)/2\rceil$ \CSWAP{}s and the sequential trees spend $3N-6$. The minimum lies at neither end. It is}
\begin{equation}
\newedit{\frac32N+\frac12\sqrt N+O(N^{1/4})\ \text{\CSWAP{}s},}
\label{eq:triple-minimum}
\end{equation}
\newedit{so the density is $\Sw=3/2$, as for all-to-all pairs, but the count exceeds $3N/2$ by a term that grows as $\sqrt N$.}

\newedit{The upper bound comes from a network that sorts the selected sites into distinct bins before gathering them. For $N\ge7$, take a simple graph with $\lfloor N/2\rfloor$ edges on the least number of vertices that can carry them, the bins, none of them isolated, for instance its first $\lfloor N/2\rfloor$ edges in lexicographic order. Each edge holds two sites, one in each of its two bins. For odd $N$, the remaining site forms a bin of its own, and $n_{\mathrm{bin}}$ counts the bins of both kinds. A first layer of $\lfloor N/2\rfloor$ \CSWAP{}s exchanges the two sites of each edge, so every site can be placed in either bin of its edge. Any three selected sites can be placed in three distinct bins. Each site has two bins to choose from, except the site with a bin of its own, which no other site can take. Three sites confined to two bins would lie on a single edge, since the graph is simple, but an edge holds only two sites. By Hall's theorem~\cite{Hall1935} the three sites therefore have distinct bins. When both sites of one edge are selected, they go to its two bins, and both bin assignments need the same setting of that edge's \CSWAP{}. Each bin then holds at most one selected site, and a swap tree per bin brings it to the root of the bin, at $N-n_{\mathrm{bin}}$ \CSWAP{}s for all bins together. The same construction on the $n_{\mathrm{bin}}$ roots leaves $n'_{\mathrm{bin}}$ roots, three of which hold the selected sites, and three sequential swap trees on these roots, as in \Cref{app:cswap-count}, bring them to $\mathcal T$. Each level on $n$ sites spends $\lfloor n/2\rfloor$ \CSWAP{}s in its edge layer and $n$ minus its number of bins in its trees, and the sequential trees spend $3n'_{\mathrm{bin}}-6$, so the network spends}
\begin{equation}
\newedit{n_{\CSWAP}(N)=\Big\lfloor\frac{3N}2\Big\rfloor-n_{\mathrm{bin}}+\Big\lfloor\frac{3n_{\mathrm{bin}}}2\Big\rfloor+2n'_{\mathrm{bin}}-6}
\label{eq:triple-network}
\end{equation}
\newedit{\CSWAP{}s, three fewer when $n'_{\mathrm{bin}}=3$, since the three roots then form $\mathcal T$. For instance, $30$ sites on six bins of five sites spend $15+24+6=45$ \CSWAP{}s, the last six on three bins of two roots. Since $n_{\mathrm{bin}}=\sqrt N+O(1)$ and $n'_{\mathrm{bin}}=O(N^{1/4})$, \Cref{eq:triple-network} gives the upper bound of \Cref{eq:triple-minimum}.}

\newedit{The enables depend only on the three selected sites, and \PREPARE{} writes each of them as a code from which the network reads every field directly. A site is named by the labels of the two bins of its edge, in increasing order, and one bit for its side; a site with a bin of its own repeats the label of that bin. On the second level each bin is a site, and its label is its code there. The code of a site thus holds $\log_2N+O(1)$ bits. In the tree of its bin, a site sits at the leaf numbered by the label of the other bin of its edge. A swap tree with sparsely numbered leaves still spends one \CSWAP{} fewer than it has leaves, because only a branching with two occupied subtrees carries one. At each level, the control generator reads the two candidate bins of each selected site from its code and chooses distinct ones by a fixed number of label comparisons. The edge layer reads one affine Boolean function of the edge label. Such a function can take any prescribed values on three distinct labels, because their two differences from the first are distinct and nonzero, hence linearly independent over $\mathbb F_2$, and elimination on their lowest nonzero bits finds its coefficients with $O(\log N)$ openings. A tree layer cannot use one such function, because every occupied bin needs its own leaf, and a separate function for each bit of the leaf would cost $O(\log^2N)$ openings. Instead, the labels are read as elements of a binary field whose degree is a power of two, and the tree of every bin reads the bits of one polynomial of degree two at the label of that bin, the polynomial that takes the leaf of each selected site at the label of its bin. Multiplication by a fixed label is linear over $\mathbb F_2$, so every enable is a parity of the three stored coefficients. These follow from the three bins and leaves by eleven multiplications and one inversion in the field, and the inversion reduces, through the norm to the subfield of half the degree, to an inversion there and three multiplications. A bilinear multiplication algorithm spends one opening for each of its products of two linear forms, and over $\mathbb F_2$ there are such algorithms with a number of products linear in the degree of the field~\cite{BalletPieltant2011}. The three sequential trees read the labels of the second level, updated by \Cref{eq:sequential-tree-controls}. The controls therefore cost $O(\log N)$ openings and as many retained control qubits in all. They serve both directions of the permutation, so the closed \SELECT{} costs $12N+O(\sqrt N)$ $T$ gates.}

\newedit{The lower bound sharpens the argument of \Cref{eq:cswap-minimum}, which follows only pairs. Run the network backwards from $\mathcal T$ and attach to every position a vector $\phi_u$ in a three-dimensional space over a field of rational functions, starting from the three unit vectors on the qubits of $\mathcal T$ and the zero vector elsewhere. A \CSWAP{} on positions $u$ and $v$ replaces $\phi_u$ and $\phi_v$ by two linear combinations of them with fresh indeterminate coefficients. A determinant of three vectors that contains exactly one of $\phi_u,\phi_v$ becomes a generic combination of its old value and the value for the triple with $u$ and $v$ exchanged, so it is nonzero exactly when one of the two was. A determinant that contains both is multiplied by a nonzero factor, and one that contains neither is unchanged. Three positions can therefore reach $\mathcal T$ together exactly when their vectors are linearly independent. A network that serves every triple ends with $N$ vectors of which no three are linearly dependent, which are $N$ points of the projective plane with no three on one line.}

\newedit{We count the points that this requires. A \CSWAP{} between a zero and a nonzero vector activates a position, as in the proof of \Cref{eq:cswap-minimum}, and creates no new point. Activations are the only steps that add a nonzero vector, and three are nonzero at the start, so there are $n_{\mathrm{act}}\ge N-3$ of them. A \CSWAP{} between two vectors that are not parallel, a mixing step, creates two new points on the line through the two old ones, distinct from each other and from every point that has appeared before. No other \CSWAP{} creates a point. With $n_{\mathrm{mix}}$ mixing steps, $3+2n_{\mathrm{mix}}$ points appear in all, and $n_{\mathrm{exc}}=3+2n_{\mathrm{mix}}-N$ of them are not among the final $N$. Each mixing line holds four of the points that appear, at most two of them final, so it contains a pair of the other $n_{\mathrm{exc}}$. No pair of points lies on two lines, so there are at most $n_{\mathrm{exc}}(n_{\mathrm{exc}}-1)/2$ mixing lines. Every final point other than the three initial ones was created on a mixing line, which holds at most two final points. Hence $N\le3+n_{\mathrm{exc}}(n_{\mathrm{exc}}-1)$\newedit{. For $N\ge4$ this requires $n_{\mathrm{exc}}\ge2$}, so $n_{\mathrm{exc}}\ge\bigl(1+\sqrt{4N-11}\bigr)/2$, and the \CSWAP{} count obeys}
\begin{equation}
\newedit{\begin{aligned}n_{\mathrm{act}}+n_{\mathrm{mix}}&\ \ge\ \frac{3N-9+n_{\mathrm{exc}}}{2}\\&\ \ge\ \left\lceil\frac{6N-17+\sqrt{4N-11}}{4}\right\rceil,\end{aligned}}
\label{eq:triple-bound}
\end{equation}
\newedit{which is $3N/2+\sqrt N/2-O(1)$ and gives the lower bound of \Cref{eq:triple-minimum}. Extra positions do not weaken it, because the points they hold at the end count among the $n_{\mathrm{exc}}$. The bin argument does not extend to four sites: four selected sites on the three edges of a triangle have only three bins between them. This construction therefore leaves the density for $n_{\mathcal T}\ge4$ at $\chi=N$ bracketed.}

\subsection{Address controls}
\label{app:controls}

The transposition bound excludes the cost of its controls. A swap tree reads single address bits and needs no further products. Address conversion and selection among tree roots can require nonlinear controls, counted by $\Npred$ in \Cref{eq:pau-closed-scaling}. The control generator computes and retains the products that the chosen permutation network needs. A toggle is needed only when consecutive trees share a field but read different addresses.

\subsubsection{Control generators}
\label{app:address-predicates}

The control generator $\mathsf G$ computes products of prepared bits before the first permutation, and $\mathsf G^\dagger$ erases them after the inverse permutation. They close by measurement once their inputs have been restored. Products include toggle carries or borrows and network controls, counted by $\Npred$, and repair predicates, counted separately by $\Nrep$. This separates control generation from the order of the system permutations: an all-to-all network reads the controls directly, whereas trees that share an address field use $\Tog$ between stages to change the field and later restore it. A product needed in only one stage may instead be computed around that stage in each direction. This adds one opening but can reduce peak workspace, as in \Cref{app:aa-network}.

The trees read one position field per group, so \PREPARE{} supplies an address of up to $\chi\lceil\log_2N\rceil$ bits. That address is charged to \PREPARE{}, while the \CSWAP{} count sets the leading term of \SELECT{}. A network on a bounded number of tree roots contributes a bounded number of controls. A logarithmic count arises in two places. The first is where the prepared address must be converted into the fields that the trees read, as for the toggle of \Cref{app:address-toggle}. The second is where the network ranges over all sites, as for the retained products of all-to-all pairs, \Cref{app:aa-network}.

\subsubsection{One address field for two trees}
\label{app:address-toggle}

When two trees read leaf indices that differ by a prepared bit, one register can serve both. The toggle $\Tog$ changes the first index into the second and restores it during the inverse permutation.

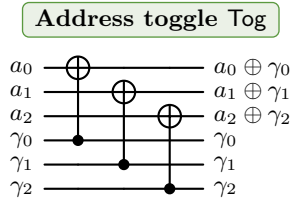
\begin{figure}[t]
\centering
\crayontitle{CrayonGreen}{\textbf{Address toggle $\Tog$}}\par\smallskip
\begin{quantikz}[row sep={0.32cm,between origins},column sep=0.28cm]
\lstick{$a_0$} & \targ{} &  &  & \rstick{$a_0\oplus\carry{0}$} \\
\lstick{$a_1$} &  & \targ{} &  & \rstick{$a_1\oplus\carry{1}$} \\
\lstick{$a_2$} &  &  & \targ{} & \rstick{$a_2\oplus\carry{2}$} \\
\lstick{$\carry{0}$} & \ctrl{-3} &  &  & \rstick{$\carry{0}$} \\
\lstick{$\carry{1}$} &  & \ctrl{-3} &  & \rstick{$\carry{1}$} \\
\lstick{$\carry{2}$} &  &  & \ctrl{-3} & \rstick{$\carry{2}$}
\end{quantikz}\par
\caption{Address toggle $\Tog$, shown for $m=3$. Adding the frozen chain $\carry{t}$ to the field $a$ steps it by $\rho=\carry{0}$, up for a carry chain and down for a borrow chain. Applying it again steps back. A wrap, when the range needs one, adds \CNOT{}s from $\carry{m}$ onto the set bits of $2^m-h$, not drawn here.}
\label{fig:address-toggle}
\label{fig:aim-address-toggle}
\end{figure}

Let the shared field $a$ hold $m$ bits and range over $h\le2^m$ values, and let a prepared bit $\rho$ set the step. Adding one to a binary number flips its lowest zero bit and every bit below it. Adding $\rho$ therefore flips bit $t$ of $a$ exactly when $\rho$ and all lower bits of $a$ are one, and subtracting $\rho$ flips it exactly when $\rho$ is one and all lower bits are zero. With $\carry{0}=\rho$, the chain $\carry{t}=\carry{t-1}a_{t-1}$ for an increment, or $\carry{t}=\carry{t-1}\bar a_{t-1}$ for a decrement, lists the bits that the step flips, and $m$ \CNOT{}s from $\carry{0},\ldots,\carry{m-1}$ onto $a$ perform it, \Cref{fig:address-toggle}.

The chain is computed once, from the initial field, at one opening per link, and then frozen. The same \CNOT{}s then undo the step, so $\Tog$ is its own inverse, and a circuit that reads both values in both permutation directions computes the chain once. The chain stays live while the system is permuted and closes by measurement, at no $T$ cost, once the field is restored.

Wrapping is where increments and decrements differ. When the step never leaves the range of the field, no wrap is needed. When it must wrap, a decrement that wraps from $a=0$ to $h-1$, rather than to $2^m-1$, needs one link more, the wrap flag $\carry{m}$, which drives \CNOT{}s onto the set bits of $2^m-h$. That flag is $\rho$ times the indicator that every bit of $a$ is zero, which is the next link of the borrow chain. An increment that wraps from $h-1$ to $0$ would instead need the indicator of $a=h-1$, which the carry chain does not contain unless $h=2^m$. A network that must wrap therefore numbers its leaves so that its step is a decrement, as the ring does, \Cref{app:nn-network}. A toggle costs $m-1$ openings, one more when $\rho$ is itself a product, one more for a wrap, and as many qubits.

\subsubsection{Controls of the sequential trees}
\label{app:sequential-tree-controls}

For supports of any fixed number $n_{\mathcal T}$ of sites, the sequential swap trees of \Cref{app:cswap-count} need only $O(\log N)$ nonlinear controls, up to all-to-all supports. \PREPARE{} writes the $n_{\mathcal T}$ selected sites, in any order.

Each tree is the swap tree of \Cref{app:swap-tree} on the $M$ leaves still in play, read by the current leaf $u$ of the next selected site. It brings that site to its root, which becomes the next qubit of $\mathcal T$, and the other leaves are renumbered in order. The content of any other leaf $v$ ends at a leaf fixed by a single bit of $v$: the lowest bit in which $v$ differs from $u$ and which, set with all lower bits cleared, keeps $v$ below $M$. The bits of $v$ above it are kept, that bit is set, and those below it are cleared. In layer $t$, only leaves whose lowest $t$ bits are zero are exchanged; one that agrees with $u$ in bit $t$ follows the exchange of $u$, and one that differs has bit $t$ set after the layer and never moves again, unless the exchange that would set that bit lies beyond the $M$ leaves.

The new leaf of every other selected site is thus read from its old one and from $u$ by three chains of products: a comparison of the prefixes of $v$ with the constant $M$, a search for the lowest admissible differing bit, and a masked copy of the bits above it. Together they hold at most $4\lfloor\log_2N\rfloor$ products. Each of the $n_{\mathcal T}(n_{\mathcal T}-1)/2$ pairs of selected sites needs one such update, so
\begin{equation}
\Npred\le 2n_{\mathcal T}(n_{\mathcal T}-1)\lfloor\log_2N\rfloor
\label{eq:sequential-tree-controls}
\end{equation}
with $n_{\mathcal T}N-n_{\mathcal T}(n_{\mathcal T}+1)/2$ \CSWAP{}s, the exact count $\sum_{t=1}^{n_{\mathcal T}}(N-t)$ of the sequential trees of \Cref{app:cswap-count} at $\chi=N$. Every product depends on the address alone, so $\mathsf G^\dagger$ erases it. For a family with $\chi$ groups, the same trees act on the $\chi$ tree roots and read the group labels that \PREPARE{} writes with the position fields, which replaces $\log_2N$ by $\log_2\chi$.

The nonlinear controls therefore stay subleading for every family of fixed support size. \newedit{For all three-site supports, the network of \Cref{app:all-triples} keeps them $O(\log N)$ at the minimum density $3/2$. Whether they also stay $O(\log N)$ at fewer than $n_{\mathcal T}$ \CSWAP{}s per site for other families with $\chi$ growing in proportion to $N$, in particular for four or more sites, is open.}

\subsubsection{Controls for any network}
\label{app:generic-controls}

\newedit{Any network that delivers every support, a minimum one included, can be controlled coherently from any prepared encoding of its $n_{\mathrm{addr}}\ge2$ valid addresses. Choose one enable pattern per valid address. A binary linear map then compresses the valid addresses injectively into $2\lceil\log_2n_{\mathrm{addr}}\rceil$ parity bits, at the cost of \CNOT{}s alone. Such a map exists because, for a random map, two given addresses collide with probability $2^{-2\lceil\log_2n_{\mathrm{addr}}\rceil}$, so the expected number of colliding pairs among the $n_{\mathrm{addr}}$ addresses is less than one half. The enable bit of each \CSWAP{} is the XOR of the equality tests between the compressed address and the compressed addresses that enable it. Each test multiplies the compressed bits, negated where needed, along a chain of temporary \AND{}s, copies the product into one accumulator qubit, and closes the chain by measurement; the same scratch serves every test. Computing the enable bit, using it and computing it again to clear the accumulator costs at most $2n_{\mathrm{addr}}(2\lceil\log_2n_{\mathrm{addr}}\rceil-1)$ openings per \CSWAP{} and direction, with at most $4\lceil\log_2n_{\mathrm{addr}}\rceil$ control qubits beyond the prepared address. With one address per support of fixed size, $n_{\mathrm{addr}}\le N^{n_{\mathcal T}}$, so the control workspace is $O(\log N)$ for every network.}

\newedit{The opening count, by contrast, grows with the number of supports. Nor can it be logarithmic for every compact encoding. For all triples of $N$ sites, a minimum network holds at most $3N-6$ \CSWAP{}s, so the minimum networks whose controls use $O(\log N)$ openings number $2^{O(N\log N)}$. The assignments of the $\binom N3$ supports to the first binary addresses number $2^{\Theta(N^3\log N)}$, and each controlled network realises at most one of them. Logarithmic control therefore depends on the address that \PREPARE{} writes, as for the trees and the toggle above.}

\subsection{Nearest-neighbour bonds}
\label{app:nn-network}
\label{app:pair-networks}
\label{app:permutation-instances}

The ring uses a nearest-neighbour network. The AIM combines this network on its bath chains with the all-to-all network of the next subsection on its impurity core.

Every bond of an open chain, or of a ring of even length, joins an even site to an odd one, so $\chi=2$, as for any bipartite bond graph. One swap tree over the odd sites brings the odd end of the selected bond to $q_1$, and one over the even sites brings the even end to $q_0$. The two trees hold $N-2$ \CSWAP{}s, the minimum of \Cref{eq:pau-chi-switches} at $\chi=2$. Their leaf indices differ by at most one, so one address field and the toggle of \Cref{app:address-toggle} serve both.

\begin{figure}[t]
\centering
\crayontitle{CrayonYellow}{\textbf{Nearest-neighbour network at $N=8$}}\par\smallskip
\begin{quantikz}[row sep={0.32cm,between origins}, column sep=0.2cm]
\lstick{$k_0$} & \ctrl{3} \gategroup[wires=4,steps=1,background,style={rounded corners=2pt,inner ysep=1pt,draw=none,fill=CrayonGreenWash,inner xsep=2pt}]{\scriptsize $\mathsf G$} &  &  &  &  &  & \ctrl{1} \gategroup[wires=4,steps=1,background,style={rounded corners=2pt,inner ysep=1pt,draw=none,fill=CrayonGreenWash,inner xsep=2pt}]{\scriptsize $\Tog$} &  &  &  &  &  & \rstick{$k_0$} \\
\lstick{$k_1$} & \control{} &  & \ctrl{4} \gategroup[11,steps=3,style={draw=none,fill=CrayonYellowWash,rounded corners=2pt,inner xsep=1pt,inner ysep=-1pt},background]{} & \ctrl{8} &  &  & \targ{} &  & \ctrl{3} \gategroup[10,steps=3,style={draw=none,fill=CrayonYellowWash,rounded corners=2pt,inner xsep=1pt,inner ysep=-1pt},background]{} & \ctrl{7} &  &  & \rstick{$k_1\oplus k_0$} \\
[0.1cm]
\lstick{$k_2$} &  &  &  &  & \ctrl{3} &  & \targ{} &  &  &  & \ctrl{2} &  & \rstick{$k_2\oplus k_0k_1$} \\
\lstick{$\carry{1}$} & \setwiretype{n} & \setwiretype{q} &  & \setwiretype{q} &  &  & \ctrl{-1} &  & \ctrl{-1} &  &  &  & \rstick{$k_0k_1$} \\
\lstick{$q_0$} &  &  &  &  &  &  &  &  & \swap{2} &  & \swap{4} &  & \rstick{$q_{k+k_0}$} \\
[0.1cm]
\lstick{$q_1$} &  &  & \swap{2} &  & \swap{4} &  &  &  &  &  &  &  & \rstick{$q_{k+\bar k_0}$} \\
\lstick{$q_2$} &  &  &  &  &  &  &  &  & \targX{} &  &  &  &  \\
\lstick{$q_3$} &  &  & \targX{} &  &  &  &  &  &  &  &  &  &  \\
\lstick{$q_4$} &  &  &  &  &  &  &  &  &  & \swap{2} & \targX{} &  &  \\
\lstick{$q_5$} &  &  &  & \swap{2} & \targX{} &  &  &  &  &  &  &  &  \\
\lstick{$q_6$} &  &  &  &  &  &  &  &  &  & \targX{} &  &  &  \\
\lstick{$q_7$} &  &  &  & \targX{} &  &  &  &  &  &  &  &  & 
\end{quantikz}\par
\caption{Nearest-neighbour network on an open chain at $N=8$, selecting $(k,k+1)$. The generator $\mathsf G$ first opens the carry $\carry{1}=k_0k_1$ at the blank two-input corner, costing $4\,T$. The first three \CSWAP{}s then select the odd endpoint using $\lfloor k/2\rfloor$. Between the trees, $\Tog$ increments this field by $k_0$ through two \CNOT{}s, using the frozen carry. The last three \CSWAP{}s select the even endpoint. After the target action, the inverse trees and toggle restore the system positions and address; only then does $\mathsf G^\dagger$ erase the carry by measurement.}
\label{fig:nn-network}
\end{figure}
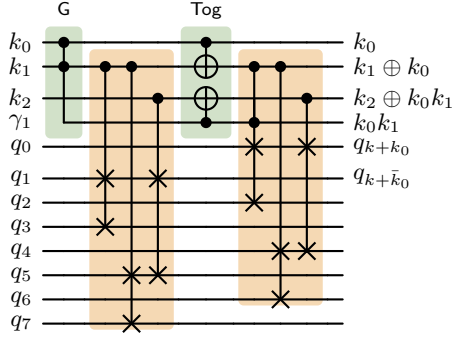

On an open chain, leaf $v$ of the odd tree holds $q_{2v+1}$ and leaf $v$ of the even tree holds $q_{2v}$. The bond $(k,k+1)$ then has its odd end at leaf $\lfloor k/2\rfloor$ and its even end at leaf $\lfloor k/2\rfloor+k_0$. The odd tree reads the high bits of $k$ directly, and the toggle increments them by $\rho=k_0$ before the even tree acts. The step never leaves the range, so no wrap is needed. \Cref{fig:nn-network} draws the network at $N=8$, with three \CSWAP{}s per tree and one opening for the carry $\carry{1}=k_0k_1$.

A ring of even $N$ also has the bond $(N-1,0)$. With the same leaves, its even end would be reached by an increment that wraps from $N/2-1$ to $0$, the wrap that the carry chain cannot supply, \Cref{app:address-toggle}. The ring therefore shifts the leaves of its even tree. If leaf $v$ holds site $2v+2$, with site $N$ read as site $0$, the bond $(j,j+1)$ has its odd end at leaf $\lfloor j/2\rfloor$ and its even end at leaf $(\lfloor j/2\rfloor-\bar j_0)\bmod(N/2)$. The step is then a decrement by $\rho=\bar j_0$, and the wrap is the last link of its borrow chain, at one opening unless $N/2$ is a power of two. The shift is a relabelling and costs no gate: the ring stores site $2v+2$ in the even qubit $q_{2v}$, so the even root is still $q_0$ and $\mathcal T=(q_0,q_1)$, as on the open chain. \Cref{app:heisenberg-ring-complete-query} draws the resulting control generator.

Under either boundary, which end of the bond lands on $q_0$ depends on the parity of the address. In \Cref{fig:nn-network}, $q_0$ receives $q_{k+k_0}$ and $q_1$ receives $q_{k+\bar k_0}$. \Cref{app:target-repairs} gives two ways to fix that order.

\subsubsection{Lattices share one address field per axis}
\label{app:lattice-addresses}

On a lattice, the swap trees need not read a separate position field per group: they can share one field per axis. The construction below shows how, and derives the lattice rows of \Cref{tab:pau-geometries}. Let the sites of a periodic square lattice of side $2h$ carry coordinates $(x,y)$, and let every tree index its sites by the halved coordinates $\lfloor x/2\rfloor$ and $\lfloor y/2\rfloor$, two fields of $m=\lceil\log_2h\rceil$ bits. Two consecutive coordinates then sit at fields that differ by at most one. As on the ring above, leaf $v$ of an even coordinate holds $2v+2$ instead of $2v$, which turns the difference into a decrement: the pair $\{x,x+1\}$ has its odd member at $a_x=\lfloor x/2\rfloor$ and its even member at $(a_x-\bar x_0)\bmod h$, whatever the parity of $x$. Every tree can therefore read one shared field per axis, decremented by a prepared bit $\rho_x$ where its member requires it. That decrement is the toggle of \Cref{app:address-toggle}, with $\rho=\rho_x$, so reading both values in both permutation directions costs $m-1+\iv{h\neq2^m}$ openings per axis, once.

The ring is the one-axis case, with $h=N/2$. Its row of \Cref{tab:pau-geometries} labels each field term by the tree root that receives it, rather than by its site, so every site is reached from its two bonds at half the coefficient. The mask then depends on the label alone, and the $j_0$-controlled orientation exchanges of \Cref{app:heisenberg-ring-complete-query} are not needed.

For plaquettes, the four groups are the coordinate parities $(o_x,o_y)$. The plaquette anchored at $(x,y)$ has the corners $\{x,x+1\}\times\{y,y+1\}$. It sets $\rho_x=\bar x_0$ and $\rho_y=\bar y_0$ for the trees of even coordinate. The trees run in the order $(1,1)$, $(0,1)$, $(0,0)$, $(1,0)$, with the $x$, $y$ and $x$ toggles between them, and the inverse permutation retraces that sequence. The two borrow chains cost $2(m-1+\iv{h\neq2^m})$ openings. With the $N-4$ \CSWAP{}s of \Cref{app:cswap-count}, this gives the plaquette row of \Cref{tab:pau-geometries}. The tree root of group $(o_x,o_y)$ receives the corner $(x+(o_x\oplus x_0),\,y+(o_y\oplus y_0))$. The anchor parities thus reflect the plaquette on $\mathcal T$, and the mask must be written in root order. A family invariant under these reflections, such as ring exchange, needs nothing, because \PREPARE{} chooses uniformly among strings of equal coefficient and a reflection only relabels them. The string $XYXY$ of the Wen plaquette model, with $X$ on the diagonal through the anchor, needs only \CNOT{}s from $x_0\oplus y_0$ onto its mask. A generic string needs its mask permuted by the two parities, at eight \CSWAP{}s.

\begin{table}[t]
\caption{Group-$1$ address of every Hubbard term, labelled by its group-$0$ mode $(x,y,\sigma)$. The shared fields $\lfloor x/2\rfloor$ and $\lfloor y/2\rfloor$ are decremented by $\rho_x$ and $\rho_y$ through the two axis toggles. The last column is flipped by a \CNOT{} from the term flag.}
\label{tab:hubbard-toggles}
\begin{ruledtabular}
\begin{tabular}{lcccc}
Term & Group-$1$ mode & $\rho_x$ & $\rho_y$ & Flipped\\
\hline
Hop along $x$ & $(x+(-1)^\tau,y,\sigma)$ & $\tau$ & $\bar y_0$ & $x_0$\\
Hop along $y$ & $(x,y+(-1)^\tau,\sigma)$ & $\bar x_0$ & $\tau$ & ---\\
$ZZ$, partner $Z$ & $(x,y,\bar\sigma)$ & $\bar x_0$ & $\bar y_0$ & $\sigma$\\
Group-$0$ $Z$ & unused & --- & --- & ---\\
\end{tabular}
\end{ruledtabular}
\end{table}

The modes $(x,y,\sigma)$ of the Hubbard model belong to the group $(x+y+\sigma)\bmod2$. Each tree indexes its modes by $(\lfloor x/2\rfloor,\lfloor y/2\rfloor,x_0,\sigma)$, which fixes $y_0$ within the group. Every term is labelled by its group-$0$ mode. A hop is also labelled by an orientation bit $\tau$, which is one for the step towards the smaller coordinate. Each hop is thus counted once, from its group-$0$ end. Each single-mode $Z$ is counted once, as a group-$0$ mode or as the on-site partner of one. The group-$0$ tree reads the label directly and holds no shifted leaves. The group-$1$ tree shifts the leaves of even $x$ and of even $y$. For its mode, a coordinate equal to that of the label is therefore read as the field decremented by the complemented low bit. A coordinate one step away is read as the field decremented by $\tau$, \Cref{tab:hubbard-toggles}. Two openings multiplex $\rho_x=\bar x_0\oplus f_x(\tau\oplus\bar x_0)$ and $\rho_y=\bar y_0\oplus f_y(\tau\oplus\bar y_0)$ from the hop flags $f_x$ and $f_y$. They stay live with the borrow chains. With the $N-2$ \CSWAP{}s, this gives the Hubbard row of \Cref{tab:pau-geometries}. Hops and $ZZ$ act alike on both roots, and each single-mode $Z$ acts on the root named by its label. The mask again depends on the label alone.

With the fermionic transposition of \Cref{app:fermionic-cswap}, consider a Jordan--Wigner order that places the two tree roots next to each other. The permutation then carries the two modes of every Hubbard term to adjacent positions. There the target operators are $XX$, $YY$, $Z$ and $ZZ$, without parity strings.

\subsection{All-to-all pairs}
\label{app:aa-network}

\begin{figure}[t]
\centering
\crayontitle{CrayonYellow}{\textbf{All-to-all network at $N=5$}}\par\smallskip
\begin{quantikz}[row sep={0.38cm,between origins},column sep=0.34cm]
\lstick{$j_2\oplus j'_2$} & \ctrl{8} &  &  &  &  & \\
\lstick{$j_1\oplus j'_1\oplus j'_2$} &  & \ctrl{5} &  &  &  & \\
\lstick{$j'_0\oplus j'_1\oplus1$} &  &  & \ctrl{4} &  &  & \\
\lstick{$j_1\oplus j_2$} &  &  &  & \ctrl{2} &  & \\
\lstick{$j_0\oplus j_1$} &  &  &  &  & \ctrl{1} & \\
\lstick{$q_0$} &  &  &  & \swap{4} & \swap{2} & \rstick{$q_j$} \\
\lstick{$q_1$} &  & \swap{1} & \swap{2} &  &  & \rstick{$q_{j'}$} \\
\lstick{$q_2$} &  & \targX{} &  &  & \targX{} & \\
\lstick{$q_3$} & \swap{1} &  & \targX{} &  &  & \\
\lstick{$q_4$} & \targX{} &  &  & \targX{} &  &
\end{quantikz}\par
\caption{Complementary-pair network at $N=5$, with the singleton $\{0\}$ and the complementary pairs $\{1,2\},\{3,4\}$. Each control row is an XOR parity of the address bits of the selected pair $(j,j')$, $j<j'$, so every \CSWAP{} is closed and needs no stored product. The first two \CSWAP{}s orient the two complementary pairs. The third is the swap tree rooted at $q_1$, over the sites $1$ and $3$, and the last two form the swap tree rooted at $q_0$, over the sites $0$, $2$ and $4$. The five \CSWAP{}s are drawn in the order that brings the selected sites in.}
\label{fig:aa-network}
\label{fig:aim-spin-up-permutation}
\end{figure}
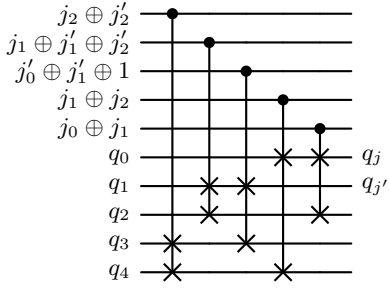

When any two sites may form a support, every group holds a single site, so $\chi=N$ and the network on the roots is the whole permutation. The complementary-pair network splits the sites into \newedit{one singleton at odd $N$ and two at even $N$} and the \emph{complementary pairs}, and selects the pair $(j,j')$, with $j<j'$. One orientation \CSWAP{} per complementary pair decides which of its two sites the swap tree rooted at $q_1$ sees and which the swap tree rooted at $q_0$ sees. One swap tree then brings the lower site $j$ to $q_0$, and the other brings the upper site $j'$ to $q_1$, \Cref{fig:aa-network}. The network spends $\lceil\tfrac32(N-2)\rceil$ \CSWAP{}s, five at $N=5$, which meets \Cref{eq:pau-chi-switches} at $\chi=N$. \newedit{At $N=5$, the singleton $\{0\}$ and the complementary pairs $\{1,2\},\{3,4\}$ need only XOR parities of the address bits of $(j,j')$ as controls. Larger networks} also read retained products of address bits, written by the control generator of \Cref{app:address-predicates} and counted below.

\newedit{At every $N$, one construction pairs neighbouring addresses and retains
\begin{equation}
\Npre(N)=\lceil\log_2N\rceil-2+\iv{N\text{ even}}=O(\log N)
\label{eq:aim-pair-network}
\end{equation}
products of address bits, except at $N=5$, where the network of \Cref{fig:aa-network} needs none and $\Npre(5)=0$. Write $m=\lceil\log_2N\rceil$ for the address width. For odd $N$, the complementary pairs are $\{2p,2p+1\}$ with $0\le2p\le N-3$, and the singleton $N-1$ joins the upper tree. The lower tree holds the even sites and reads the address without its lowest bit, $\lfloor j/2\rfloor$. The upper tree holds the odd sites and $N-1$ and reads $\lfloor j'/2\rfloor$. Both trees therefore read address bits directly. The orientation \CSWAP{} of the pair $\{2p,2p+1\}$ must exchange its two sites exactly when the pair holds $j$ at its odd site or $j'$ at its even site. Its control is
\begin{equation}
\begin{gathered}
c_p=j_0\oplus\bigoplus_{t=2}^{m}(p_{t-1}\oplus p_t)\,\psi_t,\\
\psi_m=1\oplus j_0\oplus j'_0,\qquad\psi_t=\psi_{t+1}(1\oplus j_t\oplus j'_t),
\end{gathered}
\label{eq:aa-orientation}
\end{equation}
where $p_t$ is bit $t$ of the address $2p$ and $p_m=0$. The $m-2$ products $\psi_{m-1},\ldots,\psi_2$ form one chain: $\psi_t$ is one exactly when $j$ and $j'$ have equal parity and agree in bits $t$ to $m-1$. If the parities differ, every $\psi_t$ vanishes and $c_p=j_0=1\oplus j'_0$, which is the value required on either pair. If they agree, $j$ and $j'$ lie in different pairs. Let $d\ge1$ be the highest bit in which they differ, so that $j_d=0$ and $j'_d=1$ because $j<j'$. Then $\psi_t=1$ exactly for $t>d$, and the sum telescopes to $c_p=j_0\oplus p_d$. This is $j_0$ on the pair of $j$ and $1\oplus j_0=1\oplus j'_0$ on the pair of $j'$, as required. The other orientation \CSWAP{}s act on sites that hold neither endpoint, and the singleton needs none. The network holds $\tfrac12(N-1)$ orientation \CSWAP{}s and $N-2$ tree \CSWAP{}s, $\lceil\tfrac32(N-2)\rceil$ in all.}

\newedit{For even $N$, the singletons are $0$, which can only be $j$, and $N-1$, which can only be $j'$. The complementary pairs are $\{2p,2p+1\}$ with $2\le2p\le N-4$, oriented by \Cref{eq:aa-orientation}, whose argument never used the site of the other endpoint, and the pair $\{1,N-2\}$ of the two sites left over. The lower tree holds the even sites and the upper tree the odd sites, so the pair $\{1,N-2\}$ must exchange exactly when $j=1$ or $j'=N-2$. Its control $1\oplus j'_0\oplus j_0j'_0$ vanishes only when $j$ is even and $j'$ odd. Among the pairs $(j,j')$ that involve site $1$ or $N-2$, this happens only for $(0,1)$ and $(N-2,N-1)$, where no exchange is wanted. The product $j_0j'_0$ is the one further product at even $N$. Each tree reads the address without its lowest bit, except that this exchange sends $j=1$ to the leaf of site $N-2$ and $j'=N-2$ to the leaf of site $1$. In both trees, these two leaves have the pair indices $0$ and $N/2-1$. Number the leaves of both trees by an invertible linear map of their pair index that sends $N/2-1$ to $1$, so that the two leaves differ only in the lowest bit. On such a numbering, a swap tree is built as in \Cref{app:swap-tree}: layer $t$ joins, at their smallest leaves, the two subtrees that agree above bit $t$ and differ in it, whenever both hold a leaf, so each tree still holds $N/2-1$ \CSWAP{}s. Every \CSWAP{} reads one bit of the new number, an XOR of address bits. The exception is the \CSWAP{} that joins the leaves numbered $0$ and $1$. It matters only for $j\in\{0,1,N-2\}$ in the lower tree and for $j'\in\{1,N-2,N-1\}$ in the upper tree, and it reads the lowest bit of the number XORed with $j_0$ in the lower tree and with $1\oplus j'_0$ in the upper tree, which sends each of these endpoints to its leaf. The network holds $N/2-1$ orientation \CSWAP{}s and $N-2$ tree \CSWAP{}s, again $\lceil\tfrac32(N-2)\rceil$ in all.}

\newedit{The control generator writes the products before the first permutation and erases them in reverse order after the inverse one, when their factors have been restored. Kept live across both directions, they cost $\Npre(N)$ openings at width $\Npre(N)$. At even $N$, the product $j_0j'_0$ may instead be opened around the \CSWAP{} of $\{1,N-2\}$ in each direction. This \CSWAP{} is placed farthest from the target action, so the product is never live together with the chain, and the network spends $\Npre(N)+1$ openings at width $\Npre(N)-1$. Ad hoc networks with one or two products fewer exist at some sizes, such as $N=5$, but they follow no general pattern.}

\paragraph{Fermionic signs in the all-to-all network.}
Summed over the all-to-all network, the strings of \Cref{eq:fermionic-transposition} hold $O(N\log N)$ \CZ{} gates, about $N$ per tree layer. \newedit{A parity rule reduces them to a number of Clifford gates linear in $N$, still without $T$. Order the modes for Jordan--Wigner by leaf number, alternating between the two swap trees. Every subtree then occupies consecutive modes of its tree, and every orientation \CSWAP{}, including that of $\{1,N-2\}$ at even $N$, exchanges adjacent modes and carries no string. The two swap trees act one after the other, and in the forward permutation each tree \CSWAP{} exchanges the roots of its two subtrees after both have acted, \Cref{app:swap-tree}. Every mode strictly between these two roots then either lies in the lower subtree, which has finished moving, or belongs to the other swap tree, which is idle. The string parity of the \CSWAP{} is therefore the parity of the lower subtree without its root, XOR the parity of the idle modes between the two roots. Neither parity changes while the tree acts on other qubits, so \CNOT{}s hold both in place. Two \CNOT{}s per \CSWAP{} write the parity of the merged subtree, without its root, into the upper root, which no later \CSWAP{} of the tree moves; the rule then applies again one layer up. A \CNOT{} ladder turns the idle tree into running parities, so the idle part of each string is the XOR of two of them. Each string thus takes at most three \CZ{} gates from $\wb$, and the \CNOT{}s are undone before the tree that holds them moves. Each \CSWAP{} adds a constant number of Clifford gates and each ladder is linear in $N$, so the signs of the network cost $O(N)$ Clifford gates in each direction.}

\subsection{Repairs inside a network}
\label{app:repair-count}

For a bridged gate $\lambda$ inside a network, the target action is carried back through the intervening \CSWAP{}s. Its parity change $\delta_\lambda(j,b)$ is the XOR of the $X$-mask bits that reach the two positions exchanged by $\lambda$. The $Z$ mask never enters. The repair is therefore $c_\lambda\delta_\lambda$, where $c_\lambda$ enables the gate. Affine predicates require only \CNOT{}s; nonlinear predicates are built from products and shared wherever possible. Their openings contribute $\Nrep$ in \Cref{eq:pau-pareto}. A diagonal action needs no repairs.

In a swap tree, a bridged \CSWAP{} sees a parity change only if it lies on the path of the leaf whose block the action flips. For a swap tree with a single such leaf, the repair factor of each bridged \CSWAP{} is
\begin{equation}
\rep_e=a_t\,\iv{\lfloor a/2^{t+1}\rfloor=\lfloor a_e/2^{t+1}\rfloor}.
\label{eq:aim-transport-indicator}
\end{equation}
Here $a$ is the address that the tree reads, $t$ is the layer of \CSWAP{} $e$, and $a_e$ is any leaf of the two subtrees that $e$ joins. The bracket states that $e$ lies on the path of leaf $a$, and the factor $a_t$ that $e$ is enabled, so the indicator is one exactly when $e$ moves the selected content. Every path through $e$ continues through one \CSWAP{} of layer $t+1$. \tocheck{The bracket of $e$ equals the indicator of that \CSWAP{} when bit $t+1$ of $a_e$ is one, and its bracket XOR its indicator when that bit is zero.} Built from the root layer towards the leaves, each indicator is therefore one product of $a_t$ with a parity, possibly complemented, of indicators already held. It costs one opening, and the parities cost only \CNOT{}s. In each layer the parity of all indicators is $a_t$ itself, so the last indicator of a layer costs only \CNOT{}s. The indicators of a tree thus cost one opening per \CSWAP{}, less one per layer, and they need no lookup and no quantum flip-pattern register. When the action is itself flagged, as by the hopping flag $f_\mathrm{T}$ of \Cref{case:aim}, every indicator carries the flag, and the last indicator of a layer is free only where the product of the flag with $a_t$ is already held. Corrections on distinct held work bits commute with the other inverse \CSWAP{}s, so all repairs may precede the inverse permutation.

Sharing determines the saving. If a block \CSWAP{} has $L$ lanes with the same parity change, one repair predicate serves all of them: the block pays $L$ openings plus the repair instead of $2L$. A gate that moves a single site of a pair usually needs its own path-dependent repair. Such a tree saves only those openings whose predicates follow as parities of existing products, while retaining one work bit per bridge. Both case studies therefore keep their nearest-neighbour trees closed, \Cref{app:heisenberg-ring-complete-query,app:aim-onebody-bridge}.
\section{Target actions and the preparation interface}
\label{app:target-actions}
\label{app:preparation}

\Cref{sec:interface} fixes what \PREPARE{} writes for the target action. We describe the label, the orientation and phase corrections, and the sampler that supplies them. For fixed target size and operator repertoire, the local action costs a number of $T$ gates independent of $N$; coefficient loading and any phase walks have separate costs, which can grow with $N$.

\subsection{Encoding and orienting the target action}
\label{app:target-repairs}

An explicit $XZ$ mask, \Cref{eq:target-xz}, applies one \CZ{} from $\zeta_q$ followed by one \CNOT{} from $\xi_q$ to each target qubit $q$. The action thus uses $2n_{\mathcal T}$ Clifford gates and no work qubit. The $X$ mask also determines the bridge repairs, \Cref{app:repair-count}. The coefficient sign and the phase required by the $Y$ factors are applied on the label register, as described below.

The mask adds no $T$ gate and no ancilla to \SELECT{}, as assumed in \Cref{eq:pau-closed-scaling,eq:pau-pareto}, but must be written by \PREPARE{}. Extra lookup columns are Clifford; the coherent alias sampler exchanges each output bit with its alternate at one opening per bit. The $2n_{\mathcal T}$ mask bits therefore add $8n_{\mathcal T}\,T$, independently of $N$ and the number of supports. A compressed label uses only enough bits to distinguish the operator classes, but may require decoding in the target action. Its cost $\Tact$ depends on the repertoire. If every exponent is a bit or a parity of label bits, the decoding is Clifford and $\Tact=0$.

A pair network can deliver its two sites in address-dependent order. One closed \CSWAP{} pair on $\mathcal T$ restores their order. Alternatively, the action can use the ordering bit to apply each factor where its site arrived. This requires products of the ordering bit with the relevant label bits. It pays when the slot-dependent factor is fixed, so that one product suffices, and only ties when that factor ranges over several axes. The ring keeps the exchange because its field terms use all three axes; the AIM absorbs it because its onsite factor is always $Z$, \Cref{app:heisenberg-ring-complete-query,app:aim-onebody-bridge}.

\subsection{Signs, frames and walks}
\label{app:sign-diagonal}
\label{app:frames-walks}

Coefficient signs and ordered-$XZ$ phases depend only on the label. A loaded sign bit requires a $Z$ gate\newedit{, and a loaded phase bit an $S$ gate}. A sign computed from the label requires only $Z$ and \CZ{} gates when its Boolean function is at most quadratic on the populated codes. \newedit{The phase $i^{n_Y}$ of the $Y$ factors, with $n_Y=\sum_q\xi_q\zeta_q$, is in general not Clifford in the mask bits, so \PREPARE{} loads it with the explicit mask, which is fixed by the sampled row. Its $i^2$ part joins the stored sign, and its parity is one further loaded bit, at $4\,T$. This bit vanishes when every term has an even number of $Y$ factors, as for any real Hamiltonian.} Such a correction is Clifford, needs no ancilla, and commutes with \SELECT{}, which is diagonal in the label basis. Unpopulated codes can be used to simplify the phase function.

When a term carries several factors, a \emph{parity frame}, a fixed \CNOT{} layer $\mathsf E$ applied before the first permutation and after the last, can gather its $X$ factors on fewer qubits. Only those qubits must reach a fixed target. Diagonal factors can instead be applied in place by a walk using the permutation controls. Each walk step costs one opening and closes for free because the intervening gates are diagonal. A \CSWAP{} that places such a factor is bridged without repair when only phases act between its two occurrences, \Cref{app:cswap-bridge}. The AIM interaction circuit $\mathsf I$ uses this combination, \Cref{app:aim-query-interaction}.

\subsection{The coefficient sampler}
\label{app:prepare-interface}

\PREPARE{} must give each term an amplitude set by its coefficient. One amplitude per term would add a lookup linear in the number of terms, which grows as the number of supports times the number of operator classes. Terms that share a magnitude, however, need it loaded only once, so the lookup grows only with the number of distinct coefficient rows, constant for the ring and linear in the bath for the AIM, and preparation alias-samples the distinct coefficient rows and uniform superpositions then pick one of the equal-weight terms of the sampled row. 
Loading only the distinct coefficients and spreading them over equivalent terms is standard. Ref.~\cite{Babbush2018} does so for the Hubbard and plane-wave Hamiltonians, and Ref.~\cite{Berry2019} for the symmetries of the two-electron integrals.
The signs are loaded with the magnitudes and applied by the diagonal of \Cref{app:sign-diagonal}.

The coherent alias sampler of Ref.~\cite{Babbush2018} loads an alternate row and a $\mu$-bit keep threshold. A comparison with a uniform $\mu$-bit coin sets the keep flag, which controls one \CSWAP{} per loaded bit. These gates cause the mask charge above. Since \SELECT{} leaves the label unchanged, their work bits remain valid through the query and close without repair during inverse preparation. The alias workspace stays live until then.

\newedit{The form in which the sampler writes the address also fixes the cost of the controls. Any network can be controlled from any encoding, but at an opening count that grows with the number of supports, \Cref{app:generic-controls}. Our samplers write the address in the form the trees read, so that they read address bits and the remaining products stay $O(\log N)$, \Cref{app:address-predicates}.}
\section{Heisenberg ring}
\label{app:heisenberg}

The ring query of \Cref{case:heisenberg} combines a coefficient sampler, two nearest-neighbour swap trees and a local action on a three-bit label. \Cref{tab:ring-blocks} identifies the model-specific choices. We construct the complete query of \Cref{fig:heisenberg-circuit}, then give its resource counts and the comparison conditions.

\begin{table}[t]
\caption{Blocks of the ring query, the general block that each instantiates, and the choice that the ring makes.}
\label{tab:ring-blocks}
\begin{ruledtabular}
\begin{tabular}{lll}
Block & General & Ring choice\\
\hline
$\mathsf A_\mu$, $\Unif{N}$ & \Cref{app:prepare-interface} & class table, uniform position\\
$\mathsf G_{\mathrm H}$, $\Tog$ & \Cref{app:address-toggle} & decrement by $\bar j_0$, wrap\\
$\Perm_{\jod}$, $\Perm_{\jev}$ & \Cref{app:nn-network} & even leaf $v$ holds site $2v+2$\\
$\PH$ & \Cref{app:target-repairs} & orientation exchanges\\
$\mathsf D_\mathrm{class}$ & \Cref{app:sign-diagonal} & six coupling signs\\
\end{tabular}
\end{ruledtabular}
\end{table}

\subsection{Complete query}
\label{app:heisenberg-ring-complete-query}
\label{app:target-instances}
\label{app:ring-prepare}

We take \Cref{eq:heisenberg-ring-hamiltonian} at even $N\geq4$ and set $m=\lceil\log_2(N/2)\rceil$. The label $(f,\xi,\zeta)$ distinguishes fields, $f=0$, from bonds, $f=1$, and selects their Pauli axis through $(\xi,\zeta)$. As in Ref.~\cite{DellaChiara2025}, the axis uses an ordered-$XZ$ mask; here it describes only the target register.

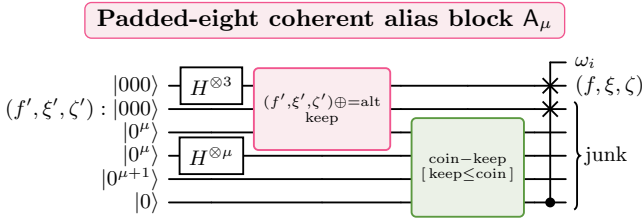
\begin{figure}[t]
\centering
\crayontitle{CrayonRose}{\textbf{Padded-eight coherent alias block $\mathsf A_\mu$}}\par\smallskip
\resizebox{\columnwidth}{!}{%
\begin{quantikz}[row sep={0.35cm,between origins}, column sep=0.16cm]
& \setwiretype{n} &  &  &  &  &  & \setwiretype{q} \rstick{$\wb_i$} \\
\lstick{$\ket{000}$} & \gate{H^{\otimes3}} & \gate[wires=3,style={draw=CrayonRose,fill=CrayonRoseFill,rounded corners=2pt, inner ysep=-3pt}]{\substack{
        (f',\xi',\zeta')\mathrel{\oplus}=\text{alt}\\
        \text{keep}}} &  &  &  & \swap{1} & \rstick{$(f,\xi,\zeta)$} \\
\lstick{$(f',\xi',\zeta'):\ket{000}$} &[-.1cm]  &[-.1cm]  &  &  &  & \targX{} &\rstick[5]{junk}  \\
\lstick{$\ket{0^\mu}$} &  &  &  & \gate[wires=4,style={draw=CrayonGreen,fill=CrayonGreenFill,rounded corners=2pt, inner ysep=-4pt}]{\substack{
        \text{coin}-\text{keep}\\
        \iv{\text{keep}\leq\text{coin}}}} &  &  &\\
\lstick{$\ket{0^\mu}$} & \gate{H^{\otimes\mu}} &  &  &  &  &  & \\
\lstick{$\ket{0^{\mu+1}}$} &  &  &  &  &  &  & \\
\lstick{$\ket0$} &  &  &  &  &  & \ctrl{-6} &
\end{quantikz}%
}
\caption{Coefficient sampler at precision $\mu$. The block-size-one clean QROAM of Ref.~\cite{Berry2019} loads the alternate label $(f',\xi',\zeta')$ and the $\mu$-bit keep threshold of the sampled code. A half comparator against a uniform $\mu$-bit coin, as in Ref.~\cite{Gidney2018}, retains their difference and sets the keep flag on the last qubit. The three class-swap lanes are bridged \CSWAP{}s, \Cref{fig:bridge-pair}(b). Their work bits $\wb_i$ hold the product of the keep flag with the mismatch of each label bit and its alternate. Those work bits and the remaining junk stay live until $\mathsf A_\mu^\dagger$. The label is unchanged in between, so they close without repair.}
\label{fig:ring-alias}
\end{figure}

\paragraph{Coefficient sampler.} The coefficients are uniform in position, so preparation separates into a class sampler and $\Unif{N}$ on $j=(j_m\cdots j_1j_0)$. The class sampler uses coherent alias sampling from Ref.~\cite{Babbush2018} at precision $\mu$, \Cref{fig:ring-alias}. Its padded table is
\begin{equation}
\begin{array}{c|cccccccc}
(f,\xi,\zeta)&000&001&010&011&100&101&110&111\\
\hline
W_{f\xi\zeta}&0&|g_Z|&|g_X|&|g_Y|&0&|J_Z|&|J_X|&|J_Y|.
\end{array}
\label{eq:ring-padded-alias-table}
\end{equation}
The zero-weight codes never reach \SELECT{}. The normalisation is $\Lnorm=N\sum_{f\xi\zeta}W_{f\xi\zeta}$.

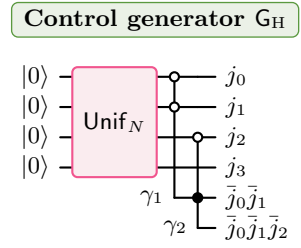
\begin{figure}[t]
\centering
\crayontitle{CrayonGreen}{\textbf{Control generator $\mathsf G_{\mathrm H}$}}\par\smallskip
\begin{quantikz}[row sep={0.4cm,between origins}, column sep=0.16cm]
\lstick{$\ket0$} & \gate[wires=4,style={draw=CrayonRose,fill=CrayonRoseFill,rounded corners=2pt, inner ysep=-4pt}]{\Unif{N}} & \ocontrol{} \wire[d][4]{q} &  & \rstick{$j_0$} \\
\lstick{$\ket0$} &  & \ocontrol{} &  & \rstick{$j_1$} \\
\lstick{$\ket0$} &  &  & \ocontrol{} \wire[d][3]{q} & \rstick{$j_2$} \\
\lstick{$\ket0$} &  &  &  & \rstick{$j_3$} \\
 & \setwiretype{n} & \lstick{$\carry{1}$}  & \control{} \setwiretype{q} & \rstick{$\bar j_0\bar j_1$} \\
 & \setwiretype{n} &  & \lstick{$\carry{2}$}  & \setwiretype{q} \rstick{$\bar j_0\bar j_1\bar j_2$}
\end{quantikz}
\caption{Control generator $\mathsf G_{\mathrm H}$ at $N=16$. $\Unif{N}$ prepares the address $j$. The two blank corners freeze the borrows $\carry{1}=\bar j_0\bar j_1$ and $\carry{2}=\bar j_0\bar j_1\bar j_2$ that drive the \CNOT{}s of $\Tog$, \Cref{fig:address-toggle}, whose lowest \CNOT{} is open-controlled on $j_0$. Open controls act on the complement of a bit. Blank two-input corners are temporary-\AND{} openings at $4\,T$ each, closed by measurement in $\mathsf G_{\mathrm H}^\dagger$ at no further $T$. At $N=16=2^{m+1}$ the register wraps on its own, so no wrap flag is drawn.}
\label{fig:ring-native-register}
\end{figure}

\paragraph{Permutation and address.} The two trees use the ring convention of \Cref{app:nn-network}: odd leaf $v$ holds site $2v+1$ in $q_{2v+1}$, and even leaf $v$ holds site $2v+2$ in $q_{2v}$, with sites modulo $N$. The odd tree brings the odd end of the bond to its root $q_1$, and the even tree brings the even end to its root $q_0$, with $N-2$ \CSWAP{}s in all. The odd tree reads $\jod=\lfloor j/2\rfloor$, while the even tree reads $\jev=(\jod-\bar j_0)\bmod(N/2)$. A shared register and the decrement toggle of \Cref{app:address-toggle} supply both. Its step bit is $\rho=\bar j_0$, used directly as an open control, and its borrows are $\carry{t}=\bar j_0\prod_{i=1}^t\bar j_i$ for $1\leq t\leq m$. Including the wrap, the address changes as
\begin{equation}
\begin{aligned}
\jev&=\jod\oplus(\carry{m-1},\ldots,\carry{1},\bar j_0)\\
&\quad\oplus\carry{m}\bigl(2^m-N/2\bigr),
\end{aligned}
\label{eq:ring-toggle-mask}
\end{equation}
and the toggle moves the shared register between the two tree addresses, \Cref{fig:address-toggle},
\begin{equation}
\Tog:\quad
\ket{j^{(\mathrm o)}}\ \longleftrightarrow\ \ket{j^{(\mathrm e)}}.
\label{eq:ring-toggle}
\end{equation}
The borrows cost $m-1$ openings, plus one for the wrap flag $\carry{m}=\bar j_0\,\iv{\jod=0}$ unless $N=2^{m+1}$. The control generator $\mathsf G_{\mathrm H}$ prepares $\Unif{N}$ and freezes these products, \Cref{fig:ring-native-register}. The second toggle restores $\jod$ before the odd inverse tree, after which the borrows close in reverse order.

\begin{figure}[t]
\centering
\crayontitle{CrayonBlue}{\textbf{The target action $\PH$}}\par\smallskip
\begin{quantikz}[row sep={0.35cm,between origins},column sep=0.25cm]
\lstick{$f$} &  &  & \gategroup[7,steps=6,style={draw=none,fill=CrayonBlueWash,rounded corners=2pt,inner xsep=2pt,inner ysep=2pt},background]{} &[-.2cm] \control{} \wire[d][6]{q} &[-.2cm]  &[-.2cm]  &[-.1cm]  &[-.15cm] \control{} \wire[d][6]{q} &  & \gategroup[7,steps=6,style={draw=none,fill=CrayonBlueWash,rounded corners=2pt,inner xsep=2pt,inner ysep=2pt},background]{} &[-.2cm] \control{} \wire[d][6]{q} &[-.2cm]  &[-.2cm]  &[-.1cm]  &[-.15cm] \control{} \wire[d][6]{q} &  &  &  \\
\lstick{$\zeta$} &  &  & \ctrl{3} & \control{} &  &  &  & \control{} &  &  &  &  &  &  &  &  &  &  \\
\lstick{$\xi$} &  &  &  &  &  &  &  &  &  & \ctrl{2} & \control{} &  &  &  & \control{} &  &  &  \\
\lstick{$j_0$} & \ctrl{1} \gategroup[3,steps=1,style={draw=none,fill=CrayonYellowWash,rounded corners=2pt,inner xsep=2pt,inner ysep=2pt},background]{} &  &  &  &  &  &  &  &  &  &  &  &  &  &  &  & \ctrl{1} \gategroup[3,steps=1,style={draw=none,fill=CrayonYellowWash,rounded corners=2pt,inner xsep=2pt,inner ysep=2pt},background]{} &  \\
\lstick{$q_0$} & \swap{1} &  & \gate{Z} &  &  &  &  &  &  & \gate{X} &  &  &  &  &  &  & \swap{1} &  \\
\lstick{$q_1$} & \targX{} &  &  &  &  &  & \gate{Z} &  &  &  &  &  &  & \gate{X} &  &  & \targX{} &  \\[.05cm]
& \setwiretype{n} &  &  &  & \setwiretype{q} & \gate{S} & \ctrl{-1} &  & \setwiretype{n} &  &  & \setwiretype{q} & \gate{S} & \ctrl{-1} &  & \setwiretype{n} &  &
\end{quantikz}
\caption{The target action of \Cref{eq:ring-sequential-target-action}. The yellow columns are the $j_0$-controlled exchange of the two targets and its inverse. The \AND{}s open the bond exponents $f\zeta$ and $f\xi$ one at a time, on the same recycled work bit. The $S$ gate on each open exponent supplies its class phase.}
\label{fig:ring-label-action}
\end{figure}
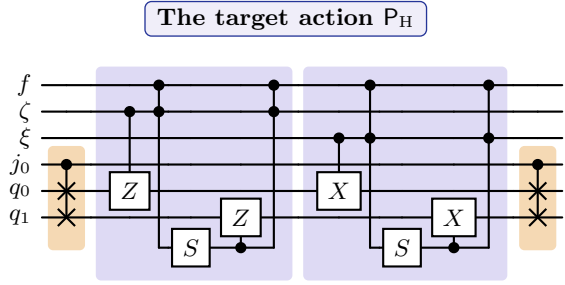

\paragraph{Target action.} The compressed label uses three bits instead of four mask bits. Decoding the bond exponents $f\xi$ and $f\zeta$ requires two products. A $j_0$-controlled exchange first places site $j$ on the first target qubit and site $j+1$ on the second, including the wraparound bond.
Between this exchange and its inverse, \Cref{fig:ring-label-action} applies $Z$ and then $X$. Each factor acts on the first target and, when $f=1$, on the second. An $S$ gate on each open bond exponent supplies the phase $i^{f(\xi+\zeta)}$. The resulting action is
\begin{equation}
\PH=i^{f(\xi+\zeta)}\bigl(X^{\xi}Z^{\zeta}\bigr)_{q_0}^{\,f\lor\bar j_0}\bigl(X^{\xi}Z^{\zeta}\bigr)_{q_1}^{\,f\lor j_0}.
\label{eq:ring-sequential-target-action}
\end{equation}
The target action spends four openings. The bond exponents $f\xi$ and $f\zeta$ cost $8\,T$, and the two orientation exchanges the other $8\,T$, \Cref{app:target-repairs}.
The two exchanges cannot be absorbed more cheaply. Without them, each axis $a\in\{\xi,\zeta\}$ needs the exponent $a(f\lor\bar j_0)$ on $q_0$ and $a(f\lor j_0)$ on $q_1$. \newedit{These four functions are linearly independent modulo affine functions of the label. Each opening multiplies two functions that are affine in the label and in earlier products, so it adds at most one new function to that span, and a circuit built from products of label bits needs four openings, the same $16\,T$.} The one-body action of the Anderson impurity model does absorb its exchange, because its onsite factor is a fixed $Z$, so that the single product $f_\mathrm{A}k_0$ suffices, \Cref{app:aim-onebody-bridge}.

\begin{figure}[t]
\centering
\crayontitle{CrayonRed}{\textbf{Class diagonal $\mathsf D_\mathrm{class}$}}\par\smallskip
\begin{quantikz}[row sep={0.6cm,between origins},column sep=0.2cm]
\lstick{$f$} & \gate{Z^{\sgn_f}} \gategroup[3,steps=3,style={draw=none,fill=CrayonRedWash,rounded corners=2pt,inner xsep=0pt,inner ysep=0pt},background]{$\mathsf D_{\sgn}$} & \control{} \wire[d][1]{q} & & & \gate{X} \gategroup[3,steps=4,style={draw=none,fill=CrayonRedWash,rounded corners=2pt,inner xsep=0pt,inner ysep=0pt},background]{$\mathsf D_Y$} & \gate{S^\dagger} & \gate{X} & \gate{Z} &  \\
\lstick{$\xi$} & \gate{Z^{\sgn_\xi\oplus\sgn_0}} & \gate{Z^{\sgn_{f\xi}}} \wire[d][1]{q} & \ctrl{1} &  &  & \gate{S} &  &  &  \\ 
\lstick{$\zeta$} & \gate{Z^{\sgn_\zeta\oplus\sgn_0}} & \gate{Z^{\sgn_{f\zeta}}} & \gate{Z^{\sgn_0}} & &  & \gate{S} &  &  & 
\end{quantikz}
\caption{The class diagonal of \Cref{eq:ring-direct-sign-diagonal,eq:ring-direct-y-phase}. The sequence $X,S^\dagger,X,Z$ applies $-iS^\dagger$ on $f$.}
\label{fig:ring-class-diagonal}
\end{figure}
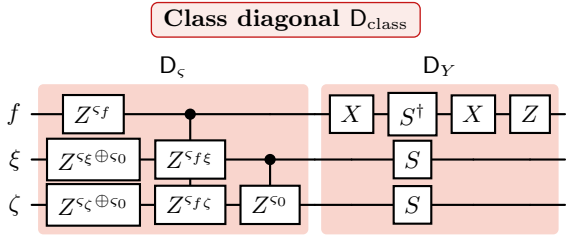

\paragraph{Class diagonal.} The coefficient signs and the $Y$ convention left over from \Cref{eq:ring-sequential-target-action} form a single diagonal on the label, of the kind described in \Cref{app:sign-diagonal}. Write $\sgn_{g_Z},\ldots,\sgn_{J_Y}$ for the negativity bits of the six couplings. The sign on the populated codes is then the quadratic form
\begin{equation}
\sgn(f,\xi,\zeta)=\sgn_0\oplus\sgn_f f\oplus\sgn_\xi\xi\oplus\sgn_\zeta\zeta
\oplus\sgn_{f\xi}f\xi\oplus\sgn_{f\zeta}f\zeta,
\label{eq:ring-sign-anf}
\end{equation}
with $\sgn_0=\sgn_{g_Z}\oplus\sgn_{g_X}\oplus\sgn_{g_Y}$, $\sgn_\xi=\sgn_{g_Z}\oplus\sgn_{g_Y}$ and $\sgn_\zeta=\sgn_{g_X}\oplus\sgn_{g_Y}$. Write $\sgn^J_\bullet$ for the same combination of the bond signs $\sgn_{J_Z},\sgn_{J_X},\sgn_{J_Y}$. The three coefficients that carry $f$ are then $\sgn_f=\sgn_0\oplus\sgn^J_0$, $\sgn_{f\xi}=\sgn_\xi\oplus\sgn^J_\xi$ and $\sgn_{f\zeta}=\sgn_\zeta\oplus\sgn^J_\zeta$. On these codes $1=\xi\oplus\zeta\oplus\xi\zeta$, so the constant term is absorbed into a \CZ{}. The sign is then realised exactly by
\begin{equation}
\mathsf D_{\sgn}=
Z_f^{\sgn_f}Z_\xi^{\sgn_\xi\oplus\sgn_0}Z_\zeta^{\sgn_\zeta\oplus\sgn_0}
\CZ_{f,\xi}^{\sgn_{f\xi}}
\CZ_{f,\zeta}^{\sgn_{f\zeta}}
\CZ_{\xi,\zeta}^{\sgn_0}.
\label{eq:ring-direct-sign-diagonal}
\end{equation}
A second factor corrects the ordered-$XZ$ convention,
\begin{equation}
\mathsf D_Y=\exp\bigl[i\pi(2\xi+2\zeta-2f-2)/4\bigr]=-iS_\xi S_\zeta S_f^\dagger.
\label{eq:ring-direct-y-phase}
\end{equation}
On the populated axis codes $\xi+\zeta-1=\xi\zeta$. The product of $\mathsf D_Y$ with the two mask phases of \Cref{eq:ring-sequential-target-action} is therefore $i^{(1+f)\xi\zeta}$. This is the phase $1$ for $X$ and for $Z$, $i$ for a site $Y$, and $-1$ for a bond $Y$. The complete class diagonal $\mathsf D_\mathrm{class}=\mathsf D_Y\mathsf D_{\sgn}$ is ancilla-free and Clifford, \Cref{fig:ring-class-diagonal}.

\paragraph{Assembly.} All \SELECT{} gates use the closed schedule. Each tree gate moves a single site, so its repair is a path-dependent product of address bits, and bridging would trade reopenings for repairs almost one for one, \Cref{app:repair-count}. The orientation exchanges also stay closed. Only the sampler exchanges are bridged.

The query runs the sampler and $\mathsf G_{\mathrm H}$, the odd tree, the toggle, the even tree, the target action and the class diagonal, then the inverses in reverse order, \Cref{fig:heisenberg-circuit}. After the inverse permutations, $\mathsf G_{\mathrm H}^\dagger$ closes the borrows and reverses the address preparation. Projecting the address and alias registers onto their initial zero states extracts the encoded block; inverse preparation alone does not reset them for every system output.

\subsection{Resources and comparison}
\label{app:heisenberg-resource-comparison}

\paragraph{Resources of the query.}
Excluding the coefficient sampler, the non-Clifford cost comes from openings and, at non-power-of-two $N$, uniform-state rotations. The forward and inverse trees contribute $2N-4$ openings, the borrows $m-1$, the wrap flag $\iv{N\neq2^{m+1}}$, and the target action four. The remaining gates are Clifford. For non-power-of-two $N$, the two uniform-state directions add $6\bigl(m-\operatorname{ctz}(N/2)\bigr)-2$ openings, where $\operatorname{ctz}$ counts trailing binary zeros. After these zeros are factored out, the amplitude-amplification construction of Ref.~\cite[Fig.~12]{Babbush2018} acts on $m-\operatorname{ctz}(N/2)$ qubits. Each direction uses two constant comparisons at one opening per active qubit, a zero-state phase predicate at one opening fewer, and two rotations. Thus, for an uncontrolled query at even $N\geq4$, before adding the coefficient sampler,
\begin{equation}
\begin{aligned}
\Nopen&=2N+m-1\\
&\quad+\iv{N\neq2^{m+1}}\bigl(6\bigl(m-\operatorname{ctz}(N/2)\bigr)-1\bigr),\\
\Ttot&=4\Nopen,\qquad \Nrot=4\,\iv{N\neq2^{m+1}},\\
\Apeak&=2m+4+\iv{N\neq2^{m+1}}\bigl(1+\iv{N/2\ \mathrm{odd}}\bigr).
\end{aligned}
\label{eq:ring-compact-power-two-resources}
\end{equation}
The Iverson bracket $\iv{N\neq2^{m+1}}$ is one for non-power-of-two $N$, so every correction vanishes at $N=2^{m+1}$, where $\Nopen=2N+m-1$. The $T$ count excludes the synthesis of the $\Nrot$ rotations. $\Apeak$ counts every non-system qubit: $m+4$ hold the address and the label, and the rest is the larger of two disjoint workspaces, the borrows, wrap flag and recycled work bit that are live during the system operations, and the uniform-state scratch, which exceeds the first by one qubit whenever $N/2$ is odd.

The remaining cost is the coefficient sampler, the standard alias construction of Ref.~\cite{Babbush2018}. It is linear in $\mu$, independent of $N$, and responsible for the only exact $T$ gates that are not openings, the eight of its measured table erasure. Including it,
\begin{equation}
\Ttot=8N+4\mu+O(\log N),\qquad
\Apeak=3\mu+O(\log N).
\label{eq:ring-leading-order-resources}
\end{equation}
At $N=2^{m+1}$ the complete query costs exactly $\Ttot=8N+4m+4\mu+40$ gates and $\Apeak=3\mu+2m+12$ qubits.

\paragraph{Accounting of the comparison.}
\Cref{fig:heisenberg-comparison} compares this query with four alternative block encodings of \Cref{eq:heisenberg-ring-hamiltonian}, all chosen to obtain the same normalisation $\Lnorm$. $\Ttot$ counts temporary-\AND{} openings at $4\,T$ each, together with every other exact $T$ gate, and $\Nrot$ counts the rotations left to synthesise. Panel~(a) of \Cref{fig:heisenberg-comparison} uses
\begin{equation}
    \Ttot+\Nrot\Trot, \qquad \Trot= \left\lceil 3\log_2\!\left(\frac{\Nrot}{\epssynth}\right) \right\rceil ,
\label{eq:rotation-synthesis-estimate}
\end{equation}
with $\epssynth=10^{-3}$ shared equally among the rotations of one method, following Ref.~\cite{Childs2018}. \newedit{$\Trot$ is the leading term of the typical-angle $T$ count of the ancilla-free synthesis of Ref.~\cite{RossSelinger2016} at accuracy $\epssynth/\Nrot$ per rotation.} The three alias-sampled constructions use $\mu=10$, which gives a coefficient error of at most $\Lnorm2^{-\mu}$. The two rotation-based constructions are exact before synthesis. Every curve is the closed form of its construction at every even $N$. For this work, the flat walk and the product-wise walk, the markers count the gates of the full decomposition, which the closed forms reproduce exactly. The remaining two rows use the gate inventories of their sources. \Cref{case:heisenberg} interprets the curves.

\paragraph{Adaptation of each comparator.}
\Cref{tab:ring-comparators} states how each comparator is adapted to \Cref{eq:heisenberg-ring-hamiltonian}. For the flat and product-wise constructions, we keep our coefficient preparation and replace only \SELECT{}. These rows therefore isolate the cost of support selection. For Refs.~\cite{DellaChiara2025,Childs2018}, we instead retain the preparation--selection trade-off of the published constructions. We omit Refs.~\cite{NibbiMendl2024,ZhangShao2026}, whose resource analyses are not given as directly comparable finite Clifford+$T$ resource counts for this model.

\begin{table*}[t]
\caption{Adaptation of the four comparators of \Cref{fig:heisenberg-comparison} to \Cref{eq:heisenberg-ring-hamiltonian}. The last column says whether a row keeps our coefficient preparation or the preparation of its source.}
\label{tab:ring-comparators}
\begin{ruledtabular}
\begin{tabular}{@{}p{0.17\textwidth}p{0.71\textwidth}l@{}}
Row & Adaptation & \PREPARE{}\\
\hline
\raggedright Flat walk, Ref.~\cite{Babbush2018} & \raggedright Generic unary iteration over the $6N$ Pauli terms, with the operator class outside the position. It needs no address arithmetic at any even $N$ and costs the $4(6N-2)$ $T$ gates of a flat walk over $6N$ terms. The signs use the Clifford diagonal of our construction. & ours\\
\raggedright Product-wise, Ref.~\cite{YOSKM2024} & \raggedright The bond-only chain example, with the site--bond distinction added. A second address $j_2=j_1+1\bmod N$ is an increment of a copy of $j_1$, followed, when $N$ is not a power of two, by a reset of the value $N$ to zero under the flag $\iv{j_1=N-1}$. The second factor-wise Pauli walk acts only on bond terms. & ours\\
\raggedright Register mask, Ref.~\cite{DellaChiara2025} & \raggedright The published open-chain Hamiltonian and whole-register mask, not modified to periodic boundaries. Logical Toffolis become recyclable four-$T$ temporary-\AND{} openings, and the exact $\pi/2$ rotations and phases are included in the $T$ ledger. & source\\
\raggedright Tree \SELECT{}, Ref.~\cite{Childs2018} & \raggedright The $6N$ Pauli terms replace the source Hamiltonian, embedded in a complete binary tree of $2^{\lceil\log_2 6N\rceil}$ leaves, between $6N$ and $12N$. The generic real-amplitude preparation and the gate inventory of the source are kept. The tree may be pruned after the populated leaves, so its cost is an upper bound. & source\\
\end{tabular}
\end{ruledtabular}
\end{table*}
\section{Anderson impurity model}
\label{app:aim-complete-query}

The AIM query of \Cref{case:aim} uses one preparation, and its \SELECT{} is the product of two circuits: $\mathsf B$ for onsite and hopping terms, and $\mathsf I$ for impurity interactions. Each is the identity outside its family, so they run in sequence. We use the chain-basis Hamiltonian of \Cref{eq:aim-fermionic-hamiltonian}, with real, spin-independent $a_{k\ell}$ and $t_{k\ell}$ and orbital-independent $U$ and $J$. Density-generated shifts are absorbed into $a_{0\ell}$ and the scalar $\eta$\newedit{, which is added back classically rather than block-encoded}. The general counts assume $n_\mathrm{imp}\geq2$ and $n_\mathrm{bath}\geq1$; detailed examples use five impurity orbitals and, where a bath size is needed, $n_\mathrm{bath}=100$.

\Cref{tab:aim-blocks} maps the two circuits to the general constructions. The one-body circuit combines nearest-neighbour trees within each chain with a bridged tree over chain fronts. The interaction circuit combines an all-to-all spin-up permutation with diagonal spin-down walks. We give the Pauli coefficients first, then preparation, selection and the complete-query comparison.

The qubit $q_{k\ell\sigma}$ represents mode $c_{k\ell\sigma}$, with $k=0$ at the impurity and $\sigma=\uparrow,\downarrow\equiv0,1$. The chain address is $2\ell+\sigma$. In the snake ordering, $k$ runs from $n_\mathrm{bath}$ to zero for spin up and back for spin down. Thus the two impurity qubits of an orbital are adjacent in Jordan--Wigner order, and impurity interactions have no bath-length parity strings. The fixed targets are $\mathcal T_\mathsf{B}=(q_{00\uparrow},q_{00\downarrow})$ and $\mathcal T_\mathsf{I}=(q_{00\uparrow},q_{00\downarrow},q_{01\uparrow},q_{01\downarrow})$.

\begin{table}[t]
\caption{Blocks of the AIM query, the general block that each instantiates, and the choice that the AIM makes.}
\label{tab:aim-blocks}
\begin{ruledtabular}
\begin{tabular}{lll}
Block & General & AIM choice\\
\hline
$\mathsf{Prep}_\mu$ & \Cref{app:prepare-interface} & $\Rrows$ rows, refinements\\
$\mathsf G_\mathsf{B}$, $\Tog$ & \Cref{app:address-toggle} & increment by $f_\mathrm{T}k_0$\\
$\Perm_k$ & \Cref{app:nn-network} & open chains, in parallel\\
$\PermC_{\ell\sigma}$, $\mathsf{Rep}_\mathsf{B}$ & \Cref{app:cswap-minimum,app:repair-count} & bridged block tree\\
$\Pb$ in $\mathsf E_\mathsf{B}$ & \Cref{app:target-repairs} & absorbs the exchange\\
$\mathsf G_\mathsf{I}$ & \Cref{app:address-predicates} & $f_\mathrm{D}\sigma$, $f_\mathrm{J}\sigma$\\
$\PermC_\uparrow$ & \Cref{app:aa-network} & all-to-all, $N=n_\mathrm{imp}$\\
$\mathsf E_\mathsf{I}$, $\Pup$, $\PermC_\downarrow$, $\Pdn$ & \Cref{app:frames-walks} & bridged without repair\\
\end{tabular}
\end{ruledtabular}
\end{table}

\subsection{Pauli form}
\label{app:aim-pauli-form}

Under Jordan--Wigner with the snake ordering, the mode $c_{k\ell\sigma}$ becomes the qubit $q_{k\ell\sigma}$, and Pauli operators carry the same index. The onsite energy $a_{k\ell}$ of \Cref{eq:aim-fermionic-hamiltonian} gives the $Z$ coefficient $-a_{k\ell}/2$, and the hopping $t_{k\ell}$ gives the coefficient $t_{k\ell}/2$ of $XX$ and of $YY$. The density terms shift the $Z$ coefficient of every impurity qubit by $-[U+(n_\mathrm{imp}-1)(2U-5J)]/4$. We fold this shift into the onsite energy: from here on, $a_{0\ell}$ in $\Lnorm$ and in the row weights denotes the shifted energy, the bare $a_{0\ell}$ of \Cref{eq:aim-fermionic-hamiltonian} plus $[U+(n_\mathrm{imp}-1)(2U-5J)]/2$. The mapping also generates a constant shift $\eta$. \newedit{It shifts every eigenvalue by the same amount, so the query block-encodes $H_{\mathrm{AIM}}^{\mathrm{JW}}-\eta$, and $\eta$ does not enter $\Lambda$.} The image splits into a one-body part $H_{\mathrm{1b}}$ and an interaction part $H_{\mathrm{int}}$,
\begin{equation}
\begin{aligned}
H_{\mathrm{AIM}}^{\mathrm{JW}}&=\eta+H_{\mathrm{1b}}+H_{\mathrm{int}},\\
H_{\mathrm{1b}}&=-\sum_{\ell,\sigma}\sum_{k=0}^{n_{\mathrm{bath}}}\tfrac12a_{k\ell}Z_{k\ell\sigma}\\
&+\sum_{\ell,\sigma}\sum_{k=0}^{n_{\mathrm{bath}}-1}\tfrac12t_{k\ell}
\sum_{y=0}^{1}P^{(y)}_{k\ell\sigma}P^{(y)}_{k+1,\ell\sigma},\\
H_{\mathrm{int}}&=\sum_{(\ell,\sigma)<(\ell',\sigma')}u_{\ell\sigma,\ell'\sigma'}Z_{0\ell\sigma}Z_{0\ell'\sigma'}\\
&+\sum_{\ell<\ell'}\sum_{y,\sigma}\kappa_{\ell\ell',y,\sigma}P^{(y\sigma)}_{\ell\ell'}.
\end{aligned}
\label{eq:aim-jw-permutation-split}
\end{equation}
Here the sum runs over pairs of distinct impurity spin orbitals, ordered by $2\ell+\sigma$. The $u_{\ell\sigma,\ell'\sigma'}$ are one quarter of the three density values of \Cref{eq:aim-fermionic-hamiltonian}, namely $U/4$, $(U-2J)/4$ and $(U-3J)/4$. Each $\kappa_{\ell\ell',y,\sigma}$ has magnitude $|J|/4$, and $P^{(0)}=X$, $P^{(1)}=Y$. The four Hund words on an orbital pair are
\begin{align}
P^{(00)}_{\ell\ell'}=XXYY,\quad P^{(01)}_{\ell\ell'}=XYYX, \nonumber \\ \quad P^{(10)}_{\ell\ell'}=YXXY,\quad P^{(11)}_{\ell\ell'}=YYXX.
\label{eq:aim-hund-words}
\end{align}
They are written on the four impurity qubits in the fixed order $(\ell\!\uparrow,\ell\!\downarrow,\ell'\!\uparrow,\ell'\!\downarrow)$. Each is the spin-flip or pair-hopping term of \Cref{eq:aim-fermionic-hamiltonian} for the coin values $(y,\sigma)$ of the row, \Cref{tab:aim-row-map}. 
In the structured form of \Cref{eq:aim-swap-form-int}, the coefficient $\alpha_{\ell\ell',b}$ of a Hund word is the corresponding $\kappa_{\ell\ell',y,\sigma}$, and that of an interorbital density word is the corresponding $u_{\ell\sigma,\ell'\sigma'}$. The intraorbital value $u_{\ell\uparrow,\ell\downarrow}=U/4$ is divided equally among the pairs whose head is $\ell$, giving $U/8$ on each of two pairs for $n_\mathrm{imp}=5$ and the whole $U/4$ on one pair for $n_\mathrm{imp}=3$.
The snake ordering keeps bath-length parity strings out of the interaction. $H_{\mathrm{1b}}$ is therefore supported on $2n_\mathrm{imp}$ nearest-neighbour chains, and $H_{\mathrm{int}}$ on the $2n_\mathrm{imp}$ impurity qubits alone.

\subsection{Preparation}
\label{app:aim-global-prepare}

Terms that differ only in spin, hopping axis, orbital pair or Hund word share a magnitude by symmetry, and the sampler loads each shared magnitude once, \Cref{app:prepare-interface}. This reduces the table from about $3N$ Pauli strings to about $N$ coefficient rows, \Cref{eq:aim-factored-lambda}. The structured comparator uses the same rows, so this reduction does not explain the advantage over that comparator.

\paragraph{Coefficient rows.} Each row carries the total magnitude of its terms. Uniform choices then select the spin, orbital or orbital pair, and Pauli word left unspecified by the row, \Cref{tab:aim-factored-coefficients}.

\par\medskip\noindent\begin{minipage}{\linewidth}
\centering
\footnotesize
\setlength{\tabcolsep}{2pt}
\begin{tabular}{llllc}
\toprule
Flag & Row type & Rows & Row weight & Refinements\\
\midrule
$f_\mathrm{A}$ & $a_{k\ell}$ & $n_\mathrm{imp}(n_\mathrm{bath}+1)$ & $|a_{k\ell}|$ & spin\\
$f_\mathrm{T}$ & $t_{k\ell}$ & $n_\mathrm{imp}n_\mathrm{bath}$ & $2|t_{k\ell}|$ & spin, axis\\
$f_U$ & $U$ & 1 & $n_\mathrm{imp}|U|/4$ & orbital\\
$f_\mathrm{opp}$ & $U-2J$ & 1 & $\binom{n_\mathrm{imp}}2|U-2J|/2$ & pair, spin\\
$f_\mathrm{same}$ & $U-3J$ & 1 & $\binom{n_\mathrm{imp}}2|U-3J|/2$ & pair, spin\\
$f_\mathrm{J}$ & $J$ & 1 & $\binom{n_\mathrm{imp}}2|J|$ & pair, word\\
\bottomrule
\end{tabular}
\captionof{table}{Coefficient rows, their flag and their total Pauli weights. The weight of a row is the total magnitude of the equal-weight terms it stands for, and the uniform refinements pick one of them: a hopping row stands for two spins and for the axis choice $XX$ or $YY$, and the word refinement picks a Hund word. The six flags $f_\mathrm{A},f_\mathrm{T},f_U,f_\mathrm{opp},f_\mathrm{same},f_\mathrm{J}$ are mutually exclusive. Each is a single live qubit that any gate may take as a control at no cost. The first two rows form the one-body family and the next four the interaction family, in which $f_\mathrm{opp}$ and $f_\mathrm{same}$ flag the density rows. \SELECT{} also uses three parities of these flags, each formed in a spare flag by \CNOT{}s: the pair flag $f_\mathrm{pair}=f_\mathrm{opp}\oplus f_\mathrm{same}\oplus f_\mathrm{J}$, the interaction flag $f_\mathrm{int}=f_U\oplus f_\mathrm{pair}$ and the density flag $f_\mathrm{D}=f_\mathrm{opp}\oplus f_\mathrm{same}$.}
\label{tab:aim-factored-coefficients}
\end{minipage}\par\medskip

For $\Lambda>0$, the normalisation and the row count are
\begin{equation}
\begin{aligned}
\Lambda={}&\sum_{\ell,k}|a_{k\ell}|+2\sum_{\ell,k}|t_{k\ell}|\\
&+\frac{n_\mathrm{imp}}4|U|+\binom{n_\mathrm{imp}}2|J|\\
&+\frac{\binom{n_\mathrm{imp}}2}{2}\bigl(|U-2J|+|U-3J|\bigr),\\
\Rrows={}&n_\mathrm{imp}(2n_\mathrm{bath}+1)+4,\\
n_R={}&\lceil\log_2\Rrows\rceil.
\end{aligned}
\label{eq:aim-factored-lambda}
\end{equation}
The $a$ and $t$ sums run over $\ell=0,\ldots,n_\mathrm{imp}-1$, with $k\leq n_\mathrm{bath}$ and $k<n_\mathrm{bath}$ respectively. The zero Hamiltonian gives a trivial zero block. At $(5,100)$ there are 1,009 rows, held without padding. \Cref{tab:aim-row-map} shows how $\mathsf B$ and $\mathsf I$ use each row.

\begin{table}[t]
\caption{Coefficient rows of the AIM and their use by $\mathsf B$ and $\mathsf I$. \PREPARE{} alias-samples one row. The row weight, \Cref{tab:aim-factored-coefficients}, includes the multiplicity of its uniform refinements. The refinements then pick one Pauli string. The last column lists what the control generator of each circuit writes. Every other control is a prepared bit or a parity of prepared bits. $\mathsf G_\mathsf{B}$ writes the endpoint address $j$, the frozen carries $\carry{t}$ of the toggle and the repair word $\rep$ of the bridged chain-front permutation. The coin $y$ picks $P^{(y)}\in\{X,Y\}$. The same pair $y,\sigma$ picks the Hund word. We write $\bar\sigma=1-\sigma$, and $f_\mathrm{D}$ flags the two density rows.}
\label{tab:aim-row-map}
\begin{ruledtabular}
\begin{tabular}{llll}
Row & Refined by & Selected Pauli & Written by $\mathsf G$\\
\hline
\multicolumn{4}{l}{One-body circuit $\mathsf B$}\\
$a_{k\ell}$ & $\sigma$ & $Z_{k\ell\sigma}$ & $j$, $f_\mathrm{A}k_0$\\
$t_{k\ell}$ & $\sigma$, $y$ & $P^{(y)}_{k\ell\sigma}P^{(y)}_{k+1,\ell\sigma}$ & $j$, $\carry{t}$, $\rep$\\
\specialrule{0.25pt}{.5pt}{1pt}
\multicolumn{4}{l}{Interaction circuit $\mathsf I$}\\
$U$ & $\ell$ & $Z_{0\ell\uparrow}Z_{0\ell\downarrow}$ & ---\\
$U-2J$ & $(\ell,\ell')$, $\sigma$ & $Z_{0\ell\sigma}Z_{0\ell'\bar\sigma}$ & $f_\mathrm{D}\sigma$\\
$U-3J$ & $(\ell,\ell')$, $\sigma$ & $Z_{0\ell\sigma}Z_{0\ell'\sigma}$ & $f_\mathrm{D}\sigma$\\
$J$ & $(\ell,\ell')$, $y$, $\sigma$ & $P^{(y\sigma)}_{\ell\ell'}$ & $f_\mathrm{J}\sigma$\\
\end{tabular}
\end{ruledtabular}
\end{table}

\paragraph{Preparation sequence.} The preparation and complete query are
\begin{equation}
\mathsf {Prep}_\mu=\mathsf F\mathsf Q_p\mathsf A_\mu\mathsf U_\mathrm{ref},
\qquad
\mathsf U_\mathrm{query}=\mathsf {Prep}_\mu^\dagger\,\SELECT\,\mathsf {Prep}_\mu.
\label{eq:aim-query-order}
\end{equation}
The blocks act from right to left: $\mathsf U_\mathrm{ref}$ prepares uniform choices, $\mathsf A_\mu$ samples a coefficient row, $\mathsf Q_p$ decodes the orbital rank into fields $\mathcal Q$, and $\mathsf F$ writes the label $\mathcal F$. The alias workspace $\mathcal A$ stays live through \SELECT{}. The clean scratch $\mathcal W$ is reused by preparation, $\mathsf B$ and $\mathsf I$, each returning it to zero. Their widths are $n_{\mathcal A}$ and $n_{\mathcal W}$. At five orbitals, the uniform register has $n_{\mathcal U}=6$ bits; only its spin coin $\sigma$ is read outside preparation.

\paragraph{Alias sampler.} The alias sampler $\mathsf A_\mu$ of Ref.~\cite{Babbush2018} puts the $n_R$-bit row address into uniform superposition over exactly $\Rrows$ values, by inequality-test preparation with one amplitude-amplification round that carries two arbitrary rotations. It then loads the fields of the selected row, the flags, $(k,\ell_\mathsf{B})$ and the sign, into $\mathcal F$ by unary iteration. The result is
\begin{equation}
\mathsf A_\mu\ket0
=\sum_{i=0}^{\Rrows-1}\sqrt{n_i/(\Rrows\,2^\mu)}\,
\ket{\mathcal F_i}_{\mathcal F}\ket{\mathrm{junk}_i}_{\mathcal A},
\label{eq:aim-alias-state}
\end{equation}
with normalised garbage states $\ket{\mathrm{junk}_i}$. With $\mu\geq1$ precision bits, the nonnegative counts $n_i$ sum to $\Rrows\,2^\mu$. The ratio $n_i/(\Rrows\,2^\mu)$ approximates the row weight divided by $\Lambda$. Padding the table to $2^{n_R}$ buckets would make the address superposition Clifford and refine this grid by a factor below two, but it would load up to twice as many rows, so it is not done. The alias workspace $\mathcal A$ is retained until inverse preparation.

\paragraph{Label.} The term label carries the fields
\begin{equation}
\begin{aligned}
\mathcal F&=(\boldsymbol f,k,\ell_\mathsf{B},\ell,\ell',\tau,
 f_\mathrm{T}y,f_\mathrm{J}y,\sgn),\\
n_{\mathcal F}&=\underbrace{\lceil\log_2(n_\mathrm{bath}+1)\rceil}_{n_k}+3\underbrace{\lceil\log_2n_\mathrm{imp}\rceil}_{n_\ell}+10.
\end{aligned}
\label{eq:aim-semantic-descriptor}
\end{equation}
Its ten single-bit entries are the six flags $\boldsymbol f$, $\tau$, the two refinement products and the sign. One-body rows specify $(k,\ell_\mathsf{B})$, where $\ell_\mathsf{B}$ is the orbital of the chain $(\ell_\mathsf{B},\sigma)$. Interaction rows specify $(\ell,\ell',\tau)$, where the orientation $\tau=f_U\tau_U$ of \Cref{tab:aim-rank-decoder} is set only on $U$ rows. The label thus reserves one orbital field per family, besides the second orbital of the pair. Inactive and not-yet-refined fields are zero. When no interaction flag is set, $\mathsf I$ acts as the identity. Labels with equal impurity orbitals receive no amplitude from \PREPARE{}, so their action does not enter the block encoding.

\paragraph{Uniform choices and rank decoder.} $\mathsf U_\mathrm{ref}$ prepares an orbital rank, a pair coin, and the operator coins $y$ and $\sigma$. The coin $y$ selects $XX$ or $YY$ and $\sigma$ selects the spin; on Hund rows the two coins select the word. At five orbitals, the rank and pair coin form $p$, uniform over ten values. The decoder $\mathsf Q_p$ maps each value to an unordered pair $(\ell,\ell')$ and an orientation, \Cref{tab:aim-rank-decoder}. Pair rows use both orbitals; $U$ rows use the head of the oriented pair. Each orbital heads two ranks, so the same uniform rank serves both kinds of row. The assignment needs one stored cubic product, built with two openings. Other impurity counts use ranks of the appropriate sizes, and unused choices are ignored.

\begin{figure*}[t]
\centering
\crayontitle{CrayonRose}{\textbf{Preparation $\mathsf {Prep}_\mu$}}\par\smallskip
{\small
\begin{quantikz}[row sep={0.5cm,between origins},column sep=0.2cm]
\lstick{$\mathcal A:\ket{0^{n_{\mathcal A}}}$} & \setwiretype{n} &[0.3cm]  & \gate[wires=4,style={draw=CrayonRose,fill=CrayonRoseFill,rounded corners=2pt}]{\mathsf A_\mu} & \gate[wires=3,style={draw=CrayonGreen,fill=CrayonGreenFill,rounded corners=2pt}]{\mathsf Q_p}\setwiretype{q} & \gate[wires=3,style={draw=CrayonGreen,fill=CrayonGreenFill,rounded corners=2pt}]{\mathsf F} &   \\
\lstick{$\mathcal F:\ket{0^{n_{\mathcal F}}}$} &  &  &  &  &  &   \\
\lstick{$\mathcal U:\ket{0^{n_{\mathcal U}}}$} &  & \gate[style={draw=CrayonRose,fill=CrayonRoseFill,rounded corners=2pt}]{\mathsf U_\mathrm{ref}} &  &  &  &   \\
 & \setwiretype{n} &  & \lstick{$\mathcal W:\ket{0^{n_{\mathcal W}}}$} & \setwiretype{q} &  & \rstick{$\ket0$}
\end{quantikz}}\par\medskip
\begin{minipage}[t]{0.33\textwidth}
\vspace{0pt}
\centering
\crayontitle{CrayonRose}{\textbf{Uniform choices $\mathsf U_\mathrm{ref}$}}\par
\vspace{1.4em}{\small
\begin{quantikz}[row sep={0.60cm,between origins},column sep=0.14cm]
\lstick{$(\ell_\mathrm{u})_0:\ket0$} & \gategroup[wires=3,steps=3,style={draw=none,fill=CrayonRoseWash,rounded corners=2pt,inner sep=0pt},background,label style={label position=above,anchor=south,yshift=-.1cm}]{$\Unif{5}$} &  & \gate{H} & \rstick[3]{$\ell_\mathrm{u}$} \\
\lstick{$(\ell_\mathrm{u})_1:\ket0$} &  & \gate{H} &  & \\
\lstick{$(\ell_\mathrm{u})_2:\ket0$} & \gate{R_y(\theta_5)} & \octrl{-1} & \octrl{-2} & \\[0.1cm]
\lstick{$\varrho:\ket0$} &  & \gate{H} &  & \\
\lstick{$y:\ket0$} &  & \gate{H} &  & \\
\lstick{$\sigma:\ket0$} &  & \gate{H} &  &
\end{quantikz}}\par
\end{minipage}
\hfill
\begin{minipage}[t]{0.32\textwidth}
\vspace{0pt}
\centering
\crayontitle{CrayonGreen}{\textbf{Rank decoder $\mathsf Q_p$}}\par
\vspace{1.4em}
{\small
\begin{quantikz}[row sep={0.42cm,between origins},column sep=0.14cm]
\lstick{$\tau_U:\ket0$} & \setwiretype{n} &  &  & \wire[d][1]{q} & \setwiretype{q} & \targ{} \wire[d][9]{q} & \targ{} \wire[d][6]{q} &  \\
\lstick{$(\ell_\mathrm{u})_0$} & \control{} \wire[d][3]{q} &  & \control{} \wire[d][3]{q} & \control{} \wire[d][7]{q} &  &  &  &  \\
\lstick{$(\ell_\mathrm{u})_1$} & \control{} &  & \control{} &  & \ctrl{5} &  &  &  \\
\lstick{$(\ell_\mathrm{u})_2$} &  &  &  &  &  & \ctrl{-3} &  &  \\
\lstick{$\wb_1:\ket0$} & \setwiretype{n} & \control{} \setwiretype{q} \wire[d][3]{q} &  & \setwiretype{n} &  &  &  &  \\
\lstick{$\varrho$} &  & \control{} &  &  &  &  & \ctrl{-5} &  \\
\lstick{$(\ell_p)_0:\ket0$} & \setwiretype{n} &  & \wire[d][1]{q} & \setwiretype{q} &  &  & \targ{} &  \\
\lstick{$(\ell_p)_1:\ket0$} & \setwiretype{n} &  & \control{} \wire[d][2]{q} \setwiretype{q} &  & \targ{} &  &  &  \\
\lstick{$(\ell'_p)_0:\ket0$} & \setwiretype{n} &  &  &  & \ctrl{1} \setwiretype{q} &  &  &  \\
\lstick{$(\ell'_p)_1:\ket0$} & \setwiretype{n} &  &  & \setwiretype{q} & \targ{} \wire[d][1]{q} & \targ{} \wire[d][1]{q} &  &  \\
\lstick{$(\ell'_p)_2:\ket0$} & \setwiretype{n} &  &  &  &  & \targ{} \setwiretype{q} & \targ{} & 
\end{quantikz}}\par
\end{minipage}
\hfill
\begin{minipage}[t]{0.31\textwidth}
\vspace{0pt}
\centering
\crayontitle{CrayonGreen}{\textbf{Label fill $\mathsf F$}}\par
\vspace{1.4em}
{\small
\begin{quantikz}[row sep={0.42cm,between origins},column sep=0.14cm]
\lstick{$\mathcal Q:\ket{0^6}$} & \control{} \wire[d][3]{q} & \control{} \wire[d][4]{q} & \control{} \wire[d][5]{q} &  \\
\lstick{$f_U$} &  & \control{} &  &  \\
\lstick{$f_\mathrm{int}$} & \control{} &  & \control{} &  \\
\lstick{$\ell':\ket{0^3}$} & \setwiretype{n} & \setwiretype{q} &  &  \\
\lstick{$\tau:\ket0$} & \setwiretype{n} &  & \setwiretype{q} &  \\
\lstick{$\ell_{0:1}:\ket{0^2}$} & \setwiretype{n} &  &  & \setwiretype{q} \\
\lstick{$f_\mathrm{T}$} & \control{} \wire[d][3]{q} &  &  &  \\
\lstick{$f_\mathrm{J}$} &  & \control{} \wire[d][3]{q} &  &   \\
\lstick{$y$} & \control{} & \control{} &  &  \\
\lstick{$f_\mathrm{T}y:\ket0$} & \setwiretype{n} & \setwiretype{q} &  &  \\
\lstick{$f_\mathrm{J}y:\ket0$} & \setwiretype{n} &  & \setwiretype{q} & 
\end{quantikz}}\par
\end{minipage}
\caption{Preparation blocks at five orbitals. The top row is $\mathsf {Prep}_\mu$ of \Cref{eq:aim-query-order}, in the order it runs. $\mathsf Q_p$ writes its six decoder fields into $\mathcal A$, and $\mathsf F$ writes the label fields of $\mathcal F$ directly. The address rows of the decoder are $p_3p_2p_1=\ell_\mathrm{u}$ and $p_0=\varrho$. Only the row-address uniform state and the lookup of the row fields borrow the clean scratch $\mathcal W$, and inverse preparation borrows it again, more deeply, for the measured erasure of the lookup. The two blank corners of the decoder build its single stored product $\wb=p_2p_1p_0$ in the two stages that a cubic monomial needs. The intermediate $\wb_1=p_2p_1$, the only work bit with a qubit of its own, is measured away once $\wb$ has been written into the field $(\ell_p)_1$, which then carries it. The stored product therefore closes in $\mathsf Q_p^\dagger$, after $\mathsf F^\dagger$ has closed the label products that read the decoder fields. The five blank corners of $\mathsf F$ are multi-bit openings, and the remaining corners are single-\CNOT{} Clifford scratch.}
\label{fig:aim-preparation-blocks}
\end{figure*}
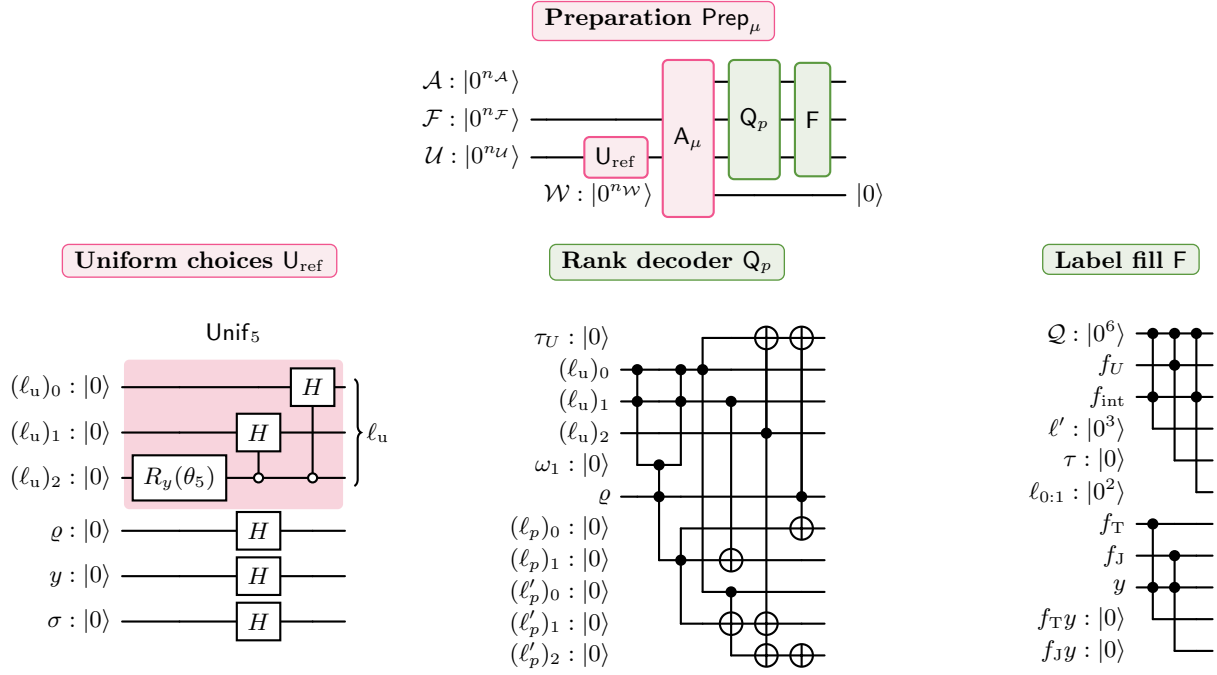

\par\medskip\noindent\begin{minipage}{\linewidth}
\centering
\begin{tabular}{r|rrrrrrrrrr}
\toprule
$p$ & 0 & 1 & 2 & 3 & 4 & 5 & 6 & 7 & 8 & 9\\
\midrule
$\ell_\mathrm{u}$ & 0 & 0 & 1 & 1 & 2 & 2 & 3 & 3 & 4 & 4\\
$\tau_U$ & 0 & 1 & 1 & 0 & 0 & 1 & 1 & 0 & 1 & 0\\
$\ell_p$ & 0 & 1 & 0 & 1 & 2 & 3 & 2 & 0 & 0 & 1\\
$\ell'_p$ & 4 & 4 & 3 & 3 & 4 & 4 & 3 & 1 & 2 & 2\\
\bottomrule
\end{tabular}
\captionof{table}{Five-orbital decoder $\mathsf Q_p$, with $p_3p_2p_1=\ell_\mathrm{u}$ and $p_0=\varrho$. Pair rows use $(\ell_p,\ell'_p)$; $U$ rows use $\ell_p$ when $\tau_U=0$ and $\ell'_p$ when $\tau_U=1$. Each orbital is selected by two ranks, so both refinements are uniform. Ranks 10--15 are unreachable. The circuit of \Cref{fig:aim-preparation-blocks} uses one constant, twelve \CNOT{}s and two openings to form the stored product $\wb=p_2p_1p_0$.}
\label{tab:aim-rank-decoder}
\end{minipage}\par\medskip

\paragraph{Label fill and cost.} $\mathsf F$ writes eight products into the zeroed label fields: the three bits of $\ell'$, $\tau$, the two bits $\ell_{0:1}$, $f_\mathrm{T}y$ and $f_\mathrm{J}y$. The five multi-bit corners of \Cref{fig:aim-preparation-blocks} show these openings. Preparation retains the six decoder fields, the label and the uniform choices. The decoder uses two openings forward; its inverse reopens $\wb_1$ once and closes both products by measurement. Including the label fill and uniform block gives
\begin{equation}
\Tref\leq8+4+4\cdot8+8+2\Trot=52+2\Trot.
\label{eq:aim-refinement-cost}
\end{equation}
Here $\Trot$ is the cost of one synthesised $R_y(\theta_5)$ at $\theta_5=2\arccos(2/\sqrt5)$, and $8\,T$ pays for the controlled Hadamards of $\Unif{5}$ and their inverses. The refinement choices that a row does not use sum to unit probability. Exact row probabilities therefore give $\bra0\mathsf {Prep}_\mu^\dagger\,\SELECT\,\mathsf {Prep}_\mu\ket0=H/\Lambda$.

\subsection{One-body circuit}
\label{app:aim-onebody-bridge}
\label{app:bridge-instances}

The one-body circuit first selects the position in every chain and then selects the chain, \Cref{fig:aim-onebody-main-circuit}. It reads $(k,\ell_\mathsf{B})$ from $\mathcal F$ and the spin coin $\sigma$, and acts on $\mathcal T_\mathsf{B}$. Decoder and closed-\CSWAP{} scratch use $\mathcal W$.

\paragraph{Swap trees and toggle.} Each chain uses the open-boundary network of \Cref{app:nn-network}, with parity-$o$ leaf $v$ holding site $2v+o$. All chains share the address $j$. The odd tree reads $\jod=\lfloor k/2\rfloor$; the even tree reads $\jev=\jod+f_\mathrm{T}k_0$. The toggle therefore increments by $\rho=\carry{0}=f_\mathrm{T}k_0$ and needs no wrap. Onsite rows leave $j$ unchanged and use the endpoint selected by $k_0$. The field width is $m=\lceil\log_2(1+\lfloor n_\mathrm{bath}/2\rfloor)\rceil$. Its $m$ carry products and the target-action product $f_\mathrm{A}k_0$ cost $m+1$ openings. At $n_\mathrm{bath}=1$, the address and carries are absent; only $f_\mathrm{A}k_0$ remains.

\paragraph{Chain-front permutation.} $\PermC_{\ell\sigma}$ is the pruned block tree of \Cref{fig:swap-tree} over the $2n_\mathrm{imp}$ chain fronts. The tilde marks a permutation of whole chains or orbitals rather than of the sites within a chain. Each leaf $a=2\ell+\sigma$ holds $(q_{0\ell\sigma},q_{1\ell\sigma})$, with address bits $a_0=\sigma$ and $a_{t+1}=(\ell_\mathsf{B})_t$. Its root is the front pair of the chain of $(0,\uparrow)$, and a final fixed $\SWAP(q_{10\uparrow},q_{00\downarrow})$ places this root pair on $\mathcal T_\mathsf{B}$; the inverse applies this swap first. The target action absorbs $\SWAP^{k_0}$. Counting both lanes of every block, the trees and chain-front permutation contain
\begin{equation}
2n_\mathrm{imp}(n_\mathrm{bath}-1)+2(2n_\mathrm{imp}-1)
=2n_\mathrm{imp}(n_\mathrm{bath}+1)-2
\label{eq:aim-onebody-general-lanes}
\end{equation}
forward \CSWAP{}s.

\begin{figure}[t]
\centering
\crayontitle{CrayonGreen}{\textbf{One-body control generator $\mathsf G_\mathsf{B}$, shown for $m=3$}}\par
\resizebox{\columnwidth}{!}{%
{\footnotesize
\begin{quantikz}[row sep={0.36cm,between origins},column sep=0.22cm]
\lstick{$f_\mathrm{T}$} &  & \control{} \wire[d][15]{q} & \control{} \wire[d][16]{q} & \control{} \wire[d][17]{q} & \control{} \wire[d][18]{q} &  & \control{} \wire[d][7]{q} &  &  &  &  \\
\lstick{$f_\mathrm{A}$} & \control{} \wire[d][13]{q} &  &  &  &  &  &  &  &  &  &  \\
\lstick{$k_0$} & \control{} &  &  &  &  &  & \control{} &  &  &  &  \\
\lstick{$k_{\geq1}$} &  &  &  &  &  & \control[style={draw=CrayonGreen,fill=CrayonGreen}]{}\wire[d][1][CrayonGreen]{q} &  &  &  &  &  \\
\lstick{$j_0$} &  &  &  &  &  & \gate[wires=3,style={draw=CrayonGreen,fill=CrayonGreenFill,rounded corners=2pt}]{\oplus k_{\geq1}} &  & \control{} \wire[d][4]{q} &  &  &  \\
\lstick{$j_1$} &  &  &  &  &  &  &  &  & \control{} \wire[d][4]{q} &  &  \\
\lstick{$j_2$} &  &  &  &  &  &  &  &  &  &  &  \\
\lstick{$\carry{0}:\ket0$} & \setwiretype{n} &  &  &  &  &  &  & \control{} \setwiretype{q} &  &  & \rstick{$f_\mathrm{T}k_0$} \\
\lstick{$\carry{1}:\ket0$} & \setwiretype{n} &  &  &  &  &  &  &  & \control{} \setwiretype{q} &  & \rstick{$f_\mathrm{T}k_0j_0$} \\
\lstick{$\carry{2}:\ket0$} & \setwiretype{n} &  &  &  &  &  &  &  &  & \setwiretype{q} & \rstick{$f_\mathrm{T}k_0j_0j_1$} \\
\lstick{$(\ell_\mathsf{B})_0$} &  & \control{} &  &  &  & \control{} \wire[d][9]{q} &  &  &  &  &  \\
\lstick{$(\ell_\mathsf{B})_1$} &  &  & \control{} &  &  &  & \control{} \wire[d][9]{q} &  & \control{} \wire[d][11]{q} & \control{} \wire[d][12]{q} &  \\
\lstick{$(\ell_\mathsf{B})_2$} &  &  &  & \control{} &  &  &  & \control{} \wire[d][9]{q} &  &  &  \\
\lstick{$\sigma$} &  &  &  &  & \control{} &  &  &  &  &  &  \\
\lstick{$f_\mathrm{A}k_0:\ket0$} & \setwiretype{n} & \setwiretype{q} &  &  &  &  &  &  &  &  & \rstick{$f_\mathrm{A}k_0$} \\
\lstick{$\rep_0:\ket0$} & \setwiretype{n} &  & \setwiretype{q} &  &  &  &  &  & \control{} &  & \rstick{$f_\mathrm{T}(\ell_\mathsf{B})_0$} \\
\lstick{$\rep_1:\ket0$} & \setwiretype{n} &  &  & \setwiretype{q} &  &  &  &  &  &  & \rstick{$f_\mathrm{T}(\ell_\mathsf{B})_1$} \\
\lstick{$\rep_2:\ket0$} & \setwiretype{n} &  &  &  & \setwiretype{q} &  &  &  &  &  & \rstick{$f_\mathrm{T}(\ell_\mathsf{B})_2$} \\
\lstick{$\rep_\sigma:\ket0$} & \setwiretype{n} &  &  &  &  & \control{} \setwiretype{q} & \control{} & \control{} &  &  & \rstick{$f_\mathrm{T}\sigma$} \\
\lstick{$\rep_{\sigma0}:\ket0$} & \setwiretype{n} &  &  &  &  &  & \setwiretype{q} &  &  & \control{} & \rstick{$f_\mathrm{T}\sigma(\ell_\mathsf{B})_0$} \\
\lstick{$\rep_{\sigma1}:\ket0$} & \setwiretype{n} &  &  &  &  &  &  & \setwiretype{q} &  &  & \rstick{$f_\mathrm{T}\sigma(\ell_\mathsf{B})_1$} \\
\lstick{$\rep_{\sigma2}:\ket0$} & \setwiretype{n} &  &  &  &  &  &  &  & \setwiretype{q} &  & \rstick{$f_\mathrm{T}\sigma(\ell_\mathsf{B})_2$} \\
\lstick{$\rep_{01}:\ket0$} & \setwiretype{n} &  &  &  &  &  &  &  &  & \setwiretype{q} & \rstick{$f_\mathrm{T}(\ell_\mathsf{B})_0(\ell_\mathsf{B})_1$} \\
\lstick{$\rep_{\sigma01}:\ket0$} & \setwiretype{n} &  &  &  &  &  &  &  &  &  & \setwiretype{q} \rstick{$f_\mathrm{T}\sigma(\ell_\mathsf{B})_0(\ell_\mathsf{B})_1$}
\end{quantikz}}%
}\par
\caption{One-body control generator $\mathsf G_\mathsf{B}$. It derives the two endpoint addresses and the predicates that the bridged chain-front permutation needs. The first column copies the upper bits of $k$ into $j$. The frozen-carry ladder writes $\carry{0}=f_\mathrm{T}k_0$ and $\carry{t}=\carry{t-1}j_{t-1}$, at one opening per bit. The remaining nine openings form the repair word $\rep$. Its subscripts name the bits of $\ell_\mathsf{B}$ multiplied into $f_\mathrm{T}$, and $\rep_\sigma=f_\mathrm{T}\sigma$ is shared by the four products below it. Each label on the right names the product that its opening holds. \Cref{fig:aim-onebody-repair-bridge} consumes these products.}
\label{fig:aim-onebody-generator}
\end{figure}
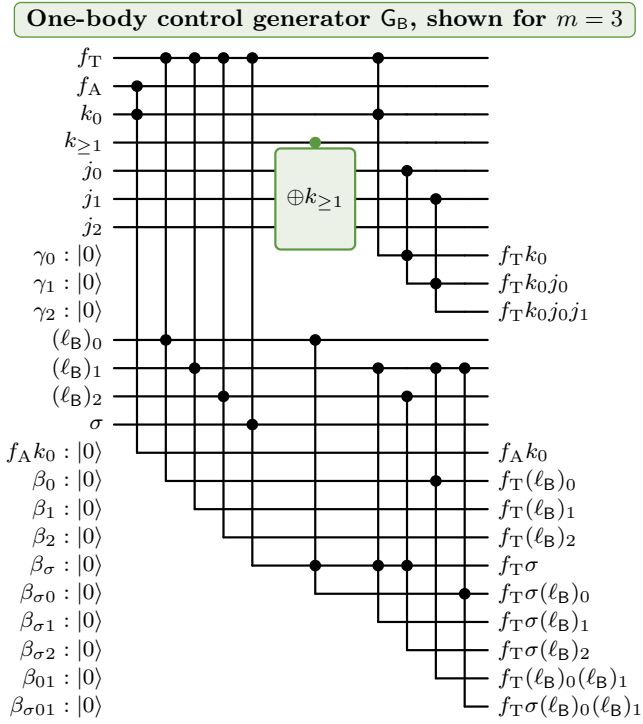

\paragraph{Control generator.} $\mathsf G_\mathsf{B}$ copies the upper bits of $k$ into $j$ and writes the carries of its toggle, the target-action bit $f_\mathrm{A}k_0$ and the nine repair predicates $\rep$ of the chain-front permutation, \Cref{fig:aim-onebody-generator}.

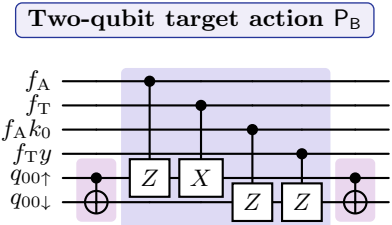
\begin{figure}[t]
\centering
\crayontitle{CrayonBlue}{\textbf{Two-qubit target action $\Pb$}}\par\smallskip
\begin{quantikz}[row sep={0.32cm,between origins},column sep=0.28cm]
\lstick{$f_\mathrm{A}$} &  & \ctrl{4}\gategroup[wires=6,steps=4,style={draw=none,fill=CrayonBlueWash,rounded corners=2pt,inner sep=0pt},background]{} &[-.15cm]  &[-.15cm]  &[-.15cm]  &  & \\
\lstick{$f_\mathrm{T}$} &  &  & \ctrl{3} &  &  &  & \\
\lstick{$f_\mathrm{A}k_0$} &  &  &  & \ctrl{3} &  &  & \\
\lstick{$f_\mathrm{T}y$} &  &  &  &  & \ctrl{2} &  & \\
\lstick{$q_{00\uparrow}$} & \ctrl{1}\gategroup[wires=2,steps=1,style={draw=none,fill=CrayonVioletWash,rounded corners=2pt,inner xsep=0pt,inner ysep=-3pt},background]{} & \gate{Z} & \gate{X} &  &  & \ctrl{1}\gategroup[wires=2,steps=1,style={draw=none,fill=CrayonVioletWash,rounded corners=2pt,inner xsep=0pt,inner ysep=-3pt},background]{} & \\
\lstick{$q_{00\downarrow}$} & \targ{} &  &  & \gate{Z} & \gate{Z} & \targ{} &
\end{quantikz}\par
\caption{One-body target action $\Pb$, the selected Pauli in the parity frame $\mathsf E_\mathsf{B}=\CNOT(q_{00\uparrow},q_{00\downarrow})$. Here $f_\mathrm{T}$ is the common $X$ exponent, and the two $Z$ exponents come from live label bits. Inside the frame, they are $f_\mathrm{A}$ on $q_{00\uparrow}$ and $f_\mathrm{A}k_0\oplus f_\mathrm{T}y$ on $q_{00\downarrow}$. Undoing the frame gives the exponents $f_\mathrm{A}\oplus f_\mathrm{A}k_0\oplus f_\mathrm{T}y$ and $f_\mathrm{A}k_0\oplus f_\mathrm{T}y$, which \Cref{fig:aim-repair-application}(b) applies directly. An onsite row thus phases the endpoint that $k_0$ selects, and a hopping row applies $XX$ or $YY$ as $y$ dictates. Every gate is Clifford, so the target action costs no $T$.}
\label{fig:aim-onebody-action}
\end{figure}

\paragraph{Target action.} $\Pb$ reads the label in the parity frame $\mathsf E_\mathsf{B}=\CNOT(q_{00\uparrow},q_{00\downarrow})$, with common $X$ exponent $f_\mathrm{T}$, \Cref{fig:aim-onebody-action}. The onsite factor is always $Z$, so $f_\mathrm{A}k_0$ directs it to the correct endpoint without an orientation exchange. No mask register is needed, and inactive rows act trivially.

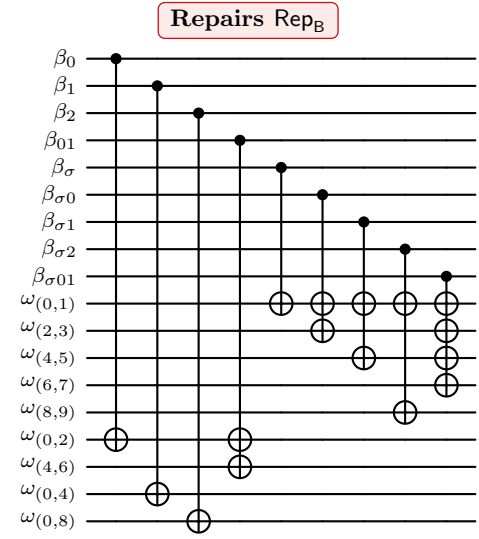
\begin{figure}[t]
\centering
\crayontitle{CrayonRed}{\textbf{Repairs $\mathsf{Rep}_\mathsf{B}$}}\par
{\footnotesize
\begin{quantikz}[row sep={0.36cm,between origins}, column sep=0.22cm]
\lstick{$\rep_0$} & \ctrl{14} &  &  &  &  &  &  &  &  &   \\
\lstick{$\rep_1$} &  & \ctrl{15} &  &  &  &  &  &  &  &   \\
\lstick{$\rep_2$} &  &  & \ctrl{15} &  &  &  &  &  &  &   \\
\lstick{$\rep_{01}$} &  &  &  & \ctrl{11} &  &  &  &  &  &   \\
\lstick{$\rep_\sigma$} &  &  &  &  & \ctrl{5} &  &  &  &  &   \\
\lstick{$\rep_{\sigma0}$} &  &  &  &  &  & \ctrl{4} &  &  &  &   \\
\lstick{$\rep_{\sigma1}$} &  &  &  &  &  &  & \ctrl{3} &  &  &   \\
\lstick{$\rep_{\sigma2}$} &  &  &  &  &  &  &  & \ctrl{2} &  &   \\
\lstick{$\rep_{\sigma01}$} &  &  &  &  &  &  &  &  & \ctrl{1} &   \\
\lstick{$\wb_{(0,1)}$} &  &  &  &  & \targ{} & \targ{} \wire[d][1]{q} & \targ{} \wire[d][2]{q} & \targ{} \wire[d][4]{q} & \targ{} \wire[d][1]{q} &   \\
\lstick{$\wb_{(2,3)}$} &  &  &  &  &  & \targ{} &  &  & \targ{} \wire[d][1]{q} &   \\
\lstick{$\wb_{(4,5)}$} &  &  &  &  &  &  & \targ{} &  & \targ{} \wire[d][1]{q} &   \\
\lstick{$\wb_{(6,7)}$} &  &  &  &  &  &  &  &  & \targ{} &   \\
\lstick{$\wb_{(8,9)}$} &  &  &  &  &  &  &  & \targ{} &  &   \\
\lstick{$\wb_{(0,2)}$} & \targ{} &  &  & \targ{} \wire[d][1]{q} &  &  &  &  &  &   \\
\lstick{$\wb_{(4,6)}$} &  &  &  & \targ{} &  &  &  &  &  &   \\
\lstick{$\wb_{(0,4)}$} &  & \targ{} &  &  &  &  &  &  &  &   \\
\lstick{$\wb_{(0,8)}$} &  &  & \targ{} &  &  &  &  &  &  &
\end{quantikz}}\par
\caption{Repairs $\mathsf{Rep}_\mathsf{B}$, driven by the products that $\mathsf G_\mathsf{B}$ of \Cref{fig:aim-onebody-generator} holds open. Each line $\wb_{(a,b)}$ stands for the two held work bits of the block \CSWAP{} that exchanges the fronts of the chains with addresses $a$ and $b$, and every $\oplus$ acts on both. All sixteen gates are \CNOT{}s, so $\mathsf{Rep}_\mathsf{B}$ costs no $T$ and its columns commute. They supply the negated address factors of \Cref{eq:aim-transport-indicator} as free control parities.}
\label{fig:aim-onebody-repair-bridge}
\end{figure}

\paragraph{Bridged chain fronts.} At five orbitals, the nine block \CSWAP{}s of $\PermC_{\ell\sigma}$ retain eighteen work bits across $\Pb$. Hopping flips both selected endpoints, so both lanes of block $e$ use the repair $f_\mathrm{T}\rep_e$ of \Cref{eq:aim-shared-repair}, with $a=2\ell_\mathsf{B}+\sigma$ in \Cref{eq:aim-transport-indicator}. The nine products $\rep_0,\rep_\sigma,\ldots$ prepared by $\mathsf G_\mathsf{B}$ already contain the factor $f_\mathrm{T}$. They supply these predicates and their complemented factors through \CNOT{} parities. The repair circuit is therefore Clifford, \Cref{fig:aim-onebody-repair-bridge}.

The tree \CSWAP{}s stay closed for the reason of \Cref{app:repair-count}: each exchanges a single qubit, so a bridged one would need its own repair, the indicator of \Cref{eq:aim-transport-indicator} restricted to the selected chain. Only about $\log_2 n_\mathrm{bath}$ of these repairs, six at $n_\mathrm{bath}=100$, follow as \CNOT{} parities of the others and of the products that $\mathsf G_\mathsf{B}$ already holds. Bridging the trees would thus save that many openings, at one live work bit per tree \CSWAP{}.

\paragraph{Cost.} At $(5,100)$, the 990 \CSWAP{}s of the swap trees, \Cref{eq:aim-onebody-general-lanes}, cost $7920\,T$ as closed pairs. The eighteen lanes of the chain-front permutation, the two \CSWAP{}s of each of its nine blocks, cost $72\,T$ as bridged pairs, \Cref{fig:aim-repair-application}(b), with $36\,T$ of shared repairs. Address arithmetic, \Cref{app:address-toggle}, is additional.

\subsection{Interaction circuit}
\label{app:aim-query-interaction}

The interaction circuit acts only on the $2n_\mathrm{imp}$ impurity qubits, where every pair of orbitals interacts. It combines the all-to-all network with the parity frame and the phase walks of \Cref{app:frames-walks}, so that only the off-diagonal part of each term is moved.

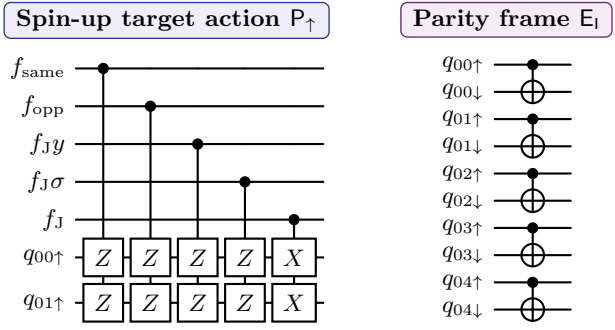
\begin{figure}[t]
\centering
\begin{minipage}[t]{0.62\linewidth}
\vspace{0pt}
\centering
\crayontitle{CrayonBlue}{\textbf{Spin-up target action $\Pup$}}\par\smallskip
{\small
\begin{quantikz}[row sep={0.5cm,between origins},column sep=0.10cm]
\lstick{$f_\mathrm{same}$} & \control{} \wire[d][5]{q} &  &  &  &  &  \\
\lstick{$f_\mathrm{opp}$} &  & \control{} \wire[d][4]{q} &  &  &  &  \\
\lstick{$f_\mathrm{J}y$} &  &  & \control{} \wire[d][3]{q} &  &  &  \\
\lstick{$f_\mathrm{J}\sigma$} &  &  &  & \control{} \wire[d][2]{q} &  &  \\
\lstick{$f_\mathrm{J}$} &  &  &  &  & \control{} \wire[d][1]{q} &  \\
\lstick{$q_{00\uparrow}$} & \gate{Z} \wire[d][1]{q} & \gate{Z} \wire[d][1]{q} & \gate{Z} \wire[d][1]{q} & \gate{Z} \wire[d][1]{q} & \gate{X} \wire[d][1]{q} &  \\[0.1cm]
\lstick{$q_{01\uparrow}$} & \gate{Z} & \gate{Z} & \gate{Z} & \gate{Z} & \gate{X} &
\end{quantikz}}\par
\end{minipage}
\hfill
\begin{minipage}[t]{0.34\linewidth}
\vspace{0pt}
\centering
\crayontitle{CrayonViolet}{\textbf{Parity frame $\mathsf E_\mathsf{I}$}}\par\smallskip
{\small
\begin{quantikz}[row sep={0.36cm,between origins},column sep=0.34cm]
\lstick{$q_{00\uparrow}$} & \ctrl{1} & \\
\lstick{$q_{00\downarrow}$} & \targ{} & \\
\lstick{$q_{01\uparrow}$} & \ctrl{1} & \\
\lstick{$q_{01\downarrow}$} & \targ{} & \\
\lstick{$q_{02\uparrow}$} & \ctrl{1} & \\
\lstick{$q_{02\downarrow}$} & \targ{} & \\
\lstick{$q_{03\uparrow}$} & \ctrl{1} & \\
\lstick{$q_{03\downarrow}$} & \targ{} & \\
\lstick{$q_{04\uparrow}$} & \ctrl{1} & \\
\lstick{$q_{04\downarrow}$} & \targ{} &
\end{quantikz}}\par
\end{minipage}
\caption{Left, spin-up target action $\Pup$. Four $Z$ controls supply $\zeta_\uparrow$ of \Cref{eq:aim-semantic-interaction-masks} to both targets, and $f_\mathrm{J}$ supplies the common $X$. Every gate is Clifford. Right, parity frame $\mathsf E_\mathsf{I}$, one \CNOT{} per orbital from its spin-up front to its spin-down front. The five \CNOT{}s act on disjoint pairs. The layer thus has depth one, is its own inverse, and costs no ancillary qubit and no $T$. Conjugating by it gives the split of \Cref{eq:aim-generic-interaction-operator}.}
\label{fig:aim-spin-up-action}
\label{fig:aim-parity-frame}
\end{figure}

\begin{figure}[t]
\centering
\crayontitle{CrayonBlue}{\textbf{Spin-down stage $\PermC_\downarrow^{-1}\Pdn\PermC_\downarrow$ at $n_\mathrm{imp}=5$}}\par\smallskip
{\footnotesize
\begin{quantikz}[row sep={0.35cm,between origins},column sep=0.20cm]
\lstick{$\zeta_{\ell\downarrow}$} &  &[-.1cm]  &  & \control{}\wire[d][7]{q}\gategroup[14,steps=17,style={draw=none,fill=CrayonBlueWash,rounded corners=2pt,inner xsep=1pt,inner ysep=0pt},background]{} &[-.3cm]  &[-.3cm] \ctrl{7} &[-.1cm]  &[-.3cm]  &[-.3cm]  &[-.3cm]  &[-.3cm]  &[-.2cm]  &[-.1cm] \ctrl{7} &[-.2cm] \control{}\wire[d][7]{q} &[-.1cm]  &[-.3cm]  &[-.3cm]  &[-.3cm]  &[-.3cm]  &[-.2cm]  &  &  &[-.1cm]  &  \\
\lstick{$\zeta_{\ell'\downarrow}$} &  &  &  &  &  &  &  &  &  &  &  &  &  &  & \control{}\wire[d][6]{q} &  & \ctrl{6} &  & \ctrl{6} & \control{}\wire[d][6]{q} &  &  &  &  \\
\lstick{$\ell_0\oplus\ell_1$} &  &  &  & \control{} &  &  &  &  &  &  &  &  &  & \control{} &  &  &  &  &  &  &  &  &  &  \\
\lstick{$\ell_1\oplus\ell_2$} &  &  &  &  &  &  & \control{}\wire[d][5]{q} &  &  &  &  & \control{}\wire[d][5]{q} &  &  &  &  &  &  &  &  &  &  &  &  \\
\lstick{$\ell'_0\oplus\ell'_1$} &  &  &  &  &  &  &  &  &  &  &  &  &  &  & \control{} &  &  &  &  & \control{} &  &  &  &  \\
\lstick{$\ell_1\oplus\ell'_1\oplus\ell'_2$}
& \ctrl{9}\gategroup[11,steps=2,style={draw=none,fill=CrayonYellowWash,rounded corners=2pt,inner xsep=1pt,inner ysep=2pt},background]{}
&  &  &  &  &  &  &  &  &  &  &  &  &  &  &  &  &  &  &  &  &  
\gategroup[11,steps=2,style={draw=none,fill=CrayonYellowWash,rounded corners=2pt,inner xsep=1pt,inner ysep=2pt},background]{}
& \ctrl{9} &  \\
\lstick{$\ell_2\oplus\ell'_2$} &  & \ctrl{9} &  &  &  &  &  &  &  &  &  &  &  &  &  &  &  &  &  &  &  & \ctrl{9} &  &  \\
\lstick{$\wb_{\mathrm{p},0}:\ket0$} & \setwiretype{n} &  &  &  & \ctrl{4}\setwiretype{q} & \targ{} & \control{} &  & \ctrl{1} &  & \ctrl{1} & \control{} & \targ{} &  & \setwiretype{n} & \ctrl{3}\setwiretype{q} & \targ{} & \ctrl{5} & \targ{} &  & \setwiretype{n} &  &  &  \\
\lstick{$\wb_{\mathrm{p},1}:\ket0$} & \setwiretype{n} &  &  &  &  &  &  & \ctrl{5}\setwiretype{q} & \targ{} & \ctrl{1} & \targ{} &  & \setwiretype{n} &  &  &  &  &  &  &  &  &  &  &  \\[.2cm]
\lstick{$q_{00\downarrow}$} &  &  &  &  &  &  &  &  &  & \gate{Z} &  &  &  &  &  &  &  &  &  &  &  &  &  &  \\[.05cm]
\lstick{$q_{01\downarrow}$} & \swap{1} &  &  &  &  &  &  &  &  &  &  &  &  &  &  & \gate{Z} &  &  &  &  &  &  & \swap{1} &  \\[.05cm]
\lstick{$q_{02\downarrow}$} & \targX{} &  &  &  & \gate{Z} &  &  &  &  &  &  &  &  &  &  &  &  &  &  &  &  &  & \targX{} &  \\[.05cm]
\lstick{$q_{03\downarrow}$} &  & \swap{1} &  &  &  &  &  &  &  &  &  &  &  &  &  &  &  & \gate{Z} &  &  &  & \swap{1} &  &  \\[.05cm]
\lstick{$q_{04\downarrow}$} &  & \targX{} &  &  &  &  &  & \gate{Z} &  &  &  &  &  &  &  &  &  &  &  &  &  & \targX{} &  &  \\[.2cm]
\lstick{$\wb_{\mathrm{c},0}:\ket0$} & \setwiretype{n} & \setwiretype{q} &  &  &  &  &  &  &  &  &  &  &  &  &  &  &  &  &  &  &  &  &  & \setwiretype{n} \\
\lstick{$\wb_{\mathrm{c},1}:\ket0$} & \setwiretype{n} &  &\setwiretype{q} &  &  &  &  &  &  &  &  &  &  &  &  &  &  &  &  &  &  & &  \setwiretype{n}  &
\end{quantikz}}\par
\caption{Spin-down stage at $n_\mathrm{imp}=5$, with the singleton $\{0\}$ and the complementary pairs $\{1,2\},\{3,4\}$. The two yellow panels are the bridged permutation $\PermC_\downarrow$ and its inverse. The blue panel between them is the phase application $\Pdn$. Each bridged \CSWAP{} of a complementary pair holds one work bit $\wb_{\mathrm{c},e}$, drawn below the system in pair order and erased by the matching corner on the right. The two phase-walk products $\wb_{\mathrm{p},e}$ sit above the system. The walks phase the spin-down front in the part that each address selects. They reuse $\wb_{\mathrm{p},0}$, and open $\wb_{\mathrm{p},1}$ only for the deeper branch. All five closings are measured and cost nothing, because only $Z$ gates separate each stored product from its corner. The walk addresses are the parities that drive the swap trees of \Cref{fig:aa-network}, with $(j,j')=(\ell,\ell')$.}
\label{fig:aim-interaction-walks}
\end{figure}
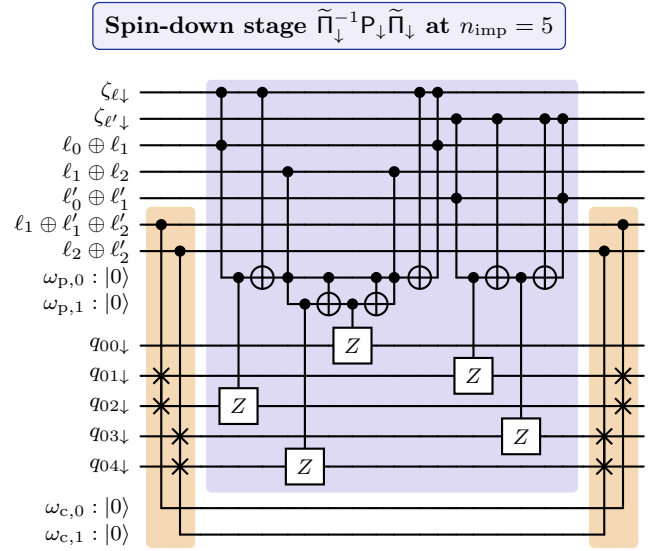

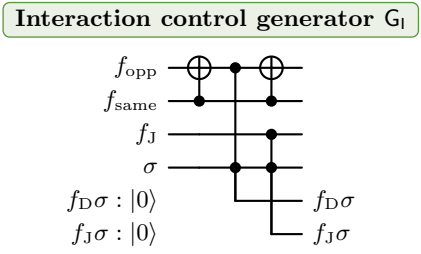
\begin{figure}[t]
\centering
\crayontitle{CrayonGreen}{\textbf{Interaction control generator $\mathsf G_\mathsf{I}$}}\par\smallskip
\begin{quantikz}[row sep={0.44cm,between origins},column sep=0.24cm]
\lstick{$f_\mathrm{opp}$} & \targ{} & \ctrl{4} & \targ{} & \\
\lstick{$f_\mathrm{same}$} & \ctrl{-1} &  & \ctrl{-1} & \\
\lstick{$f_\mathrm{J}$} &  &  & \ctrl{3} & \\
\lstick{$\sigma$} &  & \ctrl{1} & \ctrl{2} & \\
\lstick{$f_\mathrm{D}\sigma:\ket0$} & \setwiretype{n} &  & \setwiretype{q} & \rstick{$f_\mathrm{D}\sigma$} \\
\lstick{$f_\mathrm{J}\sigma:\ket0$} & \setwiretype{n} &  &  & \setwiretype{q} \rstick{$f_\mathrm{J}\sigma$}
\end{quantikz}\par
\caption{Interaction control generator $\mathsf G_\mathsf{I}$. It opens the two stored products and nothing else, and it forms $f_\mathrm{D}$ in a spare flag with a pair of \CNOT{}s. At five orbitals, the \CSWAP{} controls need no retained product, so there is no lookup and no loaded control word.}
\label{fig:aim-interaction-generator}
\end{figure}

\paragraph{Parity frame.} The frame gathers the $X$ factors on spin-up qubits and leaves the spin-down factors diagonal. It applies the self-inverse layer $\mathsf E_\mathsf{I}=\prod_\ell\CNOT(q_{0\ell\uparrow},q_{0\ell\downarrow})$, \Cref{eq:aim-interaction-frame,fig:aim-parity-frame}. The operator used by the circuit is written as
\begin{equation}
\mathsf E_\mathsf{I}P^{(\mathrm{I})}\mathsf E_\mathsf{I}
=(-1)^{f_\mathrm{J}}\underbrace{X_{0\ell\uparrow}^{f_\mathrm{J}}X_{0\ell'\uparrow}^{f_\mathrm{J}}
Z_{0\ell\uparrow}^{\zeta_\uparrow}Z_{0\ell'\uparrow}^{\zeta_\uparrow}}_{\text{spin-up fronts}}\ 
\underbrace{Z_{0\ell\downarrow}^{\zeta_{\ell\downarrow}}Z_{0\ell'\downarrow}^{\zeta_{\ell'\downarrow}}}_{\text{spin-down fronts}}.
\label{eq:aim-generic-interaction-operator}
\end{equation}
Before the frame, in target order $(\ell\!\uparrow,\ell\!\downarrow,\ell'\!\uparrow,\ell'\!\downarrow)$, $\mathsf I$ applies a common $X^{f_\mathrm{J}}$ to all four targets. It also applies the $Z$ exponents $(\zeta_{\ell\downarrow}\oplus\zeta_\uparrow,\zeta_{\ell\downarrow},\zeta_\uparrow\oplus\zeta_{\ell'\downarrow},\zeta_{\ell'\downarrow})$, built from
\begin{equation}
\begin{aligned}
\zeta_{\ell\downarrow}&=f_U\oplus\tau\oplus f_\mathrm{D}\sigma\oplus f_\mathrm{J}\sigma,\\
\zeta_{\ell'\downarrow}&=f_\mathrm{opp}\oplus f_\mathrm{J}\oplus\tau\oplus f_\mathrm{D}\sigma\oplus f_\mathrm{J}\sigma,\\
\zeta_\uparrow&=f_\mathrm{D}\oplus f_\mathrm{J}(y\oplus\sigma).
\end{aligned}
\label{eq:aim-semantic-interaction-masks}
\end{equation}
Here $f_\mathrm{D}=f_\mathrm{opp}\oplus f_\mathrm{same}$ is the density flag of \Cref{tab:aim-factored-coefficients}. Only $f_\mathrm{D}\sigma$ and $f_\mathrm{J}\sigma$ are stored, by $\mathsf G_\mathsf{I}$ of \Cref{fig:aim-interaction-generator}, and every other term is a control parity of live bits, so \Cref{fig:aim-spin-up-action,fig:aim-interaction-walks} apply the constituents directly and no mask register is built. All exponents vanish on one-body rows. The two $Z$ gates of \Cref{fig:aim-interaction-main-circuit} correct the raw Hund phase while $f_\mathrm{J}\sigma$ is live. On a $U$ row, the exponents reduce to $(1\oplus\tau,1\oplus\tau,\tau,\tau)$, with no $X$ factor. The bit $\tau$ thus selects which target slot carries the orbital that the row acts on, and the orbital of the other slot is a spectator. The $\ell$ slot of a $U$ row may therefore hold any orbital of the pair that the rank already carries.

\paragraph{Spin-up permutation.} $\PermC_\uparrow$ uses the all-to-all network of \Cref{app:aa-network} on $n_\mathrm{imp}$ spin-up fronts. In \Cref{fig:aa-network}, site $o$ is $q_{0o\uparrow}$, the selected pair is $(\ell,\ell')$, and the two tree roots are $q_{00\uparrow},q_{01\uparrow}$. It uses $\lceil\tfrac32(n_\mathrm{imp}-2)\rceil$ \CSWAP{}s. At five orbitals, the five gates need no retained control product because $\Npre(5)=0$. \newedit{At other sizes, the $\Npre(n_\mathrm{imp})$ retained products of \Cref{eq:aim-pair-network} are counted in \Cref{eq:aim-interaction-openings,eq:aim-permutation-scaling}.}

\paragraph{Spin-down stage.} $\PermC_\downarrow$ uses one bridged \CSWAP{} per complementary pair of orbitals, \Cref{app:aa-network}, to place the lower and upper orbitals on opposite sides. Phase walks then reach the selected fronts, addressed by the parities that control the swap trees of $\PermC_\uparrow$. At five orbitals, the walk for $\zeta_{\ell\downarrow}$ visits three parts of $\{0\},\{1,2\},\{3,4\}$ and that for $\zeta_{\ell'\downarrow}$ visits two, \Cref{fig:aim-interaction-walks}. Only diagonal gates act between the two occurrences of each \CSWAP{}, so its work bit closes without repair. No spin-down front is brought to a fixed qubit of $\mathcal T_\mathsf{I}$.

\paragraph{Control generator and assembly.} $\mathsf G_\mathsf{I}$ opens only the two products $f_\mathrm{D}\sigma$ and $f_\mathrm{J}\sigma$ that the interaction exponents need, \Cref{fig:aim-interaction-generator}. Inside the frame, the circuit of \Cref{fig:aim-interaction-main-circuit} realises the split of \Cref{eq:aim-generic-interaction-operator} as
\begin{equation}
\mathsf E_\mathsf{I}\big((\PermC_\uparrow^{-1}\Pup\PermC_\uparrow)\otimes(\PermC_\downarrow^{-1}\Pdn\PermC_\downarrow)\big)\mathsf E_\mathsf{I},
\end{equation}
The two stages act on disjoint qubits and commute, which permits their interleaving in \Cref{fig:aim-interaction-main-circuit}. The spin-up stage uses closed gates to bring its factors to fixed targets. The spin-down stage holds one bridged work bit per complementary pair while applying its phases in place. The red panel supplies the Hund sign correction. The bath is unchanged.

We count one opening per spin-up \CSWAP{} in each direction, one per bridged complementary pair and one per walk step. The walks and their bridges together cost exactly one opening per spin-up \CSWAP{}, so
\begin{equation}
\Nopen^{(\mathrm{I})}=3\left\lceil\tfrac32(n_\mathrm{imp}-2)\right\rceil+\Npre(n_\mathrm{imp}).
\label{eq:aim-interaction-openings}
\end{equation}
At five orbitals this is fifteen openings and $60\,T$, \newedit{besides the two retained products of $\mathsf G_\mathsf{I}$, four work bits are live at the peak.}

\paragraph{Signs.} Four sign factors, $Z_{\sgn}Z_{f_\mathrm{T}y}$ and $Z_{f_\mathrm{J}y}Z_{f_\mathrm{J}\sigma}$, with $\sgn$ the stored coefficient sign, correct the raw hopping phase $(XZ)(XZ)=-YY$ and the raw Hund phases. The constant $(-1)^{f_\mathrm{J}}$ of \Cref{eq:aim-generic-interaction-operator} is carried by the stored sign of the Hund rows. Both pairs are diagonal in the label and commute with \SELECT{}. They add no work qubit, no $T$ gate and no rotation.

\subsection{Resources}
\label{app:aim-query-resources}

The leading cost comes from the one-body trees and coefficient loading. Bridging reduces the chain-front and interaction costs while retaining work bits. Both target actions and all four sign factors are Clifford. The row-address and rank superpositions supply the rotations; we count them separately from exact $T$ gates.

\paragraph{Permutations.}
The system permutations and their repairs obey
\begin{equation}
\begin{aligned}
\Tperm\leq{}&16n_\mathrm{imp}(n_\mathrm{bath}-1)+12(2n_\mathrm{imp}-1)\\
&+12\left\lceil\tfrac32(n_\mathrm{imp}-2)\right\rceil+4\Npre(n_\mathrm{imp}).
\end{aligned}
\label{eq:aim-permutation-scaling}
\end{equation}
The first term counts the closed tree \CSWAP{}s of \Cref{eq:aim-onebody-general-lanes}. The second counts the bridged chain-front blocks with their shared repairs. The last two count the interaction family of \Cref{eq:aim-interaction-openings}. The tree \CSWAP{}s read the address bits directly. Address arithmetic costs $O(\log(n_\mathrm{bath}+1))$ further openings. By \Cref{eq:aim-transport-indicator}, no repair factor needs a lookup. Preparation adds the $\Tref$ of \Cref{eq:aim-refinement-cost}. Its only superlinear part is the rank decoder, a lookup on the $O(n_\mathrm{imp}^2)$ pair ranks.

\paragraph{Coefficient sampler.}
The remaining cost is the standard alias sampler of Ref.~\cite{Babbush2018} on the $\Rrows$ rows of \Cref{eq:aim-factored-lambda}. It consists of a uniform superposition over exactly $\Rrows$ row addresses, an uncontrolled unary-iteration lookup, a $\mu$-bit comparator and a bridged exchange of the $n_\mathrm{x}$ bits of row fields that $\mathsf A_\mu$ loads, with $n_\mathrm{x}=n_k+n_\ell+7$. The remaining $2n_\ell+3$ fields of $\mathcal F$ are written by $\mathsf F$ afterwards, and they are zero during the exchange. The alternate row fields are held at the same width. Write $\Rrows=2^{s}l$ with $l$ odd. The uniform state then costs $\Nopen^{\mathrm{unif}}=3\lceil\log_2l\rceil-1$ openings and two arbitrary rotations per direction, and nothing when $l=1$. The lookup costs $\Rrows-2$ openings, one fewer than its controlled form, because the top address bit selects the two halves directly.

The erasure of the lookup, the measured uncomputation of Ref.~\cite{Berry2019}, first measures the loaded word, then holds only the nonlinear monomials of the $\lfloor n_R/2\rfloor$ low address bits and walks the $\lceil n_R/2\rceil$ high bits by unary iteration on $\lceil n_R/2\rceil-1$ qubits. It costs $\lceil n_R/2\rceil-1$ more openings than holding the monomials of both halves at once, but about halves the workspace of the erasure, which then stays below the chain-front peak at every plotted size. The sampler therefore contributes $4(\Rrows+2^{\lfloor n_R/2\rfloor}+2^{\lceil n_R/2\rceil}-\lfloor n_R/2\rfloor-5)+8\Nopen^{\mathrm{unif}}+4\mu+4n_\mathrm{x}$ raw $T$ in all. This is linear in the row count up to the logarithmic uniform-state and square-root erasure terms, and $\Theta(n_\mathrm{imp}n_\mathrm{bath})$ at leading order.

\paragraph{Complete query.}
Adding the permutations of \Cref{eq:aim-permutation-scaling} to the sampler, one uncontrolled query with orbital-independent $U$ and $J$ costs
\begin{equation}
\Ttot=O\!\left(n_\mathrm{imp}n_\mathrm{bath}+n_\mathrm{imp}^2+\mu\right).
\label{eq:aim-query-T-scaling}
\end{equation}
Rotation synthesis adds $\Tsynth(\epssynth)$ at accuracy $\epssynth$. The peak workspace is
\begin{equation}
\Apeak=O\!\bigl(\mu+\log(n_\mathrm{imp}n_\mathrm{bath})
+n_\mathrm{imp}+\sqrt{n_\mathrm{imp}n_\mathrm{bath}}\bigr).
\label{eq:aim-query-space-scaling}
\end{equation}
It counts the carries, the temporary products of \SELECT{} and the lookup outputs kept live, against one clean scratch. That scratch is reused by coefficient loading, rank decoding, $\mathsf B$, $\mathsf I$ and lookup erasure. The interaction family contributes only $O(n_\mathrm{imp})$ \CSWAP{}s and $O(\log n_\mathrm{imp})$ stored products to either bound.

\subsection{Comparison}
\label{app:aim-babbush-comparison}

We compare one uncontrolled complete query, comprising \SELECT{} and both directions of \PREPARE{}, at the same model size, coefficient precision $\mu$ and normalisation. All counts include the rotation-synthesis estimate of \Cref{eq:rotation-synthesis-estimate} at $\epssynth=10^{-3}$. Workspace excludes the system register. \Cref{fig:aim-query-resources} compares three constructions with ours.

\paragraph{Comparator constructions.}
The \emph{flat} comparator applies unary iteration from Ref.~\cite{Babbush2018} to the full list of $L=3N+4n_\mathrm{imp}^2-7n_\mathrm{imp}$ non-identity Pauli strings. The \emph{structured} comparator uses the coefficient factoring and indexed selection of that reference, with the same $\Rrows$ rows as \Cref{eq:aim-factored-lambda}. The coefficients do not depend on spin, hopping axis or orbital pair, so uniform refinements supply these choices. A uniform spin bit controls \CSWAP{}s that substitute the spin-down copies of the selection fields. A uniform coin controls a \CSWAP{} of the two hopping indices, and its \SELECT{} applies $YY$ when they arrive in decreasing order and $XX$ otherwise. The orbital pair is a uniform $a$ with $b=a+e+1\bmod n_\mathrm{imp}$, for an offset $e$ uniform over the other $n_\mathrm{imp}-1$ orbitals. Its one-body selection uses two selected-Majorana passes and an indexed $Z$. Hund exchange and pair hopping lie outside the four term types that this selection admits. Each factors into one two-qubit Pauli per orbital, selected by additional impurity passes. In total, the comparator uses $3N-5$ one-body openings and six passes of $n_\mathrm{imp}-1$ openings. \newedit{The one-body block also exchanges its two $n_N$-bit index registers under a single control bit and later reverses the exchange, with $n_N=\lceil\log_2N\rceil$; each direction costs one opening per exchanged qubit pair.} The comparator \SELECT{} thus costs $4(3N+6n_\mathrm{imp}+2n_N-11)$ raw $T$. No pass visits all orbital pairs.

The \emph{sparse} comparator follows Refs.~\cite{Berry2019,Lee2021THC}, tabling the $d=N+n_\mathrm{imp}(n_\mathrm{imp}-1)$ nonzero integrals of the sparse $(pq|rs)$ representation. Its \SELECT{} applies the Majorana pair twice, costing about $16N$ rather than the leading $12N$ of the structured comparator. Measured lookup erasure offsets this extra selection cost. The structured comparator instead reverses its lookup at full cost, as in Ref.~\cite{Babbush2018}\newedit{, which predates the measured erasure of Ref.~\cite{Berry2019}}. Each comparator is built only from the methods of the references it follows, without the constructions of this paper. \newedit{Every \CSWAP{} of a comparator, including those of its alias sampler, is charged one opening per exchanged qubit pair, as in this paper.}

\paragraph{Coefficient loading.}
Our preparation and the structured comparator alias-sample the $\Rrows$ coefficient rows of \Cref{eq:aim-factored-lambda}, the onsite and hopping coefficients of $H_{\mathrm{1b}}$, and the few parameters of $H_{\mathrm{int}}$. The sparse comparator alias-samples its $d$ nonzero integrals, and the flat comparator one amplitude per Pauli string. The flat lookup cost depends on which entries of its alias table repeat, and hence on the coefficient values, so it is counted at every plotted size rather than bounded. The structured lookup loads row fields, alternates and the keep word together. It also loads spin-down copies of the selection fields to implement the snake ordering, adding workspace.

\paragraph{Cost and workspace.}
As the bath grows at fixed orbital count, our complete query asymptotically needs three fifths of the $T$ gates of the cheapest comparator, 40\% fewer. This is the ratio of the leading terms in \Cref{fig:aim-query-resources}(c), $24n_\mathrm{imp}n_\mathrm{bath}$ here against $40n_\mathrm{imp}n_\mathrm{bath}$ for the sparse and structured comparators. \newedit{In the same spirit as the structured comparator the measured erasure of Ref.~\cite{Berry2019}, it is possible to build a variant that lowers its leading term to $32n_\mathrm{imp}n_\mathrm{bath}$, so the saving becomes one quarter; the flat comparator, improved in the same way, remains the most expensive.} All four costs are linear in bath size at fixed orbital count, so the reduction is a constant factor. At five orbitals, the cheapest comparator is sparse at bath size $100$ and structured at $1000$, with the two costs agreeing to within two per cent. The saving is larger at small baths, where subleading terms matter, and our construction costs less at every plotted size. At five orbitals and bath sizes $100$ and $1000$, our query needs \newedit{just over a third} of the $T$ gates of the flat comparator.

Rotation counts remain bounded as the model grows: six here, four for flat and sparse, and eight for structured. Panel~(c) omits these bounded synthesis costs.

Panel~(b) shows the workspace tradeoff. At every plotted size, our peak is set by the one-body chain-front permutation, with its bridged lanes and repair word live. At five orbitals and bath sizes $100$ and $1000$, our peak lies within seven per cent of the sparse comparator's, an eighth to a fifth below the structured comparator's, and about twice the flat comparator's. The reduction in $T$ count does not imply a uniform reduction in workspace.
 
\end{document}